\documentclass[authoryear,sort&compress,final,3p,times,english]{elsarticle}
\usepackage{lineno}
\usepackage{amsmath}
\usepackage{hyperref}
\usepackage{cleveref}
\usepackage{amsthm}
\usepackage{amsfonts}
\usepackage{natbib}
\usepackage{bibunits} 
\usepackage[bf]{caption}
\usepackage{caption}
\usepackage[T1]{fontenc}
\usepackage[utf8]{inputenc}
\usepackage{babel}
\usepackage[export]{adjustbox}
\usepackage{a4,graphicx}
\usepackage{subfig}
\usepackage{float}
\usepackage{array}
\usepackage{setspace}
\usepackage{geometry}
\usepackage{array}
\usepackage{tikz}
\usepackage{xcolor}
\usepackage{color}
\usepackage{tabularx}
\usetikzlibrary{shapes,shapes.geometric,arrows,fit,calc,positioning,automata}
\usetikzlibrary{automata,positioning}
\usepackage{varwidth}
\usepackage{mathrsfs}
\usepackage{epstopdf}
\usepackage[toc,page]{appendix}

\usepackage{algorithm}
\usepackage{algorithmicx}
\usepackage{multicol}
\usepackage{arydshln}
\usepackage{caption}
\usepackage{mathtools}
\usepackage{algpseudocode}
\hypersetup{colorlinks=true,linkcolor=green,filecolor=blue,urlcolor=blue}
\makeatletter
\def\ps@pprintTitle{%
	\let\@oddhead\@empty
	\let\@evenhead\@empty
	\def\@oddfoot{\reset@font\hfil\thepage\hfil}%
	\let\@evenfoot\@oddfoot}

\def\E{\mathbb E}
\def\R{\mathbb R}
\def\Var{\mathbb Var}
\def\Cov{\mathbb Cov}
\def\du{du}
\newcommand{\1}{{\mathbf 1}}
\DeclarePairedDelimiter\abs{\lvert}{\rvert}
\DeclarePairedDelimiter\norm{\lVert}{\rVert}

\DeclareFontEncoding{LS1}{}{}
\DeclareFontEncoding{LS2}{}{\noaccents@}
\DeclareFontSubstitution{LS1}{stix}{m}{n}
\DeclareFontSubstitution{LS2}{stix}{m}{n}

\DeclareMathAlphabet\mathscr{LS1}{stixscr}{m}{n}
\SetMathAlphabet\mathscr{bold}{LS1}{stixscr}{b}{n}
\DeclareMathAlphabet\mathcal{LS2}{stixcal}{m}{n}
\SetMathAlphabet\mathcal{bold}{LS2}{stixcal}{b}{n}

\DeclareMathOperator*{\argmin}{arg\,min}
\DeclareFontFamily{U}{BOONDOX-calo}{\skewchar\font=45 }
\DeclareFontShape{U}{BOONDOX-calo}{m}{n}{ <-> s*[1.05] BOONDOX-r-calo}{}
\DeclareFontShape{U}{BOONDOX-calo}{b}{n}{ <-> s*[1.05] BOONDOX-b-calo}{}
\DeclareMathAlphabet{\mathcalboondox}{U}{BOONDOX-calo}{m}{n}
\SetMathAlphabet{\mathcalboondox}{bold}{U}{BOONDOX-calo}{b}{n}
\DeclareMathAlphabet{\mathbcalboondox}{U}{BOONDOX-calo}{b}{n}

\begin{document}
\begin{frontmatter}
\title{Large-scale modelling and forecasting of ambulance calls in northern Sweden using spatio-temporal log-Gaussian Cox processes}
\author[1]{Fekadu L. Bayisa}
\author[1]{Markus \AA dahl}
\author[1]{Patrik Ryd\'{e}n}  
\author[1,2]{Ottmar Cronie$^{*}$}
\address[1]{Department of Mathematics and Mathematical Statistics, Ume{\aa} University, 901 87, Ume{\aa}, Sweden}
\address[2]{Biostatistics, School of Public Health and Community Medicine, Institute of Medicine, University of Gothenburg, Sweden}
\cortext[author] {Corresponding author.\\\textit{E-mail address:} ottmar.cronie@umu.se/ottmar.cronie@gu.se}

\begin{abstract}
Although ambulance call data typically come in the form of spatio-temporal point patterns, point process-based modelling approaches presented in the literature are scarce. In this paper, we study a unique set of Swedish spatio-temporal ambulance call data, which consist of the spatial (GPS) locations of the calls (within the four northernmost regions of Sweden) and the associated days of occurrence of the calls (January 1, 2014 – December 31, 2018). Motivated by the nature of the data, we here employ log-Gaussian Cox processes (LGCPs) for the spatio-temporal modelling and forecasting of the calls. To this end, we propose a K-means clustering based bandwidth selection method for the kernel estimation of the spatial component of the separable spatio-temporal intensity function.  The temporal component of the intensity function is modelled using Poisson regression, using different calendar covariates, and the spatio-temporal random field component of the random intensity of the LGCP is fitted using the Metropolis-adjusted Langevin algorithm. Spatial hot-spots have been found in the south-eastern part of the study region, where most people in the region live and our fitted model/forecasts manage to capture this behaviour quite well. Also, there is a significant association between the expected number of calls and the day-of-the-week and the season-of-the-year. A non-parametric second-order analysis indicates that LGCPs seem to be reasonable models for the data. Finally, we find that the fitted forecasts generate simulated future spatial event patterns that quite well resemble the actual future data.  
\end{abstract}
\end{frontmatter}

\section{Introduction}
To obtain a desired outcome of the prehospital care in a given country, for instance, keeping the mortality low, the ambulance response times should be as small as possible \citep{blackwell2002response, o2011role, pell2001effect}. 
Ideally, any given country would have the financial means to let the number of active ambulances, at any given time of the day and in any given region, be so large that the higher quantiles of the empirical response time distribution would be very small. Unfortunately, this is not/rarely the case and many countries struggle to make the most of their existing resources. Even Sweden, for example, which is a prime example of a well-fare state with large resources at hand, struggles with making its state financed prehospital care meet different desired response time goals; in 2018, the Swedish prehospital had about 660 ambulances that responded to approximately 1.2 million calls per year, and cost more than 4 billion SEK annually. Some of the challenges that the Swedish prehospital care is facing include an ageing population, an increasing population and urbanisation. A particular challenge is Northern Sweden, where there are large rural areas with few and relatively elderly inhabitants who need fast access to prehospital care. Here the hospitals are located in cities, which often are far from the rural areas in question. Hence, the ambulances need to be positioned not only in the cities but also on other places so that the region of interest gets a high-quality prehospital care with acceptable response times throughout the region. The placement of ambulances and how they are scheduled over time within a region can be thought of as an optimisation problem, where the objective is to minimise the response times (e.g.~the median response time in the region) given some practical and economical constrains.  To be able  to find solutions to these issues, it is crucial to understand the evolution of the spatio-temporal risk of the occurrence of an ambulance call, at any given time and place. However, given the discussion above, it should be clear that the ambulance/emergency alarm call frequency throughout Sweden is both quite complex and dynamical, with ever changing conditions in both space and time. 

The data at hand, which consist of the emergency alarm calls in the four (northernmost) Swedish regions Västerbotten, Norrbotten, Västernorrland and Jämtland-Härjedalen, can be described by a spatio-temporal point pattern \citep{baddeley2015spatial,diggle2010spatio,diggle2013statistical,gonzalez2016spatio}, i.e.~it is of the form $\{(s_i,t_i)\}_{i=1}^{n}$ where $s_i$ is the spatial (GPS) location of the $i$th call and $t_i$ is the associated call time (here obtained as the date). Although there is a vast literature on prehospital care optimisation (see e.g.~\citet{aringhieri2017emergency} and the references therein), surprisingly very little effort has been spent on actually modelling the calls by means of point processes, which are the models used for point patterns. \citet{zhou2015spatio} and \citet{zhou2016predicting} have introduced a time-varying Gaussian mixture model and kernel warping method for spatio-temporal point process modelling of ambulance data, which managed to predict their ambulance call data quite well. 
Our aim here is to propose an alternative to the approach of  \citet{zhou2015spatio} and \citet{zhou2016predicting}, 
and we will justify its appropriateness for modelling ambulance call data using point process-based non-parametric statistical tools. 
The model family chosen, log-Gaussian Cox processes, have been extensively used for modelling data where there is an underlying latent risk which gives rise to the data. More specifically, one assumes that the latent risk is governed by an unobserved Gaussian random field, which captures spatial and temporal dependence/interaction. In particular, such models have been extensively used to model and predict various non-infectious disease event datasets \citep{diggle2013statistical} -- we argue that such data types are close in nature to spatio-temporal ambulance call data.

The primary goal of this work is to describe the spatio-temporal dynamics of the calls with the main focus of the study being to identify hotspot-regions within the spatial study region as well as generating a short-term forecast-model, which allows us to simulate spatially and temporally realistic future emergency alarm call locations.
These simulated alarms can in combination with realistic simulations of the prehospital care (e.g.~simulation of how alarms are dispatched, driving times and treatment times) be used to optimally allocate prehospital resources and design dispatching strategies. Since the location and time of each emergency alarm call are known, a spatio-temporal point process provides the right framework to model the space-time dynamics in the emergency alarm call data. 
As will be verified by our non-parametric analysis and due to the nature of our data, inhomogeneous log-Gaussian Cox processes (LGCPs) \citep{moller1998log} tend to be particularly suited to model the space-time dynamics of the emergency alarm calls -- LGCPs take spatial and/or temporal clustering/aggregation into account. 
The LGCP frameworks further allow us to carry out forecasting of the spatial structure for future time periods and, consequently, simulate "realistic" future emergency alarm scenarios. 
Following \citet{diggle2013statistical}, the space-time intensity function of the LGCP considered is expressed as a product of two deterministic baseline functions and one stochastic log-Gaussian random field component. The deterministic components represent purely spatial and purely temporal variations while the stochastic component deals with the unexplained space-time variation; the employed approach is semi-parametric in the sense that the spatial baseline function is taken to be a non-parametric spatial density estimate whereas the temporal one is taken to be a temporal intensity estimate which is based on temporal covariates. 
The spatial density estimate, which is a normalised spatial intensity estimate \citep{cronie2018non,moradi2019resample,van2012estimation}, is obtained through kernel intensity estimation and we here propose a new clustering approach to select the associated bandwidth. As our data are discrete in time, with one time unit representing one day, in the case of the temporal intensity estimation we employ a Poisson regression model which is based on temporal covariates such as day-of-the-week. To fit the stochastic component of the model, minimum contrast estimation \citep{baddeley2015spatial,diggle2013statistical,moller2003statistical} and Metropolis-adjusted Langevin algorithm \citep{roberts1996exponential, roberts2002langevin} are the natural ways to proceed. Finally, the forecasting is carried out by exploiting the Ornstein-Uhlenbeck approximation of \citet{brix2001spatiotemporal}.

The structure of the article is as follows. Section \ref{Data} briefly presents the ambulance call data, Section \ref{s:LGCP} provides spatio-temporal log-Gaussian Cox processes and Section \ref{sm} provides an overview of the statistical inference used in this work.  We present the results of the study in section \ref{EASTD} and discuss the implications of the results as well as provide a precise summary of the work in section \ref{dis}.

\section{Data}\label{Data}

\begin{figure}[H]
 	\centering
 	\includegraphics[height=8cm, width=8cm]{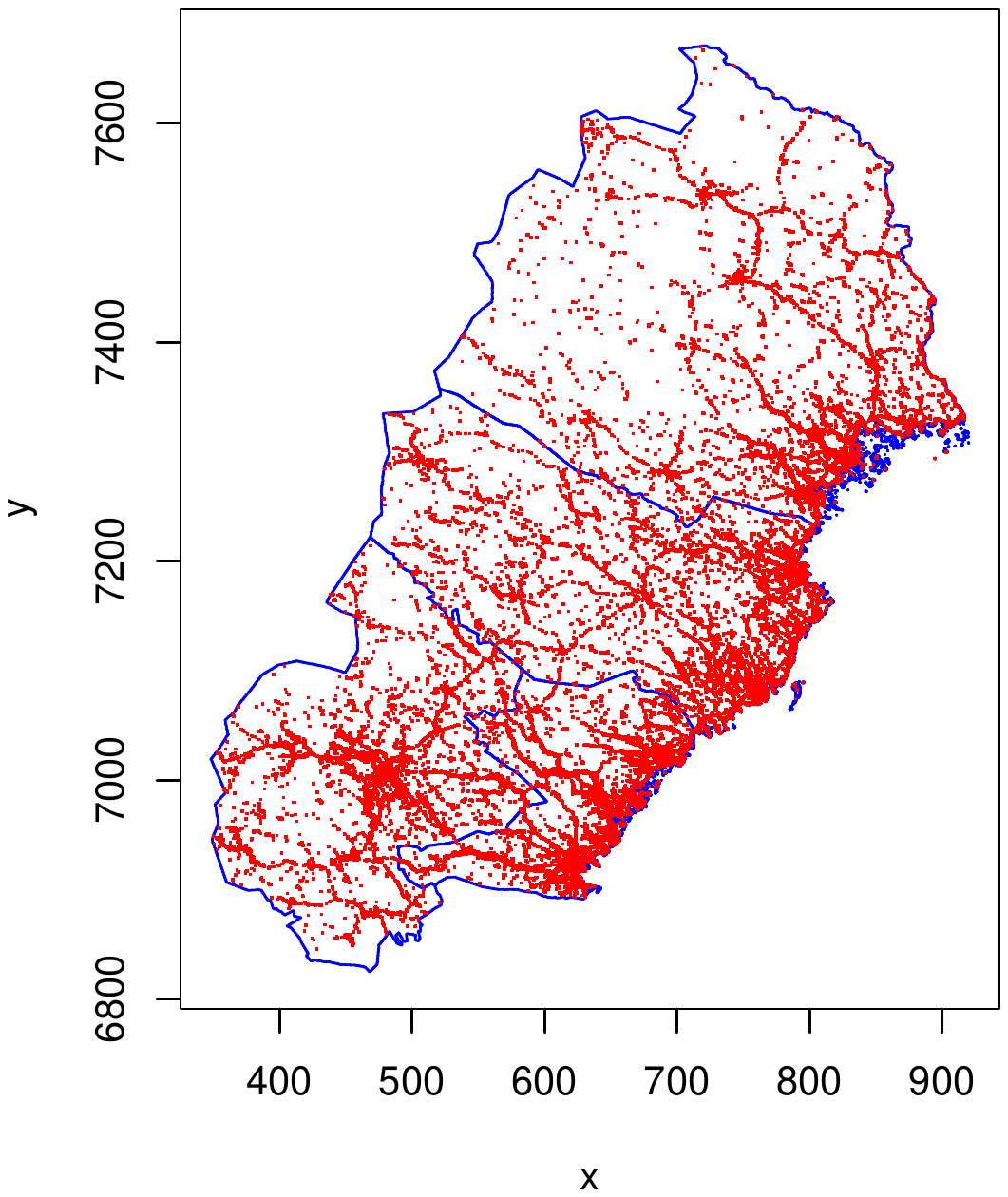}
    \includegraphics[height=8cm, width=8cm]{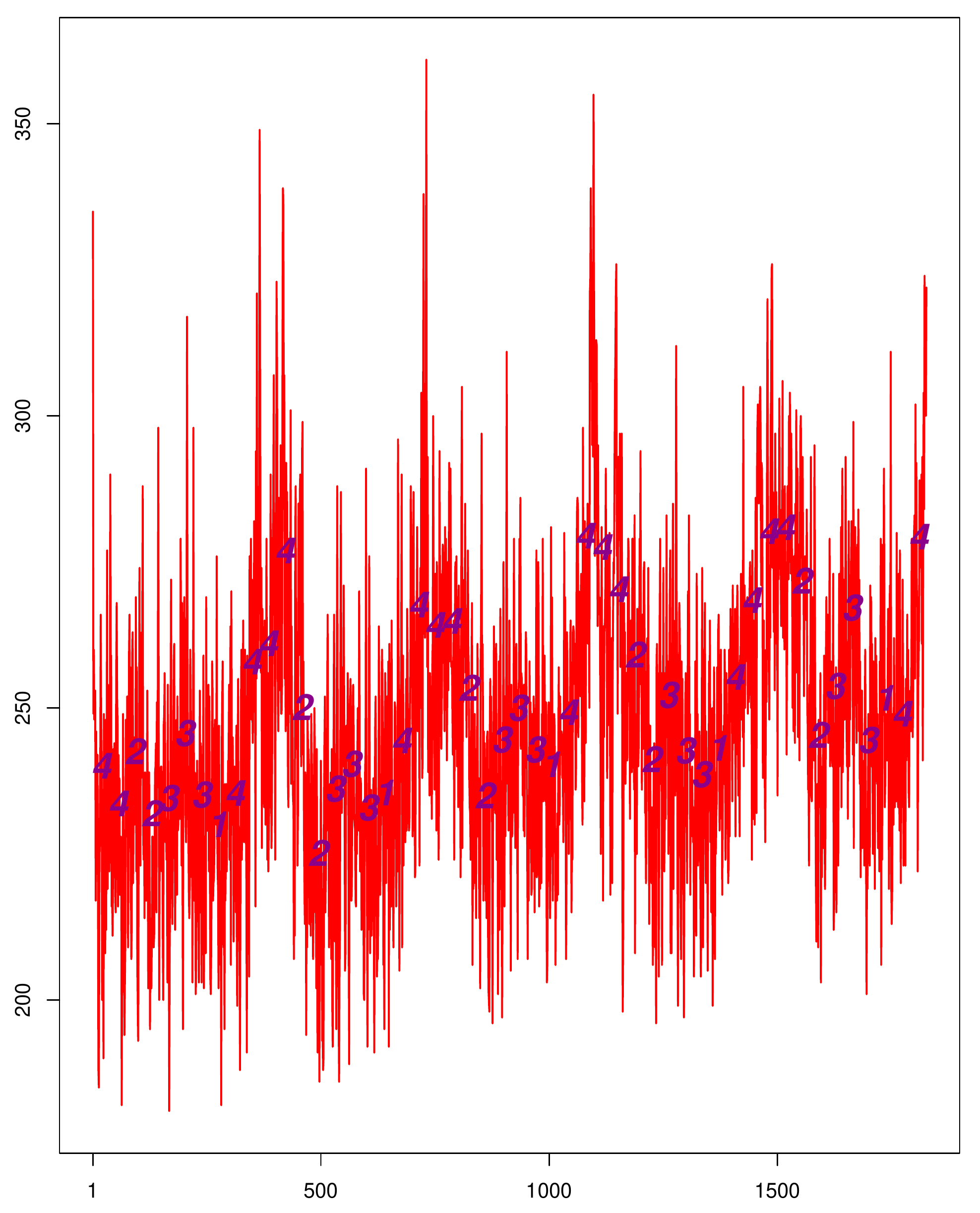}
 	\caption{The spatial and temporal components of the spatio-temporal emergency alarm call data. The left panel shows the spatial locations of the calls while the right panel presents the call counts over the study region as a function of time. The scales of measurements for the spatial locations and the temporal count data are meter (scaled by 1000) and day, respectively. The numbers in the right panel plot indicate the smoothed seasons-of-a-year, which are Fall (1),  Spring (2), Summer (3), and Winter (4), of the emergency alarm calls; the first time point corresponds to January 1, 2014}. 
 	\label{Or21}
 \end{figure}

The ambulance/emergency alarm call data $\{(s_i,t_i)\}_{i=1}^{n}$, where $s_i\in R\subseteq\R^2$ is a spatial (GPS) location of an event and $t_i\in [T_{0},T_{1}]$ is the date of occurrence of the event, have been collected in northern Sweden. The data consist of $n=444~283$ events, i.e.~ambulance/emergency alarm calls, and the temporal domain  $[T_{0},T_{1}]$ 
ranges from January 1, 2014 to  December 31, 2018. 
The left-hand panel in Figure \ref{Or21} displays the spatial locations of the calls while the right-hand panel presents the call count over time. From the left-hand panel of the figure, we can see that the calls are unevenly distributed over the four northern regions of Sweden, i.e.~Norrbotten, V\"{a}sterbotten, V\"{a}sternorrland, and J\"{a}mtland-Härjedalen, which account for 9\% of Sweden’s population, or 900 000 people, and 50\% of the country's total area, which corresponds to roughly the area of Great Britain. In addition, as expected, the calls tend to be located in populated areas, which in turn lie along the road network -- this will make the inference that we propose quite challenging. For convenience, the data have been scaled by dividing the xy-coordinates of the alarm call locations by 1000. 

The data, which may be viewed as a point pattern in $\R^3$, are what one commonly refers to as a spatio-temporal point pattern \citep{Diggle2003statistical,gonzalez2016spatio}. Note that since we will use spatio-temporal point processes for the modelling, we will model daily accumulated data as if they were in fact observed on a temporal continuum, i.e.~on intervals between midnight and midnight on consecutive dates, we will in effect treat the calls occurring during a given day as though they were uniformly distributed over that day. Of course, if one additionally would have access to an (empirical) distribution for how the calls distribute in time over each/an arbitrary day of a year, then this could be used as an additional layer to generate continuous event times. 

\section{Log-Gaussian Cox processes}\label{s:LGCP}
Let $\mathscr{y}=\left\{(s_i,t_i)\right\}_{i=1}^{N}\subseteq R\times[T_{0},T_{1}]$ be a spatio-temporal point process in $R\times[T_{0},T_{1}]$, where $R\subseteq \mathbb{R}^2$ is a bounded spatial domain and  $[T_{0},T_{1}]\subseteq \mathbb{R}$ is a bounded temporal domain \citep{diggle2013statistical,gonzalez2016spatio}. Heuristically, $\mathscr{y}$ is a random sample with a random total point count, with possible dependence between the points in the sample. Formally, it may be characterised by the collection $N(A)$, $A\subseteq R\times[T_{0},T_{1}]$, giving the random point counts of $\mathscr{y}$ in all (Borel) subsets $A$. 

If we find indications of clustering/aggregation in spatio-temporal data and the data are such that the spatio-temporal interactions intuitively seem to come from some latent risk, which is the case with our ambulance call data (underlying factors such as weather and time of the day are possible drivers for the variations in the call risk), a reasonable family of models to employ is spatio-temporal log-Gaussian Cox processes (LGCPs), which are particular instances of spatio-temporal Cox processes  \citep{diggle2013statistical}. 
Let $\Lambda=\{\Lambda(s, t)\in\mathbb{R}^{+}\mid (s, t)\in R\times[T_{0},T_{1}]\}$ be a spatio-temporal random field, referred to as the random intensity process, such that (with probability one) $\iint_A\Lambda(s, t)ds dt<\infty$ for any bounded subset $A$ of $R\times[T_{0},T_{1}]$. A spatio-temporal Cox process with random intensity $\Lambda$ is generated by conditioning on $\Lambda=\rho=\{\rho(s,t):(s, t)\in R\times[T_{0},T_{1}]\}$ and then, in turn, generating a Poisson process with intensity function $\rho$: 
\begin{align}\label{dfc2}
N\left(A\right)
\mid \Lambda = \rho\sim Poisson\left(\iint_{A}\rho\left(s, t\right)dtds \right)
,
\qquad A\subseteq R\times[T_{0},T_{1}]
.
\end{align}
Its first- and second-order intensity functions, defined according to 
$\lambda(s,t)=
\lim_{\abs{\delta s \times \delta t}\to 0} 
\E\left[N\left(\delta s\times \delta t\right)  \right] /\abs{\delta s \times \delta t}$ and 
$\lambda^2((s, t), (s', t')) 
= 
\lim_{\abs{\delta s \times \delta t},\abs{\delta s' \times \delta t'}\to 0}
\E\left[
N\left(\delta s\times\delta t\right) 
N\left(\delta s'\times\delta t'\right)
\right]/(\abs{\delta s \times \delta t}\abs{\delta s' \times \delta t'})$, $(s, t), (s', t')\in R\times[T_{0},T_{1}]\}$, are given by $\lambda(s,t)=\E[\Lambda(s, t)]$ and $\lambda^2((s,t),(s',t'))=\E[\Lambda(s, t)\Lambda(s',t')]$. 

A spatio-temporal LGCP is a spatio-temporal Cox process with the property that the logarithm of its stochastic intensity function is a Gaussian process \citep{moller1998log}. Following e.g.~\cite{diggle2005point}, in the case of an LGCP, we specifically have that 
\begin{align}\label{key3w2}
\Lambda(s,t) = \lambda_{0}(s)\lambda_{1}(t)\exp\left\{Z(s,t)\right\},
\end{align}
where $\lambda_{0}(s)$ and $\lambda_{1}(t)$ can be defined as
\begin{align*}
\lambda_{0}(s) = 
\lim_{\abs{\delta s}\to 0} \frac{\E\left[ N\left(\delta s\times[T_{0},T_{1}]\right)  \right]}{\abs{\delta s}}, 
\quad s\in R,
&&
\lambda_{1}(t) &= \lim_{\abs{\delta t}\to 0} \frac{\E\left[ N\left(R\times\delta t\right)  \right] }{\abs{\delta t}}, 
\quad t\in[T_{0},T_{1}],
\end{align*}
and $Z = \{Z(s,t)\in\mathbb{R}\mid (s,t)\in R\times[T_{0},T_{1}]\}$ is an (unobservable) spatio-temporal Gaussian random field/process with mean and the covariance functions 
\begin{align*}
\mu(s,t)=\E[Z(s,t)] 
\quad\text{and}\quad 
C((s_{i},t_{i}),(s_{i'},t_{i'}))
=
{\rm \Cov}\left(Z(s_{i},t_{i}), Z(s_{i'},t_{i'})\right).
\end{align*}
The distribution of the spatio-temporal Gaussian random  process $Z$ is completely determined by the functions $\mu(s,t)$ and $ C((s_i,t_i),(s_{i'},t_{i'}))$. For any fixed tuple  $((s_{1},t_{1}),\ldots,(s_{n},t_{n}))\in(R\times[T_{0},T_{1}])^n$, $n\geq1$, by definition, $(Z\left(s_{1},t_{1}\right), \ldots , Z\left(s_{n},t_{n}\right))$ follows a multivariate Gaussian distribution with mean vector $(\mu(s_{1},t_{1}),\ldots,\mu(s_{n},t_{n}))$ and covariance matrix given by the $n$-by-$n$ matrix $[C((s_{i},t_{i}),(s_{i'},t_{i'}))]_{i,i'=1,\ldots,n}$. 
The spatio-temporal random process $Z$ represents unobservable space–time Gaussian noise that affects the occurrence of events and its inference provides information about clustering of the events. The properties of $Z$ determine the properties of its corresponding LGCP. For instance, if $Z$ is stationary and/or isotropic, then its corresponding LGCP is also stationary and/or isotropic \citep{moller1998log}. 

Defining $\sigma^2(s,t)=\Var(Z(s,t))=C((s,t),(s,t))$, the first- and second-order intensity functions are given by \citep{moller1998log}
\begin{align}\label{IntensitiesLGCP}
\lambda(s,t)&=\lambda_{0}(s)\lambda_{1}(t)\E[\exp\left\{Z(s,t)\right\}] =\lambda_{0}(s)\lambda_{1}(t)\exp\{\mu(s,t)+\sigma^2(s,t)/2\},
    \\
\lambda^2((s,t),(s',t'))&=\lambda_{0}(s)\lambda_{0}(s')\lambda_{1}(t)\lambda_{1}(t')\E[\exp\left\{Z(s,t)+Z(s',t')\right\}]
    \notag
    \\
    &=\lambda_{0}(s)\lambda_{0}(s')\lambda_{1}(t)\lambda_{1}(t')
    \exp\{\mu(s,t) + \mu(s',t') + \sigma^2(s,t)/2 + \sigma^2(s',t')/2 + C((s,t),(s',t'))\}
    .
    \notag
\end{align}
In this work, we assume that $\mu(s,t)= -0.5\sigma^{2}$ and $\sigma^2(s,t)=\sigma^{2}>0$ so that $\E[\exp\{Z(s,t)\}]=1$ and we note that the constant variance $\sigma^{2}$ determines the point-wise variability of $Z$ and it scales the log-intensity. 
This assumption implies that $\lambda(s,t)=\E\left[\Lambda(s,t)\right]=\lambda_{0}(s)\lambda_{1}(t)$, and we obtain that the LGCP is separable, i.e. $\lambda(s,t) = \lambda_{0}(s)\lambda_{1}(t)$,  \citep{diggle2013statistical}. We further have that the deterministic temporal intensity component $\lambda_{1}(t)\geq 0$ represents the expected number of events in $R$ that occur in an infinitesimal neighbourhood of $t$ and, consequently, the fixed spatial intensity component $\lambda_{0}(s)\geq 0$ is a density, i.e.~$\int_{R}\lambda_{0}(s)ds=1$. The above (separable) construction helps to provide separate spatial and temporal meaning to the expectation of the stochastic intensity. We further assume that $Z$ is stationary and isotropic, whereby the covariance function depends only on the separation lags between the evaluation points. 
 
In addition, we assume a separable covariance structure, i.e.~the covariance between two points in space-time can be factored into purely spatial and purely temporal components: 
 \begin{align}\label{eqq}
C((s,t), (s',t')) =
C(s'-s,t'-t) = \sigma^{2}r_{\phi}\left(\norm{s'-s}\right)r_{\theta}\left(\abs{t'-t}\right),
\end{align}
where $\norm{\cdot}$ represents the Euclidean norm in $\R^2$, $\abs{\cdot}$ denotes the absolute value, $r_{\phi}\left(\cdot\right)$ represents the spatial correlation function,  $r_{\theta}\left(\cdot\right) $ represents the temporal correlation function, and $\phi$, $\sigma^{2}$, and $\theta \in \mathbb{R}^{+}$ are fixed parameters (to be estimated). We interpret temporal dependence as the covariance between the number of events in the study region at two points in time. Several forms of isotropic correlation functions exist, e.g.~the correlation functions corresponding to  exponential and Mat\'{e}rn covariance functions \citep{cressie1999classes}; in the former case, $\phi$ and $\theta$ are scale parameters which control the rate of decay of the spatial and temporal correlation functions. 
In the case of our ambulance data, we assume that the spatial and temporal correlation functions are given by the exponential forms  $r_{\phi}\left(u\right) = \exp\{-u/\phi\}$ and $r_{\theta}\left(v\right) = \exp\{-v/\theta\}$, respectively. Here, the scale parameters $\phi$ and $\theta$ control the smoothness of the underlying (marginal) fields and indicate predictability, i.e.~essentially how close to each other two events have to be "to influence each other".

Assuming stationarity and isotropy for the covariance function, and consulting the first- and the second-order intensity functions in \eqref{IntensitiesLGCP}, it follows that the pair correlation function and the (inhomogeneous) $K$-function \citep{baddeley2000non,gabriel2009second} of the LGCP with driving covariance function \eqref{eqq} are given by 
\begin{align}
g(s'-s, t'-t) = \frac{\lambda^2((s, t), (s', t'))}{\lambda(s,t)\lambda(s',t')}& =\frac{\lambda^2(s'-s,t'-t)}{\lambda(s,t)\lambda(s',t')}, \label{PairFun}\\
&= \exp\{C(s'-s,t'-t)\} =\exp\left\{\sigma^{2}r_{\phi}\left(\norm{s'-s}\right)
r_{\theta}\left(\abs{t'-t}\right)\right\},\notag\\
K\left(a, b\right) &= 2\pi\int_{0}^{a}u\int_{0}^{b}g\left(u, v\right)dv du, \label{KFun}\\
&= 2\pi \int_{0}^{a} u \int_{0}^{b}
\exp\{\sigma^{2}r_{\phi}(u)r_{\theta}(v)\}dv du.\notag
\end{align}
Note that isotropy here refers to the spatial covariance function being a function of the norm $\|s'-s\|$. Treated as the restriction of a point process on $\R^3$, the assumed LGCP is second-order intensity-reweighted stationary (SOIRS) provided that the first-order intensity is positive on $R\times[T_{0},T_{1}]$ and the spatio-temporal pair-correlation function $g((s, t), (s', t'))  = g(s'-s, t'-t)$ \citep{baddeley2000non,gabriel2009second}. For any inhomogeneous spatio-temporal Poisson point process with intensity function $\lambda(s,t)>0$, we have that $K\left(a, b\right) = 2\pi a^{2}b$, so using this as a benchmark we obtain that: if $K\left(a, b\right) - 2\pi a^{2}b>0$, we have indications of clustering/aggregation for spatial lags less than $a$ and temporal lags less than $b$, and $K\left(a, b\right) - 2\pi a^{2}b<0$ instead indicates inhibition/regularity. For a Poisson process with intensity function $\lambda(s,t)>0$, we have that $g((s, t), (s', t'))=1$. Hence, if the pair correlation function of an arbitrary spatio-temporal point process satisfies $g((s, t), (s', t'))>1$, then there is clustering/aggregation between points located around $(s, t)$ and $(s', t')$. Reversely, $g((s, t), (s', t'))<1$ instead indicates inhibition \citep{gabriel2009second}.

\section{Statistical inference}  \label{sm}   
We next turn to the statistical inference based on a realised spatio-temporal point pattern $\mathbf{y} =\left\{(s_i,t_i)\right\}_{i=1}^{n}$ generated by the spatio-temporal point process $\mathscr{y}=\left\{(s_i,t_i)\right\}_{i=1}^{N}\subseteq R\times[T_{0},T_{1}]$. Notice that, for simplicity, we are using the notation $(s_{i}, t_{i})$ interchangeably for the stochastic event and the observed event.

\subsection{Spatial kernel intensity estimation}\label{s:IntensityEstimation}
We have used kernel smoothing based on a quartic kernel to describe the purely spatial intensity function $\lambda_{0}(s)$, which generates a spatially smooth estimate of the local intensity of events. More precisely, we consider the kernel 
$\kappa(u)=(1 - u^{2}/2)^{2}\1\{\abs{u}\leq \sqrt{2}\}$ 
and the estimator of the purely spatial intensity function is given by  
\begin{align}\label{eq1}
\hat\lambda_0(s)
=\hat\lambda_0(s;h) = \frac{1}{h}\sum_{i = 1}^{n}\kappa\left(\frac{ s-s_{i}}{h}\right),
\qquad s\in R.
\end{align}
In most situations it is necessary to consider edge effects, which are boundary problems due to  unobserved information outside the study region -- uncorrected, the statistical analysis ignores how objects outside the study area may affect/interact with objects within the study region \citep{griffith1980towards}, in the estimation of the spatial intensity function. The methods described in \cite{diggle1985kernel} and \cite{berman1989estimating} may be applied to incorporate edge correction factors into the spatial intensity estimator. Note, however, that in the case of our ambulance data (recall Figure \ref{Or21} and a map of Sweden and its border), edge effects may not actually be significantly present since few people live near the boundary of northern Sweden. Moreover, to the west is a different country, Norway, and Sweden has the Baltic sea as boundary to the east.

Although it is often claimed that the choice of kernel used is irrelevant (in comparison to the bandwidth choice), this is not entirely true \citep{cronie2018non} and we note, in particular, that kernels with bounded supports, such as the quartic kernel, may generate intensity estimates which depend on sparse values of the quartic kernel in the study region, as opposed to using e.g.~the Gaussian kernel which has unbounded support. This means that kernels with bounded supports may help to balance over and under-smoothing effects of kernels with unbounded support. The upside is that we, as in the case of our ambulance data, may better control where we place the mass of the estimated intensity. 

\subsubsection{Machine learning technique based bandwidth selection}\label{s:Bandwidth}
Selecting an appropriate bandwidth $h$ in practice is quite challenging. In optimal bandwidth selection, the main challenge of the kernel smoothing method is to optimise the balance between under- and over-smoothing. In the context of point processes, we are also faced with the additional challenge that the points of the underlying point process may be dependent. A range of different methods has been proposed in the literature (see e.g.~\citet{baddeley2015spatial,davies2018tutorial} for overviews), and most noteworthy are perhaps the recent method of \citet{cronie2018non} and the Poisson processes likelihood cross-validation method of \citet{Load99}, see the illustration in Section \ref{EASTDD}.

We here take on an approach which is different from the point process derived methods mentioned above, namely to exploit the idea of $K$-means clustering.  $K$-means clustering is a machine learning technique commonly employed to identify groups or clusters of data points in a multidimensional space. 
Let $\{s_{1}, \ldots, s_{n_{0}}\}$ be a realisation of the (well-defined) projection $\mathscr{y}_S=\{s_i\}_{i=1}^{N}\subseteq R\subseteq\R^2$ of a spatio-temporal point process $\mathscr{y}\subseteq R\times[T_{0},T_{1}]$ onto $R$. The data are partitioned into $K\geq1$ clusters and each cluster is composed of data points whose inter-point distances are smaller than their distances to points outside of the cluster. After taking clustering information into account, the average of the standard deviations of the clusters can be taken as an optimal estimate of the bandwidth. We have summarised this approach in Algorithm \ref{alg:Oro}.

\newcommand{\sfunction}[1]{\textsf{\textsc{#1}}}
\algrenewcommand\algorithmicforall{\textbf{foreach}}
\algrenewcommand\algorithmicindent{.8em}
\begin{algorithm}[!htpb]
\caption{$K$-means clustering for bandwidth selection}
\label{alg:Oro}
\begin{algorithmic}[1]
\State Initialize $K$ centroids of the clusters: $\varpi^{\left(v\right) }_{k}$, $v = 0$ and $k = 1, 2, 3, \ldots, K$.
\State Optimize  $1$-of-$K$ class indicator variable $\iota_{ik}$ for each observation $i = 1, 2, 3, \ldots, n_{0}$: \label{kmeans1}
\begin{align*}
\zeta^{(v)}_{i} = \argmin_{q\in\left\lbrace 1, \ldots, K\right\rbrace } \big \| s_{i} - \varpi^{\left(v\right)}_{q} \big \|^{2}, \quad \iota^{(v)}_{ik} =
\1\left\{k = \zeta^{(v)}_{i}\right\}, \quad k = 1, 2, 3, \ldots, K;
\end{align*} 
\State Update the centroids of the clusters (the mean locations of the clusters) $\varpi_{k}$:\label{kmeans2}
\begin{align*}
\varpi^{\left(v+1\right)}_{k} 
= \left\lbrace \sum_{i = 1}^{n_{0}} \iota^{\left(v\right)}_{ik}s_{i}\right\rbrace 
\Bigg/
\sum_{i = 1}^{n_{0}}\iota^{\left(v\right)}_{ik}, 
\quad k = 1, 2, 3, \ldots,K;
\end{align*}
\State Repeat steps \ref{kmeans1} and \ref{kmeans2} for $v = 1, 2, 3, \ldots$, unless $\displaystyle\sum_{k=1}^{K}\left\| \varpi^{(v+1)}_{k} - \varpi^{(v)}_{k} \right\|^{2}\le \varepsilon$ \label{km2}. 
\State Denote the optimal class indicator, class of a data point,  and centroid of a cluster by $\hat{\iota}_{ik}$, $\hat{\zeta}_{i}$, and $\hat{\varpi}_{k}$, respectively. 
\State Compute an optimal selection of the bandwidth by  
\begin{eqnarray*}
\hat{h} 	= 
\sqrt{\frac{1}{2K}\sum_{k=1}^{K}\sum_{i=1}^{n} \frac{1}{n_{k}}\1\left\{\hat{\zeta}_{i} = k\right\} \left\| s_{i} - \hat{\varpi}_{k} \right\|^{2}}, \quad n_{k} = \sum_{i=1}^{n}\1\left\{\hat{\zeta}_{i} = k\right\}, \quad k = 1, 2, 3, \cdots, K.
\end{eqnarray*}
\end{algorithmic}
\end{algorithm}
There are two key steps in the K-means algorithm. Data points are re-assigned to clusters (see step \ref{kmeans1}), the cluster means are re-computed (see step \ref{kmeans2}) and these steps are repeated in turn until the sum of the Euclidean norm of  the successive difference of each centroid of the clusters is smaller than a user specified value. The other options to control the convergence of the algorithm are to repeat steps \ref{kmeans1} and \ref{kmeans2}  until there is no further change in the assignments of data points to clusters or until some maximum number of iterations is achieved.   In our case, we have used $\varepsilon = 10^{-5}$.

We have used the former and set $\varepsilon = 10^{-5}$ in step \ref{km2} in the algorithm. A few things need to be highlighted here. The \citet{cronie2018non} method and the Poisson processes likelihood cross-validation method are both based on integral expressions, where the integrals run over the whole spatial domain $R$. Since essentially all of the ambulance calls have spatial locations which are close to the Swedish road network (see Figure \ref{Or21}), we have big holes in $R$ where there are no events. This is exactly the scenario where the aforementioned methods do not tend to  work well \citep{cronie2018non}. The $K$-means clustering idea, on the other hand, operates on a more local scale and thus works better for our ambulance data. 
Note that the $K$-means clustering idea, similarly to the Poisson processes likelihood cross-validation method, does not fully compensate for dependence and, consequently, densely packed groups of events are interpreted as regions with high intensity function values, when they in fact may be the results of aggregation; e.g., in the case of an LGCP we are essentially trying to reconstruct the random intensity function rather than estimating the actual intensity function. Moreover, we have to choose the smoothing parameter $K$ by means of visual inspection but this we found to be doable in the case of the ambulance data (see Section \ref{EASTD} for details).  In Section \ref{EASTDD} we have compared the K-means clustering based bandwidth selection method with the state of the art in spatial kernel intensity estimation bandwidth selection, in the context of our ambulance data.

\subsection{Poisson regression based temporal intensity modelling}\label{s:TemporalIntensityModelling}
Recall from Section \ref{s:LGCP} that, by assumption, $\int_R\lambda_0(s)ds=1$ so that the temporal component $\lambda_{1}(t)$ reflects the expected number of events in the spatial study region at time $t$. Following \citet{diggle2005point}, it seems to make sense to use a Poisson regression approach to model the temporal intensity component of the stochastic intensity function $\Lambda(s,t)$. Day-of-the-week, season-of-the-year, and the periodic nature of the data can be used as covariate information for modelling the expected number of events, i.e.~the emergency alarm calls, over time, and we here define the seasons as Spring (months 3-5), Summer (months 6-8), Fall (months 9-10) and Winter (months 11-12 and 1-2). Hence, the effects of different seasons on the expected number of calls over the study region can be understood from the proposed regression model. Let $N_{R}(t)$ be the expected number of emergency calls in $R$ at time $t$; under a Poisson distribution assumption, the mean and variance of $N_{R}(t)$ are equal and here given by $\lambda_{1}(t)$.   
The obtained Poisson regression model can be given by
\[
\log\lambda_{1}(t) 
= \alpha_{0} 
+ 
\sum_{i=1}^{7} \alpha_{i}\1\{d\left(t\right)=i\}
+ 
\sum_{j=1}^{4} \gamma_{j}\1\{sn\left(t\right)=j\}
+ \beta_{1}\sin\left(\tau t\right)
+ \beta_{2}\cos\left(\tau t \right)
+ \beta_{3}\sin\left(2\tau t \right)
+ \beta_{4}\cos\left(2\tau t\right)
+ \delta t,
\] 
where  $\left\lbrace \alpha_0, \alpha_1, \cdots, \alpha_7,  \gamma_1,  \cdots, \gamma_4, \beta_1,  \cdots, \beta_4, \delta\right\rbrace $ are the set of parameters,  $d\left(t\right)$ is the day-of-the-week at time $t$, $sn\left(t\right)$ is the season-of-the-year at time $t$, $\delta$ is the trend parameter, $\tau = 2\pi/365$, and $\1\{d\left(t\right)=i\}$ and $\1\{sn\left(t\right)=i\}$ are indicators of the day-of-the-week and the season-of-the-year at time $t$. The days-of-a-week, from Monday to Sunday, have been numbered ranging from 1 to 7 while the four seasons-of-a-year, Spring, Summer, Fall, and Winter, have been numbered ranging from 1 to 4, respectively.  The annual periodicity of the events can be taken into account in the model setting through the parameters $\beta_{i}$, $i = 1,\ldots,4$.

\subsection{Non-parametric second-order summary statistic estimation}\label{EstimationKfunction}

In the current context of LGCPs and their statistical inference, second-order summary statistics and their non-parametric estimators play a crucial role. Not only do they help in the non-parametric estimation of the underlying spatio-temporal dependence structures of the data, but they also play a crucial role in the minimum contrast-based parameter estimation for the latent Gaussian field $Z$ in an LGCP's random intensity function. 

There are some issues that may arise in the estimation of spatio-temporal inhomogeneous $K$-functions or pair correlation functions. 
First off, they contain the true, unknown intensity function, but this we may replace by an intensity estimate. However, such a replacement should be done with caution, as it often adds bias to the estimation \citep{baddeley2000non}. This may for instance be the case when we estimate both a varying intensity function and a second-order characteristics using the same point pattern. We can overcome such problem by modelling the intensity parametrically \citep{diggle2007second, mrkvivcka2014two, arbia2012clusters, liu2007characterizing, moller2003statistical}. 

A further problem is related to edge effects, i.e.~the influence of unobserved events outside the study region on the estimation of spatio-temporal interaction structures. 
In comparison to the purely spatial setting, edge effects are more difficult to overcome here where the dimension is greater than two \citep{baddeley1999spatial}. 
Various edge-correction methods have been proposed in the spatial setting (see e.g.~\citet{
baddeley1999spatial,
Cressie1993statistical,
Diggle2003statistical,
illian2008statistical,
law2009ecological,
li2007comparison,
RipleyBrian1988statistical,
goreaud1999explicit,
haase1995spatial,
pommerening2006edge,
yamada2003empirical}) 
and in the spatio-temporal context, \cite{gabriel2014estimating} has extended three classical spatial edge-correction factors to the spatio-temporal setting while \cite{cronie2011some} proposed methods for parametric models.

Following \cite{gabriel2009second,gabriel2014estimating} and 
assuming SOIRS with isotropy as in \cite{gabriel2009second}, 
we employ the  unbiased non-parametric estimator
\begin{align}\label{kfun}
\hat{K}\left(r, t\right) 
= 
\frac{1}{|R\times T_{*}|}
\sum_{(s_{1},t_{1})\in\mathscr{y}\cap R\times[T_{0},T_{1}]}
\;
\sum_{(s_{2},t_{2})\in\mathscr{y}\cap R\times[T_{0},T_{1}]\setminus\{(s_{1},t_{1})\}}
\frac{\1\{\norm{s_{2}-s_{1}} \le  r\}
\1\{\abs{t_{2}-t_{1}} \le t\}
}{\lambda\left(s_{1}, t_{1}\right)\lambda(s_{2}, t_{2})}
\omega_R(s_{1},s_{2})
\omega_{T_{*}}(t_{1},t_{2}),
\end{align}
of the $K$-function in equation \eqref{KFun}. 
Here the notation $\1\{\cdot\}$ is used for indicator functions,  $\omega_R(s_{1},s_{2})$ is Ripley's edge correction factor in $R$ \citep{ripley1977modelling}, $\omega_{T_{*}}(t_{1},t_{2})$ is the one-dimensional analogue of Ripley's edge correction factor in $T_{*} = [T_{0},T_{1}]$, and $|R\times T_{*}|$ is the volume of the spatio-temporal region $R\times T_{*}$. Note that, as an alternative, one could of course consider any of the higher-order summary statistics of \citet{cronie2015aj}.

Let $\mathscr{y}(t)=\{s_i:(s_i,t_i)\in\mathscr{y}, t_i=t\}$, $t = 1, 2, \ldots,T$ be a spatial point process time series of the spatiotemporal point process $\mathscr{y}=\left\{(s_i,t_i)\right\}_{i=1}^{N}\subseteq R\times[T_{0},T_{1}]$, where the event times are non-negative integers as in the case of our ambulance data, $0 \leq T_{0} \leq 1$ and $1 < T \leq T_{1}$. Assuming separability and following \citet{davies2013assessing}, the purely spatial non-parametric pair correlation function estimator, which can be considered as a  spatial marginal obtained by averaging over time, can be given by
\begin{align}
\label{TempAvSummaries}
\hat{g}(u) =
\frac{1}{T}\sum_{t=1}^{T}
\hat{g}_t(u) =
\frac{1}{2\pi u|R|T}
\sum_{t=1}^{T}
\frac{1}{\lambda^{2}_{1}(t)}
\sum_{s_{1}\in\mathscr{y}(t)}
\sum_{s_{2}\in\mathscr{y}(t)\setminus\{s_{1}\}}
\frac{\kappa_{s}\left(u -\norm{s_{1}-s_{2}};h_s\right) }{\lambda_{0}(s_{1})\lambda_{0}(s_{2})}
\omega_{R}\left(s_{1}, s_{2}\right),
\end{align}
where $u$ is a spatial lag, $0 < u \leq u_{max}$,  $\omega_{R}$ is Ripley's isotropic edge correction with respect to the region $R$, $\kappa_{s}$ is a univariate smoothing kernel with bandwidth $h_{s}$, and $|R|$ is the area of the spatial region $R$.
Note that the above forms are derived from the fact that the expected number of events with (integer) event time $t$ is given by $\lambda_{1}(t)\int_R \lambda_{0}(s)ds$. For the selection of the bandwidth $h_{s}$ in the smoothing kernel of the pair correlation function estimation in \eqref{TempAvSummaries}, we use Stoyan's rule of thumb \citep{stoyan1994fractals}. Note that, in practice, in the estimator above we have to plug in estimates of the intensity functions. On the other hand, the theoretical pair correlation function for our LGCP can be given by
\begin{align}\label{Theoretical}
g\left(u\right) = \exp\left\{\sigma^{2}r_{\phi}\left(u\right)\right\},
\end{align}
where $u$ is a spatial lag.

We can also describe the temporal dependence between different (spatial point process) components of the series $\mathscr{y}(t)$, $t=1,\ldots,T$. More specifically, we can determine the covariance between the number of events in the spatial domain $R$ at two points in time. By \citet{diggle2005point} and  \citet{davies2013assessing}, and recalling the count function $N_t(A)$, $A\subseteq R$, of $\mathscr{y}(t)$, $t=1,\ldots,T$, the temporal covariance between $N_{t}(R)$ and $N_{t-v}(R)$ can be given by 
\begin{align}\label{eq3}
C\left(t, v; \theta,\sigma^{2},\phi\right) &=
\Cov\left(N_t\left(R\right),N_{t-v}\left(R\right)\right),\\
&=
\lambda_{1}(t)\lambda_{1}(t-v)
\left(
\int_{R}\int_{R}
\lambda_{0}(s_{1}) \lambda_{0}(s_{2})
\exp\left\{\sigma^{2}r_{\phi}\left(u\right)r_{\theta}\left(v\right)\right\} ds_{1}ds_{2} - 1
\right)
,
\quad v>0,
\notag
\end{align}
and an empirical auto-covariance function is a natural non-parametric estimator for the temporal covariance \eqref{eq3} and it is given by 
\begin{eqnarray}\label{PAIR}
	\hat{C}\left(t, v\right) = N_t\left(R\right)N_{t-v}\left(R\right)-\lambda_{1}(t)\lambda_{1}(t-v).
\end{eqnarray}

\subsection{Parameter estimation for the Gaussian random field parts}\label{s:LGCPfitting}

We next turn to the  statistical inference for the whole LGCP model. The key here is to note that we estimate 
$\boldsymbol\lambda_{0} = \{\lambda_{0}(s)\mid s\in R\}$ 
and 
$\boldsymbol\lambda_{1} = \{\lambda_{1}(t)\mid t\in [T_{0},T_{1}]\}$ separately, non-parametrically and parametrically, respectively, so we are only left with estimating the covariance functions of the assumed LGCP, which are governed by the parameters $\sigma^{2}$, $\phi$, and $\theta$. In other words, we are here only concerned with estimating the parameters related to latent Gaussian random field $Z$ in the formulation of the random intensity of the LGCP.

It is common practice to exploit a regular grid discretisation $\{(s_{i},t_{i})\}^{m}_{i=1}\subseteq R\times[T_{0},T_{1}]$ of the study region \citep{moller1998log, brix2001spatiotemporal, diggle2005point}, which yields a discretised version $\mathbf{z} = \{z(s_{i},t_{i})\}^{m}_{i=1}$ of the Gaussian process $Z$ over the grid (recall Section \ref{s:LGCP}). 
Conditionally on $Z$, 
the number of events located within the grid cell centred at $(s_{i},t_{i})$ may be treated as an observation of a Poisson random variable $x(s_i,t_i)$. Assuming that the stochastic intensity function $\Lambda$ is constant over the cells (in practice this may approximately be achieved by assuming that the cells are small), conditionally on the collection $\{\Lambda(s_i,t_i)\}^{m}_{i=1}$, the collection $\{x(s_i,t_i)\}^{m}_{i=1}$ follows a multivariate Poisson distribution. Note that vector-equivalents may be obtained by ordering the cells. The event times of our ambulance call data are recorded daily and we have thereby integer encoded them. Hence, we may consider time slices $\mathbf{x}_{t} = \text{lexo}\{\{x(s_i,t)\}^{m}_{i=1}\}$, $t\geq1$, which are column vectors obtained by lexicographical ordering (lexo) of the cell-aggregated data at time $t$, of three-dimensional $\{x(s_i,t_i)\}^{m}_{i=1}$. Here the short notation $\mathbf{x}_{1:t}$ represents the cell-counts at time slices $1, 2, 3, \ldots,t$. The notations $\mathbf{z}_{t}$ and $\mathbf{z}_{1:t}$, which pertain to the discretisation of the random field $Z$, are interpreted in a similar manner as the notations $\mathbf{x}_{t}$ and $\mathbf{x}_{1:t}$.

The natural starting-point for parameter estimation is maximum likelihood estimation. However, in general, the likelihood of an LGCP is analytically intractable since it is expressed as a high-dimensional integral with respect to the distribution of the unobserved Gaussian process $Z$. Under the current parametrisation, the likelihood of the LGCP can be expressed as
\begin{align*}
\mathcal{L}(\sigma^{2}, \theta, \phi, \boldsymbol\lambda_{0}, \boldsymbol\lambda_{1}\mid \mathbf{y}) &=\E_{\boldsymbol\Lambda}[\mathcal{L}
(\sigma^{2}, \theta, \phi, \boldsymbol\lambda_{0}, \boldsymbol\lambda_{1}\mid \mathbf{y}, \boldsymbol\Lambda)]
=
\E_{\boldsymbol\Lambda}\left[\exp\left\lbrace \int_{0}^{T}\int_{W}\left( 1-\Lambda(s,t)\right) dsdt\right\rbrace \prod_{i=1}^{n}\Lambda(s_{i},t_{i})\right], 
\end{align*}
where $\mathbf{y} = \{(s_i,t_i)\}_{i=1}^{n}\subseteq R\times[T_{0},T_{1}]$ is an observed  spatio-temporal point pattern. In principle, importance sampling can be used to carry out maximum likelihood estimation of LGCP \citep{geyer1994convergence,moller1998log}. 
However, in order to use importance sampling for maximum likelihood estimation of the LGCP, we need to sample from the conditional distribution of the whole of $\mathbf{z}_{1:m}$, given all the observations $\mathbf{x}_{1:m}$, and this is unfortunately infeasible in big data settings such as the current one \citep{brix2001spatiotemporal}. 

Instead, a minimum contrast estimation approach using the pair-correlation function, i.e.~based on equations \eqref{TempAvSummaries} and \eqref{Theoretical}, is used to estimate  the variance and spatial scale parameters, $(\sigma^{2}, \phi)$, of the spatial component of the separable covariance function. In a similar vein, minimum contrast estimation with the empirical auto-covariance function, i.e.~based on equations \eqref{eq3} and \eqref{PAIR}, is exploited to estimate the temporal scale parameter $\theta$ of the temporal component of the separable covariance function. We refer to \cite{davies2013assessing} for a complete overview of minimum contrast estimation of covariance function parameters in spatial and spatio-temporal LGCPs.

\subsection{Simulation of the Gaussian random field}
The discretisation $\mathbf{z}$ of the unobservable Gaussian process $Z$ on the study region follows a multivariate Gaussian distribution. It follows that simulation can be used to do inference for the unobservable random vector $\mathbf{z}$, by sampling from the conditional distribution of the unobservable random vector given its corresponding observable random vector \citep{brix2001spatiotemporal}. 
Letting $\left[\cdot \right]$ denote the probability distribution of a random quantity, 
the conditional distribution of the unobserved time series of random vectors $\mathbf{z}_{1:t}$, given its corresponding observed time series of random vectors $\mathbf{x}_{1:t}$, satisfies
\begin{align}\label{rwe}
\left[\mathbf{z}_{1:t}\mid \mathbf{x}_{1:t}\right]\propto \left[\mathbf{x}_{1:t}\mid \mathbf{z}_{1:t}\right]\left[\mathbf{z}_{1:t}\right].
\end{align}
Note that here $[\mathbf{x}_{1:t}\mid \mathbf{z}_{1:t}]$ essentially describes the distribution of a (discretised) Poisson process, given that its intensity is obtained through a realisation of a (discretised) random field, and $[\mathbf{z}_{1:t}]$ represents the unconditional distribution of the (discretised) field. 
Hence, to infer about $\mathbf{z}$, we need to be able to sample  $\mathbf{z}_{1:t}$ from the conditional distribution $\left[\mathbf{z}_{1:t}\mid \mathbf{x}_{1:t}\right]$.  We may approximate the right hand side of \eqref{rwe} by $[\mathbf{x}_{v:t}\mid \mathbf{z}_{v:t}][\mathbf{z}_{v:t}]$, where $t-v$ is a small integer, since remote past observations tend to have a negligible effect on the current state. The distribution $\left[\mathbf{z}_{v:t}\right]$ can be evaluated using the estimates of the parameters $\sigma^{2}$, $\phi$, and $\theta$.

In the Supplementary material we provide details on the Fast Fourier transform for covariance matrix computation and the Metropolis-adjusted Langevin algorithm for simulation.

\subsubsection{Forecasting}\label{s:Forecast}
In this work, the emergency alarm call point patterns at the last six time points, i.e.~days, are reserved and we forecast their corresponding point patterns to visually evaluate the 
applicability of the spatio-temporal log-Gaussian Cox process model. We also assess the spatial dependence between the forecasted emergency alarm calls and the true (unknown) calls, reflected by the corresponding observed point patterns. 

Furthermore, we aim at forecasting the emergency alarm call locations beyond the last time point $T$ of data observation. 

Using the last time point $T$, the random intensity function in equation \eqref{key3w2} can be rewritten as 
\begin{align}\label{keysafuu}
\Lambda(s,T+\Delta)
=
\lambda_{0}(s)\lambda_{1}(T+\Delta)\exp\left\{Z(s,T+\Delta)\right\},
\end{align}
where $\Delta$ is the number of time units (days in the case of the ambulance data) beyond the last time point $T$. Let $q = MN$, where $M$ and $N$ are the number of rows and columns of the discretisation of the spatial domain, see expression \eqref{extendedcentriod} in the Supplementary material. On the extended lattice (with dimension $q\times q$), the time series  $ \mathbf{z}^{ext}_{t}$, $t=1, 2, 3, \ldots,T$, is assumed to be a stationary Gauss–Markov process. It follows that this is a discrete-time sample of an Ornstein–Uhlenbeck process  \citep{brix2001spatiotemporal}, which solves the stochastic differential equation
\begin{align}\label{MMsobduu}
d\mathbf{z}^{ext}_{t} = \boldsymbol\theta\left(\boldsymbol\mu-\mathbf{z}^{ext}_{t}\right)dt + \boldsymbol\sigma d\boldsymbol\epsilon_{t},
\end{align}
where $\boldsymbol\theta$ is a $q \times q$ invertible real matrix, $\boldsymbol\mu$ is a $q$-dimensional long-run mean, $\boldsymbol\sigma$ is a $q \times q$ positive
semidefinite matrix, $\boldsymbol\epsilon_{t}$ is a $q$-dimensional Wiener process \citep{jacobsen1993brief} and $t = 1, 2, 3, \ldots,T$. The diffusion coefficient $\boldsymbol\sigma$ measures the size of the noise,
$\boldsymbol\mu$ is the asymptotic mean level for the process $\mathbf{z}^{ext}_{t}$ and $\boldsymbol\theta$ determines how fast the process reacts
to perturbations. Solving \eqref{MMsobduu} and using the stationary Gauss–Markov process property of the process, 
we have that 
\begin{align}\label{key987}
\mathbf{z}^{ext}_{T+\Delta}
\mid
\mathbf{z}^{ext}_{T}
\sim
\mathcal{N}\left(
\boldsymbol\varphi(\Delta) \mathbf{z}^{ext}_{T} + (\mathbf{I}-\boldsymbol\varphi(\Delta))\boldsymbol\mu, 
\int_{T}^{T+\Delta} \boldsymbol\varphi(T+\Delta-u)
\boldsymbol\sigma
\boldsymbol\sigma' \boldsymbol\varphi(T+\Delta-u)'
du
\right),
\end{align}
where $\boldsymbol\varphi\left(\Delta\right) = \exp\{-\boldsymbol\theta \Delta\}$ and $\mathbf{I}$ is $q \times q$ identity matrix. Given the observed data $\mathbf{x}_{\nu:T}$, the forecast distribution of $\mathbf{z}_{T+\Delta}$ can be formulated as 
\begin{eqnarray*}
\left[\mathbf{z}^{ext}_{T+\Delta}\mid \mathbf{x}_{\nu:T}\right] = \int\left[\mathbf{z}^{ext}_{T+\Delta}\mid \mathbf{z}^{ext}_{T}\right]\left[\mathbf{z}^{ext}_{T}\mid \mathbf{x}_{\nu:T}\right]d\mathbf{z}^{ext}_{T},
\end{eqnarray*}
and based on equation \eqref{key987}, the expected value of the forecast distribution $\mathbf{z}^{ext}_{T+\Delta}\mid \mathbf{x}_{\nu:T}$ can be obtained as follows: 
\begin{eqnarray}
\E\left[\mathbf{z}^{ext}_{T+\Delta} \mid \mathbf{x}_{\nu:T}\right]  
&=& \E\left[
\E\left[
\mathbf{z}^{ext}_{T+\Delta}\mid \mathbf{z}^{ext}_{T}\right]
\mid \mathbf{x}_{\nu:T}\right],
\notag\\
&=&
\boldsymbol\varphi\left(\Delta\right)
\E\left[\mathbf{z}^{ext}_{T} \mid \mathbf{x}_{\nu:T}\right] 
+ 
\left(1-\boldsymbol\varphi\left(\Delta\right) \right)\boldsymbol\mu,
\label{e334r}
\end{eqnarray} 
Using the expected values in equations \eqref{key987} and \eqref{e334r}, the variance of the forecast distribution $\mathbf{z}^{ext}_{T+\Delta}\mid \mathbf{x}_{\nu:T}$ may be obtained as 
\begin{eqnarray}\label{unityp}
\Var\left( \mathbf{z}^{ext}_{T+\Delta}\mid \mathbf{x}_{\nu:T}\right) 
= 
\boldsymbol\varphi\left(\Delta\right)
\Var\left(\mathbf{z}^{ext}_{T}\mid \mathbf{x}_{\nu:T}\right)
\boldsymbol\varphi\left(\Delta\right)' 
+ \int_{T}^{T+\Delta} \boldsymbol\varphi\left(T+\Delta-u\right)\boldsymbol\sigma\boldsymbol\sigma' \boldsymbol\varphi\left(T+\Delta-u\right)'\du.
\end{eqnarray}
Simplifying assumptions need to be imposed to estimate the forecast distribution. One of the key assumptions is to assume the parameter matrices $\boldsymbol\theta$ and $\boldsymbol\sigma$ to be $\mathbf{I}\theta$ and $\mathbf{I}\sigma$, and  \citet{taylor2013lgcp} have implemented the estimation of the forecast distribution under these simplified assumptions. Using these assumptions, equations \eqref{key987}, \eqref{e334r}, and \eqref{unityp} can be simplified and Monte Carlo estimation can be used to estimate the forecast expectation in equation \eqref{e334r}. Let $\mathbf{z}^{ext}_{T}{(i)}$ denote the $i^{th}$ sample obtained during the simulation of the Gaussian random field, which  can be taken as a draw from $[\mathbf{z}^{ext}_{T}\mid \mathbf{x}_{\nu:T}] \propto [\mathbf{x}_{\nu:T}\mid \mathbf{z}^{ext}_{T}][\mathbf{z}^{ext}_{T}]$ once the chain has reached stationarity. Using equation \eqref{key987}, we can then generate a draw $\mathbf{z}^{ext}_{T+\Delta}{(i)}$ from $[\mathbf{z}^{ext}_{T+\Delta}\mid \mathbf{z}^{ext}_{T}]$ given the sample $\mathbf{z}^{ext}_{T}{(i)}$. Considering a burn-in time $b$, the mean of the samples  $\mathbf{z}^{ext}_{T+\Delta}{(i)}$, $i = b, b+1, \ldots, B$, is an unbiased estimator of the expectation $\E[ \mathbf{z}^{ext}_{T+\Delta}\mid \mathbf{x}_{\nu:T}]$. Once we estimate the expected value in equation \eqref{e334r} and predict the expected number of emergency alarm calls by the estimated Poisson regression model, equation \eqref{keysafuu} can be computed to simulate a  spatio-temporal inhomogeneous Poisson process.

\section{Spatio-temporal ambulance call data analysis}\label{EASTD}

We have finally reached the point where we apply the above LGCP framework to the spatio-temporal ambulance call data in Section \ref{Data}. 
Recall that the main task here is to simulate realistic future emergency alarm calls in both time and space, where time is expressed in days. 
This was achieved using equation \eqref{key987} with components estimated as described below. At the end of this section we describe how the fitted model was used to simulate future alarm calls.

\subsection{Comparing bandwidth selection methods for the ambulance data} \label{EASTDD}
Applying a kernel intensity estimator with a quartic kernel to our ambulance data, we have compared the K-means clustering based bandwidth selection method with some existing bandwidth selection methods. Some of the existing bandwidth selection methods we have explored are the maximal smoothing principle of  \citet{terrell1990maximal}, a mean-squared error criterion based cross-validation method \citep{berman1989estimating}, likelihood cross-validation method \citep{Load99}, the Cambell formula-based criterion of \citet{cronie2018non}, isotropic fixed bandwidth selection based on likelihood cross-validation and unbiased least squares cross-validation \citep{davies2018fast}. Unlike many existing fixed bandwidth selection methods, such as those of \citet{davies2018tutorial},  \citet{cronie2018non} and \cite{Load99}, K-means clustering based bandwidth selection is fast computationally and works well for big data, such as our ambulance data. Bootstrapping can also be used to select bandwidth; recently  \citet{davies2018tutorial} have proposed a bootstrapping approach for bandwidth selection. However, this approach is not feasible for our ambulance data as it requires very large quantities of memory at its default values of the algorithm. 

To select the number of clusters, $K$, to use, we have turned to visual inspection; we found that estimates with $K>5$ tended too much to accentuate regions which (visually) have a low spatial intensity. 
In practice, to compute the spatial intensity component we use the function \texttt{kernel2d} in the \textsf{R} package \texttt{splancs}. The bandwidth selection method has been carried out using our own implementation in \textsf{R}.

\begin{figure}[H]
\centering
\includegraphics[width=0.75\textwidth, height=0.45\textwidth]{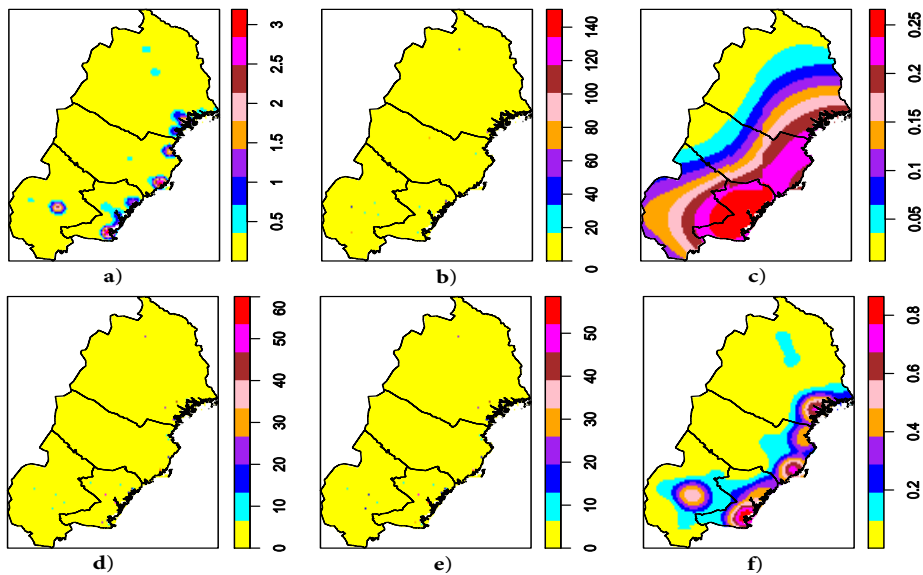}
\caption{The roles of different bandwidth selection methods, using quartic kernel intensity estimation for the ambulance data in Figure \ref{Or21}: the maximal smoothing  principle (a), mean-square error based cross-validation (b), the Cronie and Van Lieshout criterion (c), Poisson likelihood cross-validation (d), unbiased least squares cross-validation (e) and K-means clustering (f). Note that what interests us here are the relative scales rather than the raw scales, since the final spatial intensity estimate will be scaled to a spatial density function; note that the values in the plots above have been multiplied by 1000 for ease of visualisation. }
\label{Or3567456}
\end{figure}

Figure \ref{Or3567456} illustrates a comparison between the different methods.  Based on the observed data in Figure \ref{Or21}, we argue that the K-means clustering based bandwidth selection performs best in balancing the over- and under-smoothing of the spatial intensity of the ambulance call events.
 
\subsection{Non-parametric estimation of the spatial component of the stochastic intensity function}
\label{s:SpatialIntensityData}

The first thing that we do is to obtain a non-parametric estimate of the spatial intensity component $\lambda_0(s)$, $s\in R$, which we do in accordance with Section \ref{s:IntensityEstimation}. More specifically, we employ the quartic kernel estimator \eqref{eq1}, where we select the bandwidth using the $K$-means clustering based method outlined in Section \ref{s:Bandwidth}; the outcome, which is visually illustrated in the left panel of Figure \ref{Or3567} (or the last panel of Figure \ref{Or3567456}), suggests that hotspot (or high intensity) regions (red/pink coloured) have been estimated in the south-eastern part of the study region. 
In addition to visual inspection motivating the choice $K=5$, this was also the choice which gave rise to the simulated forecasted point patterns that most resembled the true data (see the part on forecasting below).

\subsection{Non-parametric spatio-temporal interaction analysis}

\begin{figure}[H]
	\centering
	\includegraphics[width=0.35\textwidth, height=0.35\textwidth]{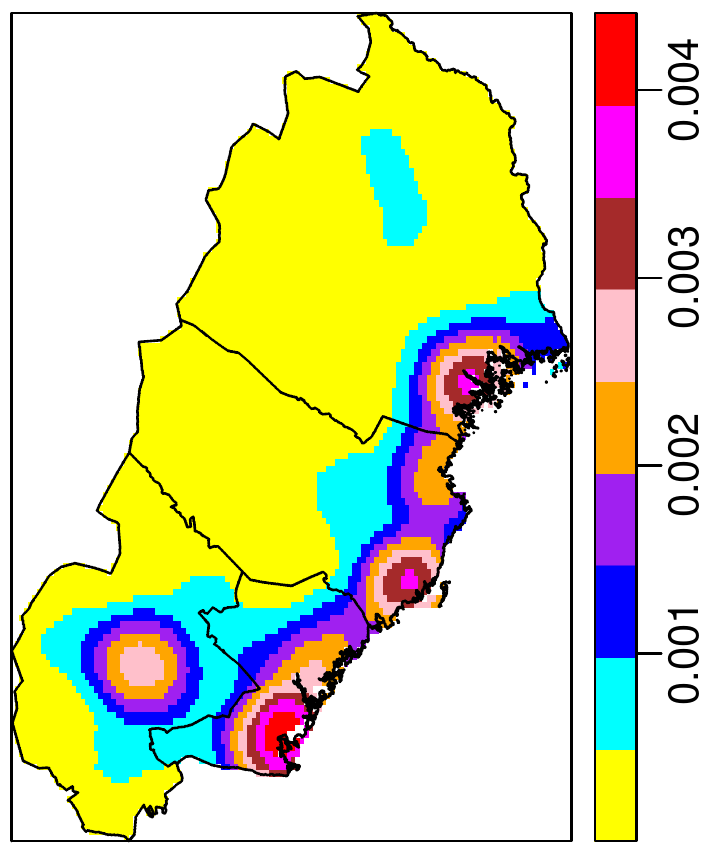}
	\includegraphics[width=0.35\textwidth, height=0.375\textwidth]{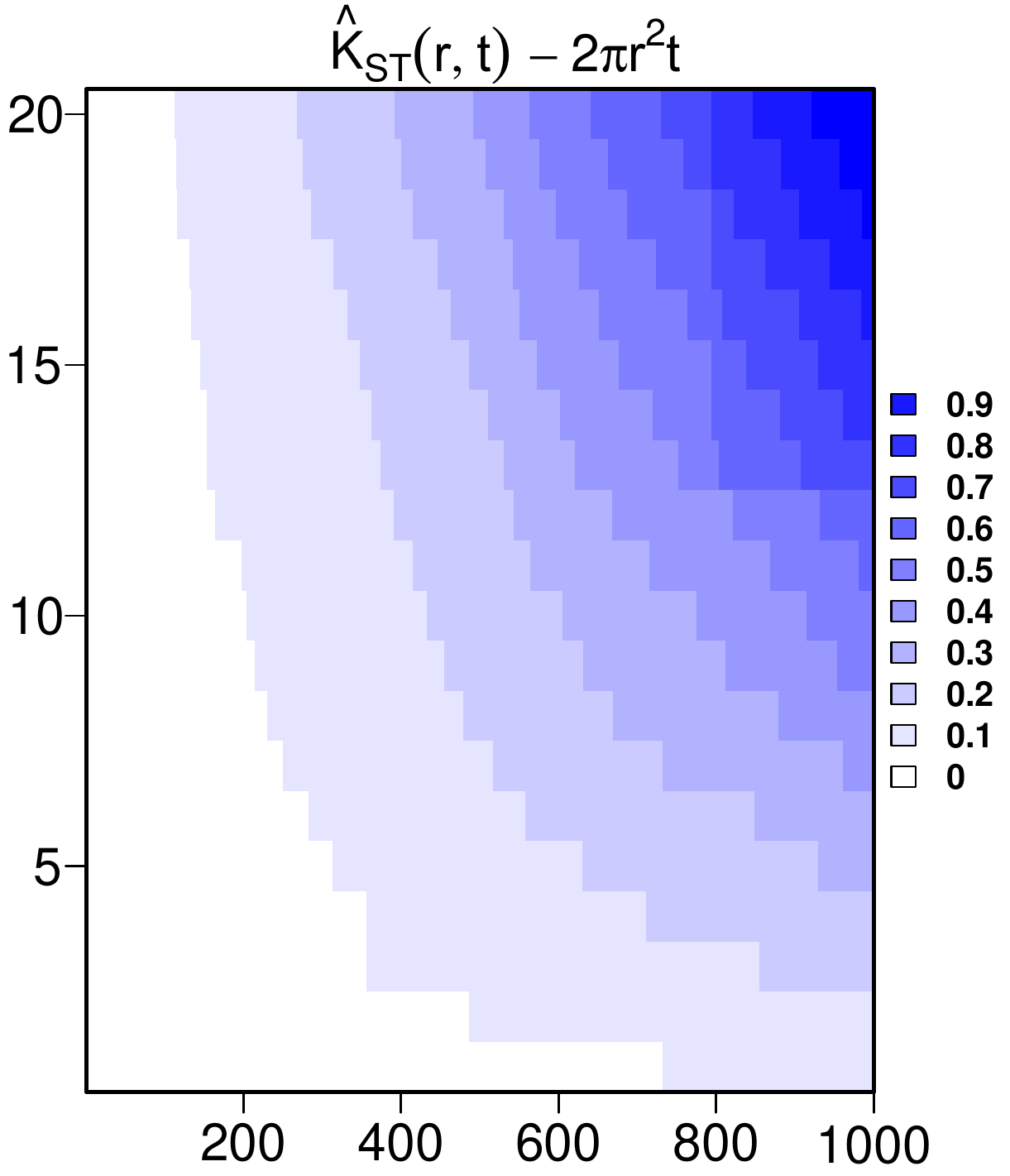}
	\caption{The plots of the estimated spatial kernel intensity (left) and inhomogeneous $K$-function (right) for the ambulance call data. In the right plot, the x-axis label denotes spatial lags in meters while the y-axis label represents temporal lags in days.}
	\label{Or3567}
\end{figure}

Before choosing a particular model to fit to a given point pattern, it is paramount to carry out a non-parametric analysis of spatio-temporal interaction, to better understand the data and to see what type of model may be appropriate for the data. Here the idea is to quantify the interaction (clustering/aggregation or inhibition/regularity) present at/within different inter-point distances. As previously indicated, we here employ the inhomogeneous $K$-function estimator $\hat{K}(r,t)$ in \eqref{kfun}, which is compared with the theoretical $K$-function $K\left(r, t\right) = 2\pi r^{2}t$ obtained for an inhomogeneous spatio-temporal Poisson process (with any intensity function), to see whether there is excess spatio-temporal interaction in the underlying spatio-temporal point process.  Recall that we plug an intensity estimate $\hat\lambda(s,t)$ into expression \eqref{kfun}, which we here let be given by a product of the spatial intensity estimate obtained in Section \ref{s:SpatialIntensityData} and a temporal kernel estimate with an Epahnechnikov kernel. Given the scale of the data, the effects of the miss-specification that our data are discrete in time while \eqref{kfun} is, in fact, defined for temporally continuous data are negligible.  The right panel of Figure \ref{Or3567} shows a plot of $\hat{K}(r, t) - 2\pi r^{2}t$, where all values have been scaled by the maximum value of the estimator $\hat{K}(r,t)$ in order to obtain values between 0 and 1. We see that there is aggregation present in our data; the right panel of the figure shows that space-time clustering subsists in our emergency alarm call data. For instance, there seems to exist space-time clustering in our data for ranges within 0.9 $km$ and 19 days. There is a caveat, however, which is that since we assume first-order separability, $\hat{K}(r,t)$ reflects both spatio-temporal aggregation and non-separable first-order effects. It should further be noted that there may be stronger temporal interactions present, which operate on time lags smaller than one time unit, i.e.~one day. To verify spatio-temporal clustering  more formally, we have applied the $K$-function based Monte-Carlo test procedure of \citet{diggle1995second}, which is found in the \textsf{R} package \texttt{splancs}, using the function \texttt{stmctest}. Note that this test is based on stationary $K$-functions but since the space and time lags/ranges considered are small, we deem it justified to still apply the test. The test, which is a Monte-Carlo test, is based on repeatedly randomly rearranging the recorded time stamps of the events and using  the sum of residuals based on $K$-functions to evaluate  the existence of spatio-temporal clustering in the data. One may formally test the null hypothesis that there is lack of space-time clustering against the alternative that there in fact is space-time clustering present. However, we here we do not formally carry out such a test, but merely look for indications of space-time clustering. Accordingly, 86\% of the 200 randomly rearranged datasets yielded test statistic values which were smaller than the test statistic value obtained for the original (unshuffled) data -- a large value of the test statistic is an indication of clustering. We view this as evidence of the presence of space-time clustering, thus supporting the above argument. Hence, these results advocate for employing an LGCP-model (a special case of an LGCP is a Poisson process).  

\subsection{Estimation of the temporal intensity component of the stochastic intensity function}

We next turn to modelling the temporal evolution of the calls, which is done in accordance with Section \ref{s:TemporalIntensityModelling}. 
To fit the Poisson regression model for the temporal component $\lambda_{1}(t)$, we use the iteratively reweighted least squares method and the obtained results can be found in Table \ref{tab:Poissonlog}; in practice, we use the function \texttt{glm} in the \textsf{R} package \texttt{stats}. We see that our model is neither over nor under-estimating as the median deviance residual is close to zero, see Table \ref{tab:Poissonlog}. By looking at the residual deviance we see that the inclusion of covariates in the Poisson regression model generates a better fit as compared to how well we would have performed with a model that only includes the intercept (reflected by the null deviance). The table also shows that all covariates are significant, thus reflecting that the daily expected number of calls has a strong relationship with the covariate day-of-the-week as well as with  season-of-the-year, e.g.~summer and fall. Of course, one could naturally think of further calendar-based refinements here.

 \begin{table}[!htpb]
 	\caption{Estimated Poisson regression model for the temporal component of the spatio-temporal intensity function.}
 	\label{tab:Poissonlog}
 	\hspace*{-0.2cm}
 	\centering
 	\begin{tabular}{l|c|c|c|c}
 		\hline 
 		Variables &estimate&std.error ($\times10^{-3}$)&z-value& $P\left(>\abs{z}\right)$  \\\hline
 		Monday	&5.433&5.840&930.261&0.000\\
 		Tuesday &5.396&5.875&918.443&0.000\\
 		Wednesday&5.398&5.867&920.032&0.000\\
 		Thursday&5.403&5.862&921.831&0.000\\
 		Friday	&5.447&5.818&936.104&0.000\\
 		Saturday&5.491&5.743&956.086&0.000\\
 		Sunday	&5.442&5.813&936.217&0.000\\
 		Summer	&0.021&9.522&2.191&0.028\\
 		Fall	&0.019&7.939&2.349&0.019\\
 		Cos(wt)	&0.059&4.663&12.550&0.000\\
 		Sin(wt)	&0.035&3.920&8.852&0.000\\
 		Cos(2wt)&0.024&3.160&7.681&0.000\\
 		Sin(2wt)&0.019&2.614&7.357&0.000\\	
 		time/day of event&0.000&0.003&20.404&0.000\\\hline
 		&\multicolumn{3}{|c|}{Computed value}& Degrees of freedom\\\hline
 		Median deviance residual&\multicolumn{3}{|c|}{-0.023}&\\
 		Null deviance&\multicolumn{3}{|c|}{4002192}&1826\\
 		Residual deviance&\multicolumn{3}{|c|}{3282}&1812\\
 		Akaike information criterion&\multicolumn{3}{|c|}{16689}&\\\hline
 	\end{tabular}
 \end{table}
 
\begin{figure}[H]
\centering
\includegraphics[width=0.45\textwidth]{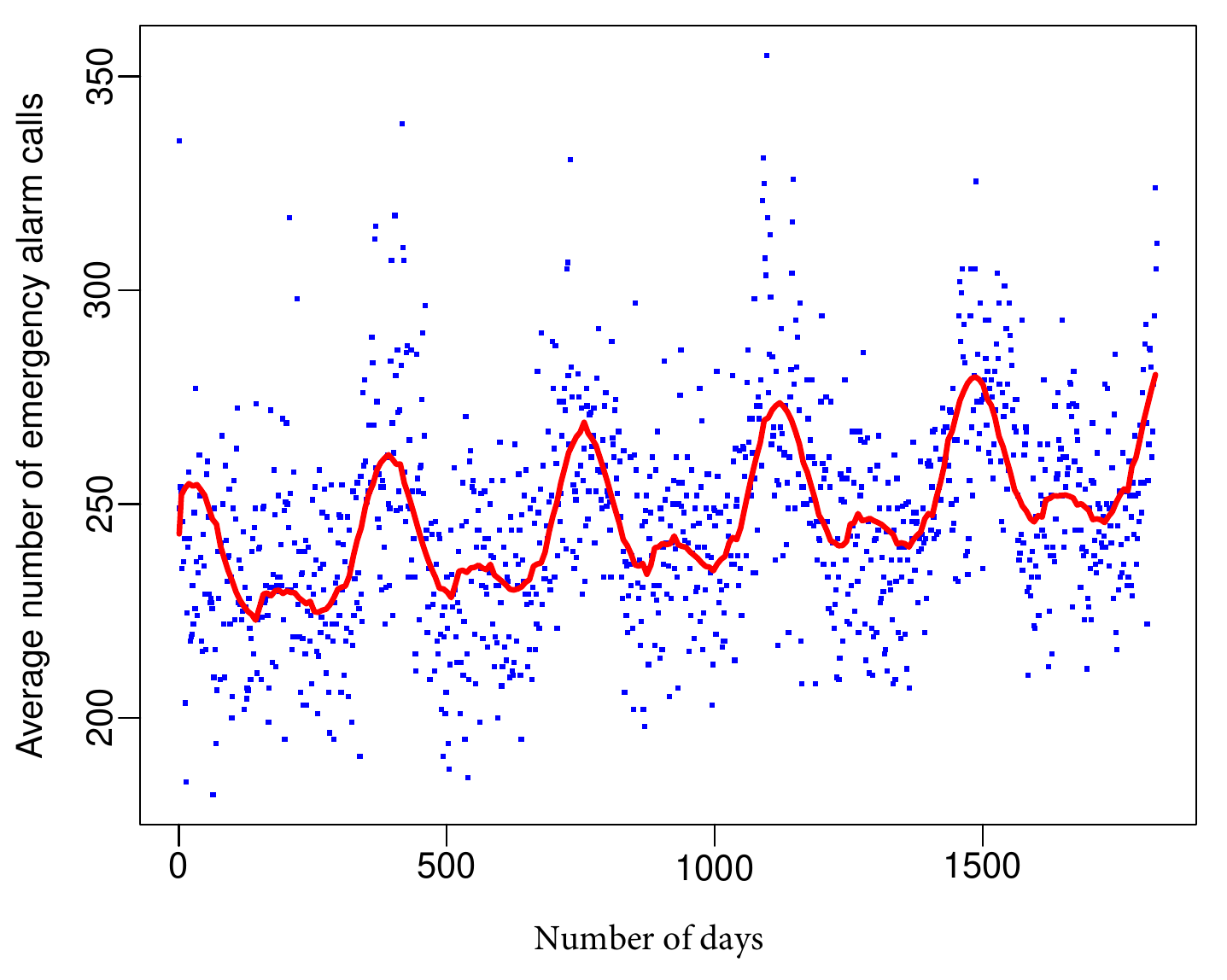}
\includegraphics[width=0.45\textwidth]{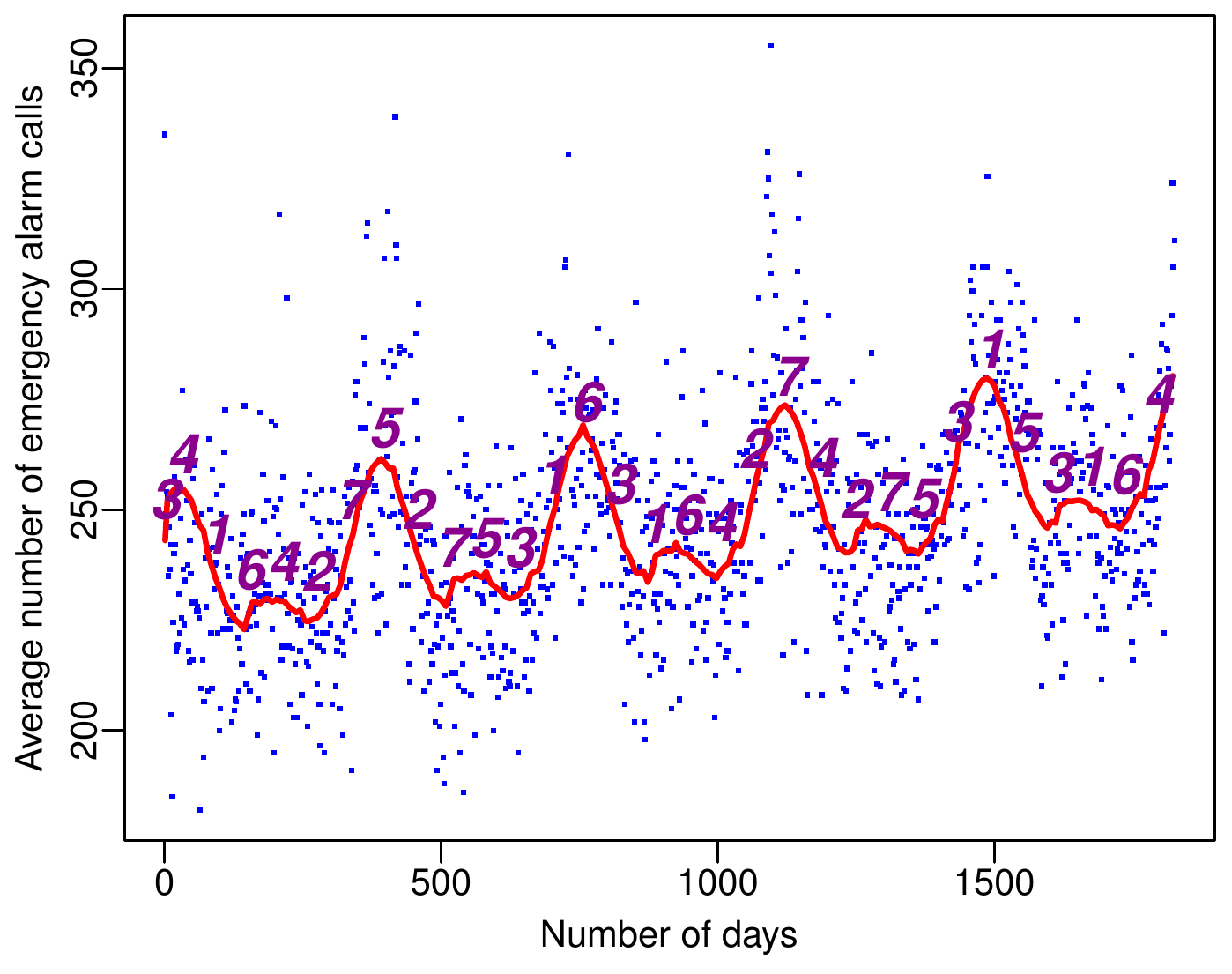}\\
\includegraphics[width=0.45\textwidth]{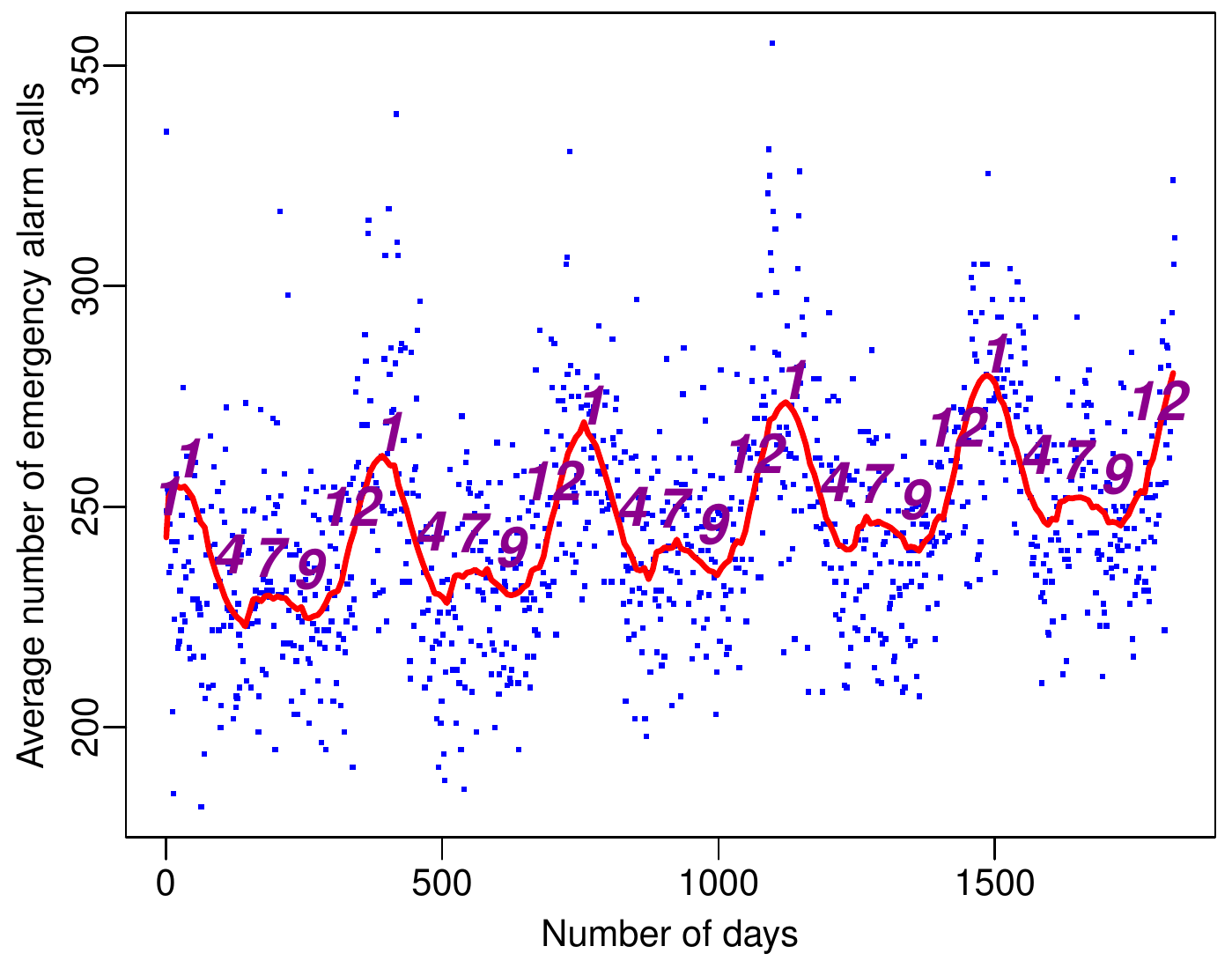}
\includegraphics[width=0.45\textwidth]{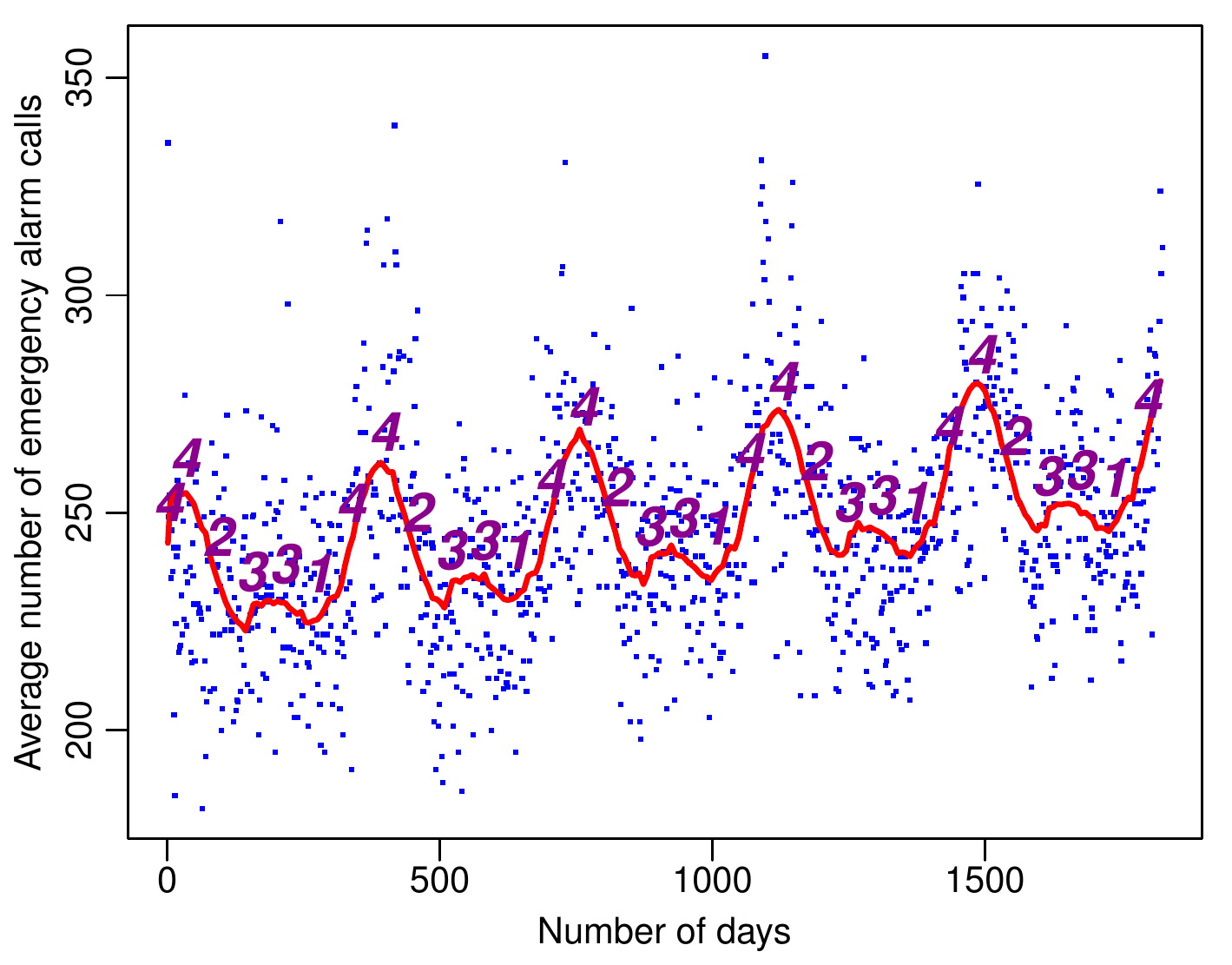}
\caption{The plot in the left panel of the first row represents the smoothed data (solid dots) and the fitted (solid lines) expected number of emergency alarm calls. The remaining plots are identical to the left plot in the first row but the numbers in the plot indicate the smoothed days (the right plot in the first row), smoothed months (the left plot in the second row) and smoothed seasons (the right plot in the second row) of the smoothed intensity at the indicated time index (number of days). Days-of-a-week: Monday (1), Tuesday (2), Wednesday (3), Thursday (4), Friday (5), Saturday (6), and Sunday (7). Months-of-a-year: January (1), February (2), $\ldots$, December (12). Seasons-of-a-year: Fall (1),  Spring (2), Summer (3), and Winter (4).}
\label{fig:FittedGLM}
\end{figure}

A plot of the fitted Poisson regression model and the observed number of emergency alarm calls is shown in Figure \ref{fig:FittedGLM}. We can here visually verify that the fitted model has captured the pattern in the observed data to a large degree. However, the model seems to under estimate the intensity during high and low intensity periods, which would suggest that a more flexible model and/or the inclusion of further/other covariates would be warranted. The deviating behaviour (see Table \ref{tab:Poissonlog}) of summer and fall can be visually detected by the offsets observed in the harmonic behaviour in Figure \ref{fig:FittedGLM}. It should finally be noted that holidays, which may have a tendency of driving up the intensity, have not been incorporated as covariates here.
In the presence of other covariates in the model setting, the daily expected number of calls has a strong positive relationship with the day-of-the-week. Moreover, in the presence of other covariates in the model, season-of-the-year, e.g.~Summer and Fall, are positively associated with the daily expected number of calls, see Table \ref{tab:Poissonlog}. The results in the figure suggest that the peaks of the expected number of emergency alarm calls occur in the winter season. However, the season-of-the-year covariate Winter is not significantly associated with the expected number of emergency alarm calls. This may be due to fact that the winter season in northern Sweden is the longest season, thus having a highly smoothing effect on the local information in our data and hence, the presence of days-of-the-week may in effect exclude the season-of-the-year covariate Winter in the model fitting. From Figure \ref{fig:FittedGLM} (top right panel) we can deduce that the expected number of calls tends to vary between different weekdays. Moreover, we see that the expected number of calls tends to peak in the winter, in particular in the month of January (see Figure \ref{fig:FittedGLM}, bottom panels). Given the fact that the coefficients corresponding to the different weekdays are roughly the same, the variation in the expected number of calls during shorter periods are reflected through the fitted trigonometric covariates. Variation corresponding to longer periods are reflected by the seasons. Note finally that, according to the fitted model, in the winter, the variations in the expected number of calls are explained by the weekdays and the trigonometric covariates.

\subsection{Covariance fitting,  forecasting and simulation of future ambulance call locations}

One of the main aims of forecasting is to obtain a dynamical model which step-wisely takes in new data and outputs updated forecasts, i.e.~on-line spatio-temporal risk-mapping. Hence, we here leave out datasets (test data) at the last six time points in the fitting of the model to the remaining data (training data) and then generate the forecast models for the last six time points to simulate realistic "future" spatial event locations.

By employing the minimum contrast approach outlined in Section \ref{s:LGCPfitting}, which deals with the estimation of the variance parameter $\sigma^2$, the spatial correlation function $r_{\phi}\left(u\right) = \exp\{-u/\phi\}$ and the temporal correlation function $r_{\theta}\left(v\right) = \exp\{-v/\theta\}$, 
we obtain the estimates $(\hat\sigma^{2}, \hat\phi, \hat\theta)
=
(4.933, 3494.705, 0.182)$; 
in practice, the estimation is carried out using the function \texttt{minimum.contrast.spatiotemporal} from the \textsf{R} package \texttt{lgcp}  \citep{taylor2013lgcp}. Hence, given the current separable setting, we have indications that there are intra-day dependencies at play, given the fitted spatial dependence structure (as the temporal covariance component is fitted conditionally on the estimated spatial one, see \cite{davies2013assessing}), and the variance parameter estimate indicates that we have more structure than in a Poisson process. At the same time, the fairly large spatial scale parameter estimate indicates that there is long-range dependence in space.

\begin{figure}[H]
\centering
\includegraphics[width=0.14\textwidth]{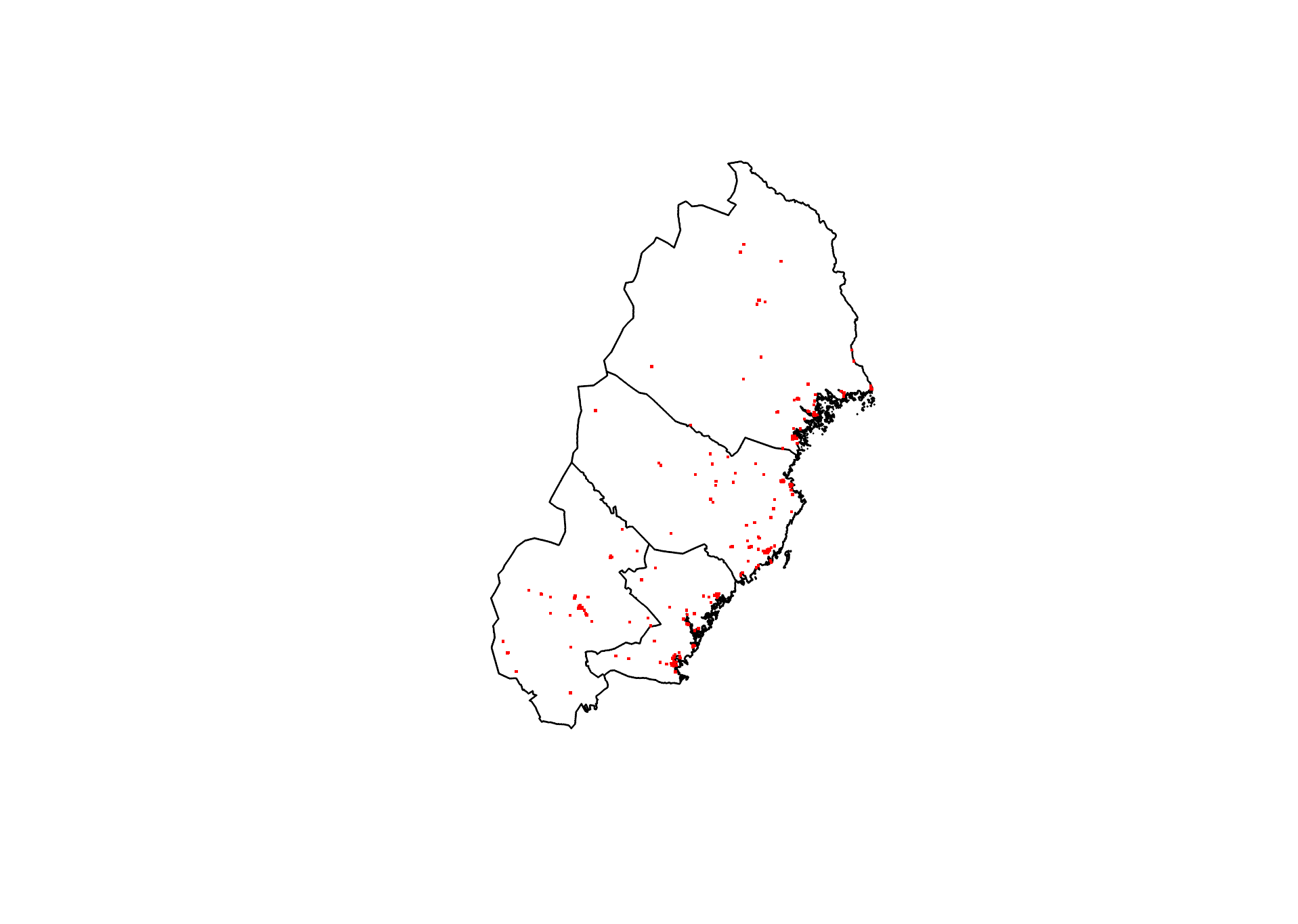}
\includegraphics[width=0.14\textwidth]{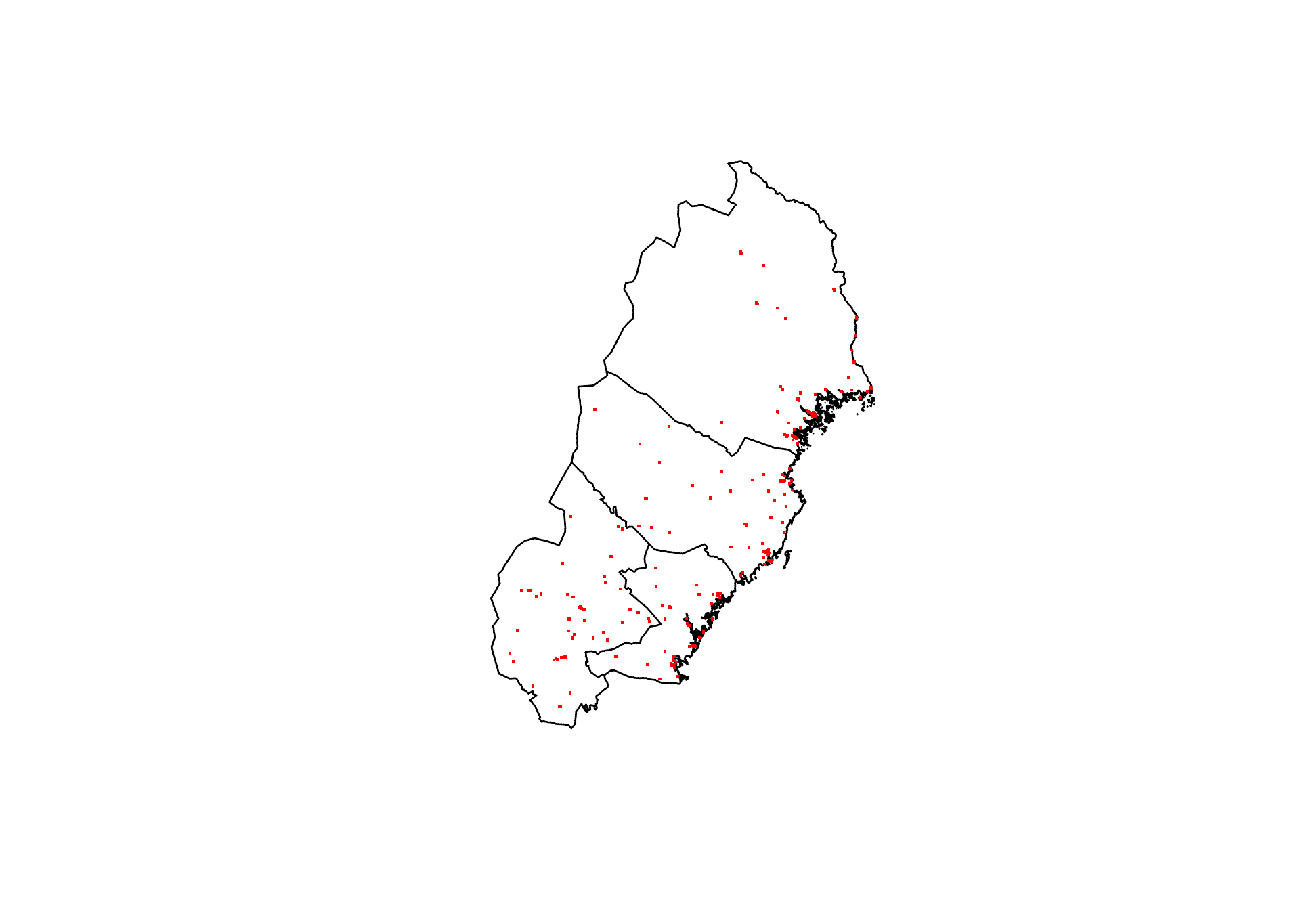}
\includegraphics[width=0.14\textwidth]{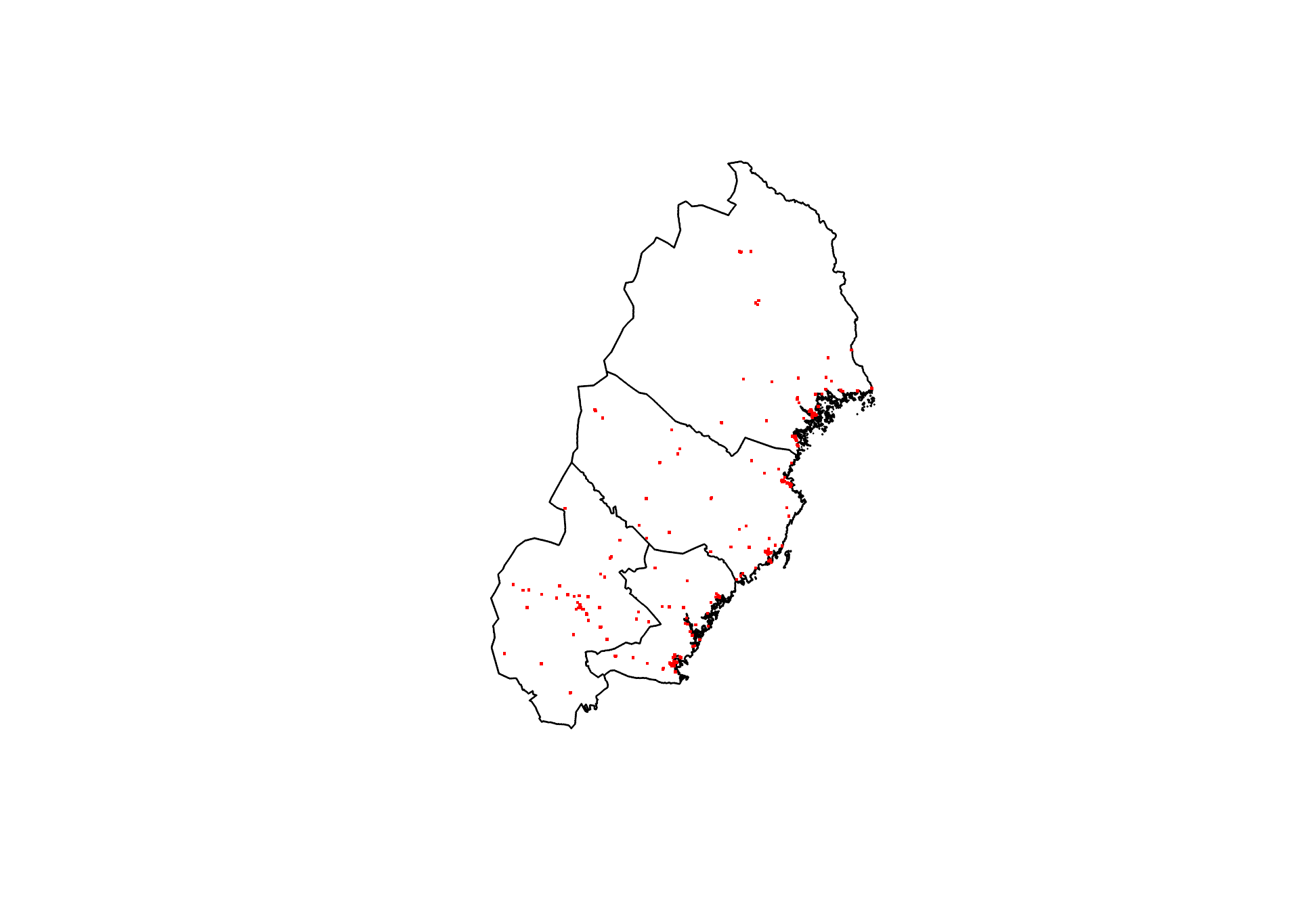}
\includegraphics[width=0.14\textwidth]{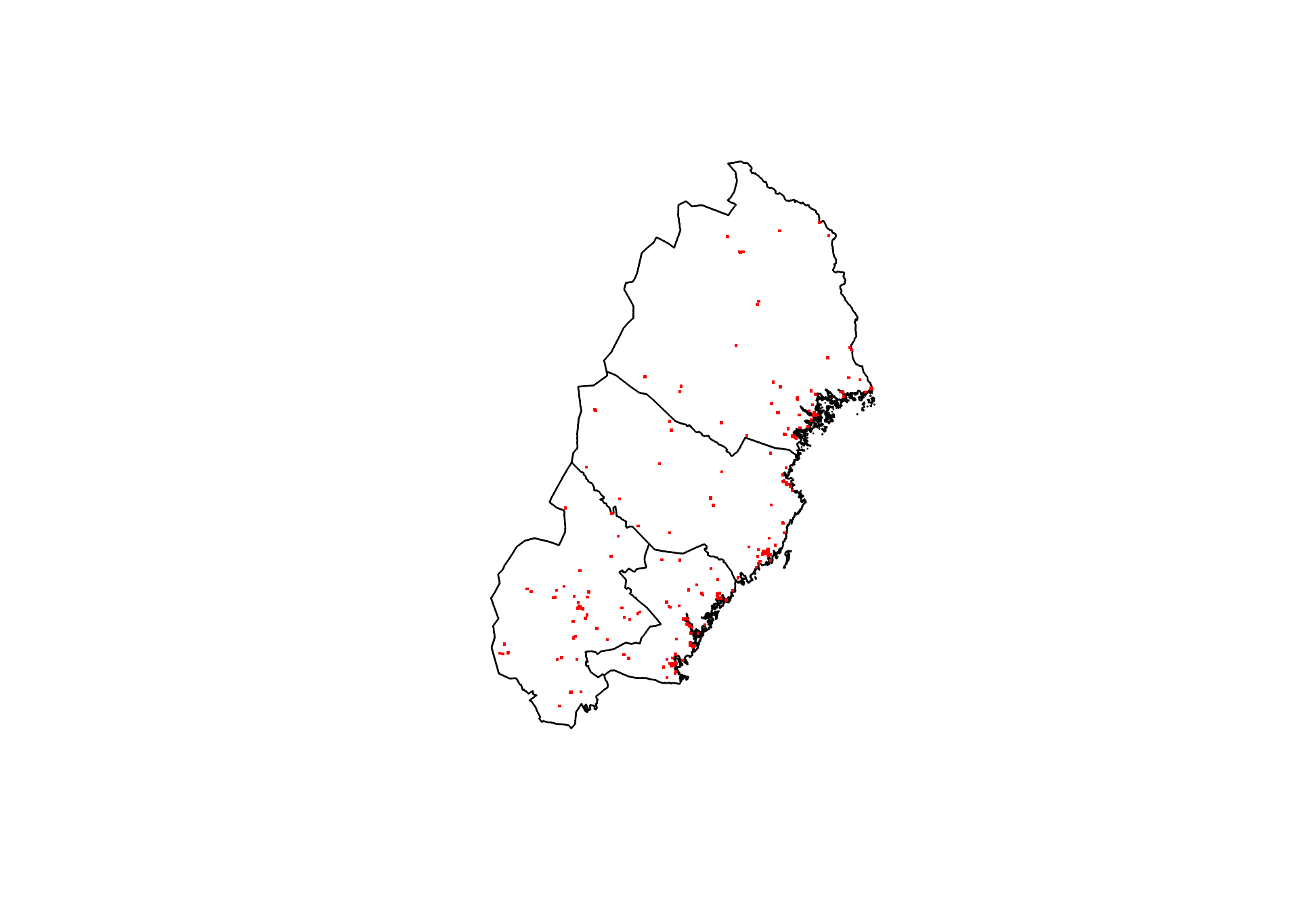}
\includegraphics[width=0.14\textwidth]{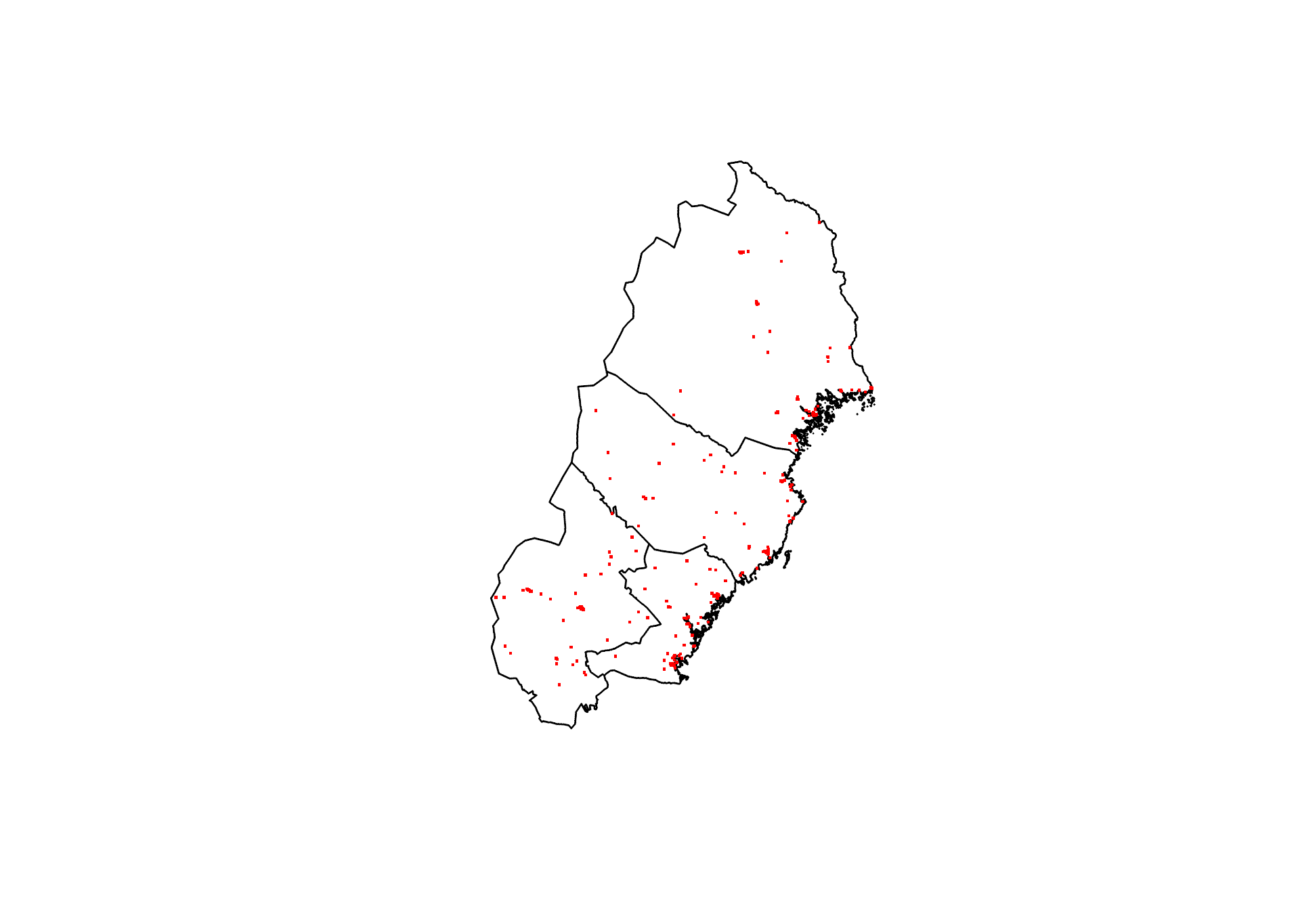}
\includegraphics[width=0.14\textwidth]{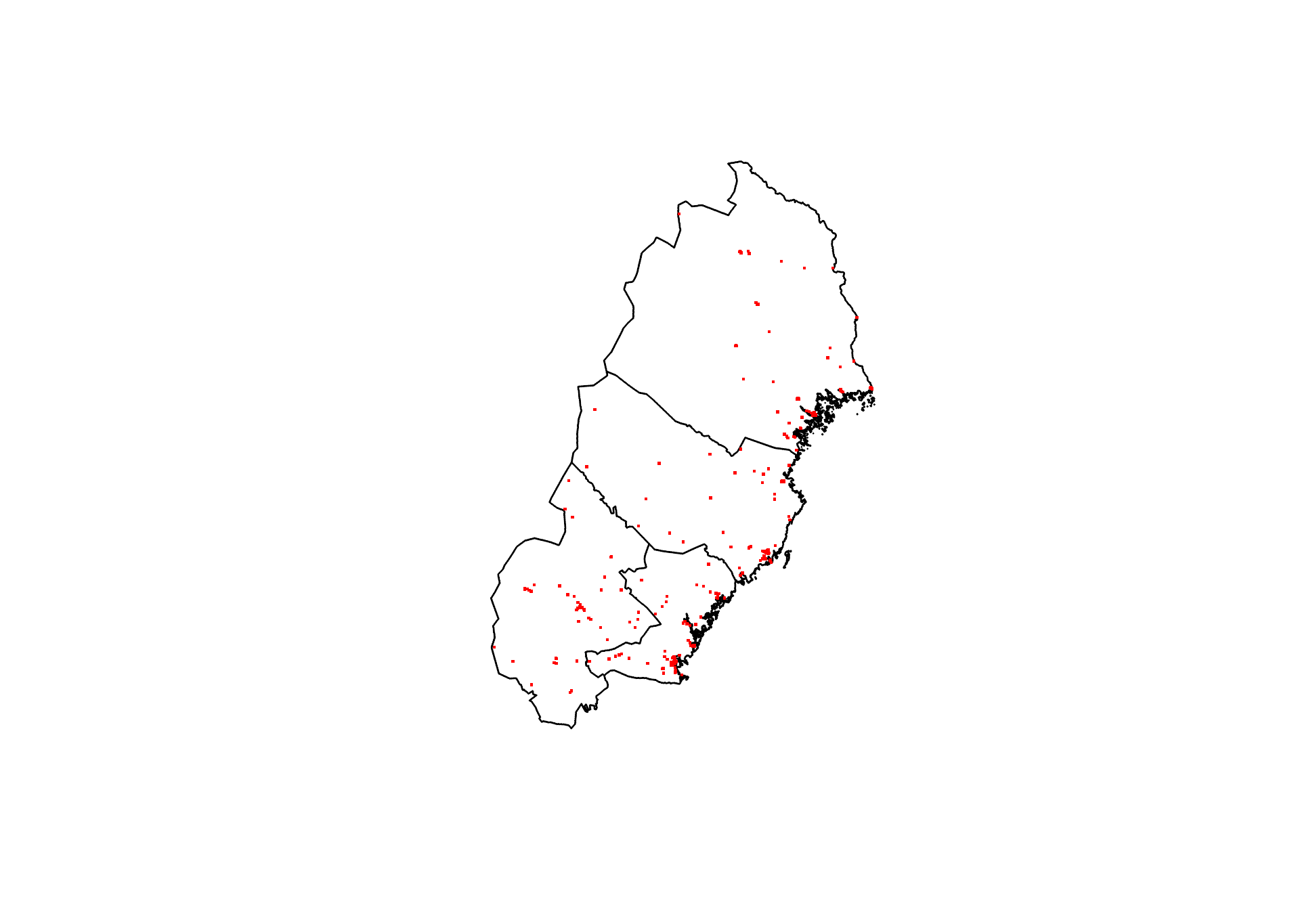}\\
\includegraphics[width=0.14\textwidth]{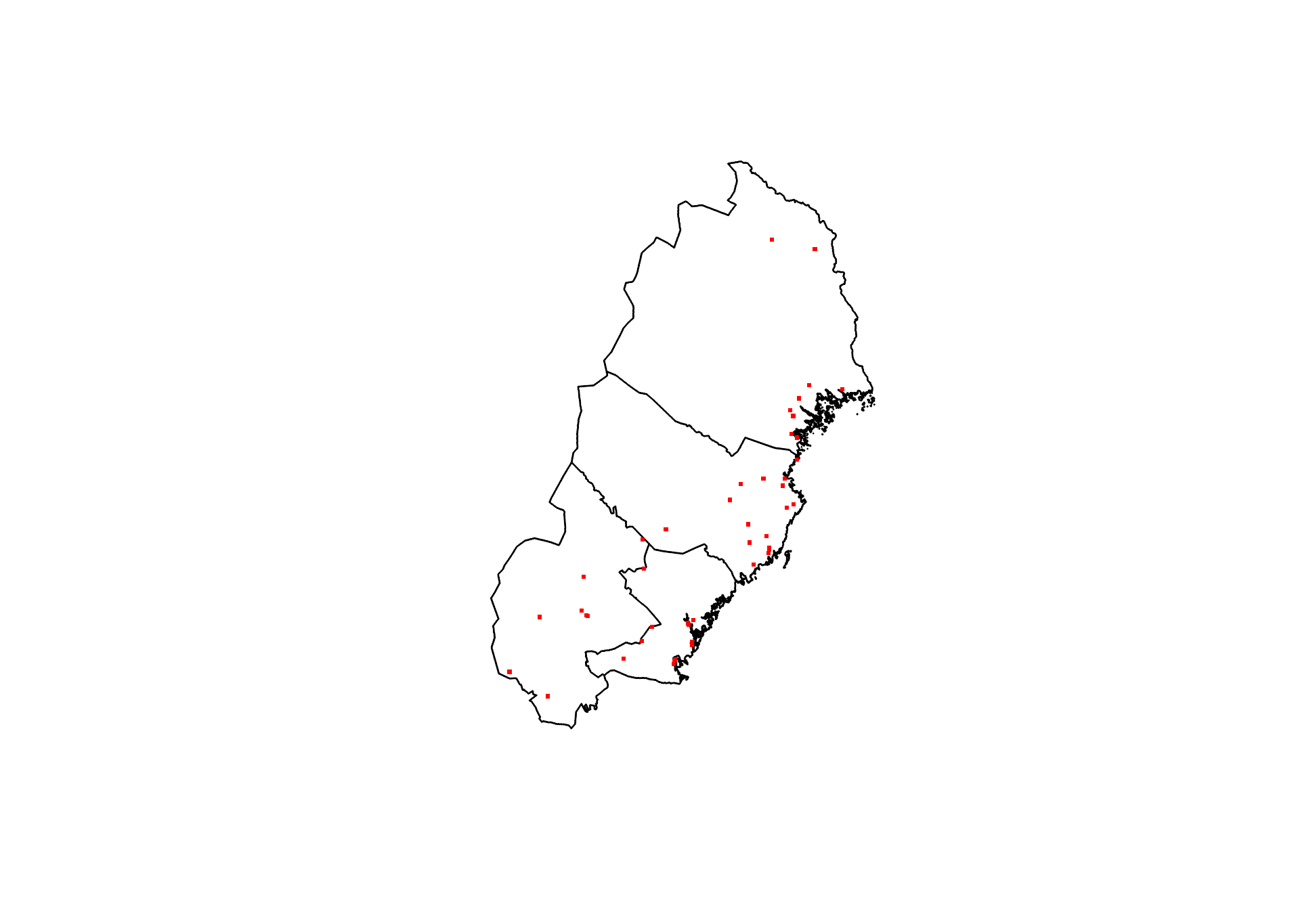}
\includegraphics[width=0.14\textwidth]{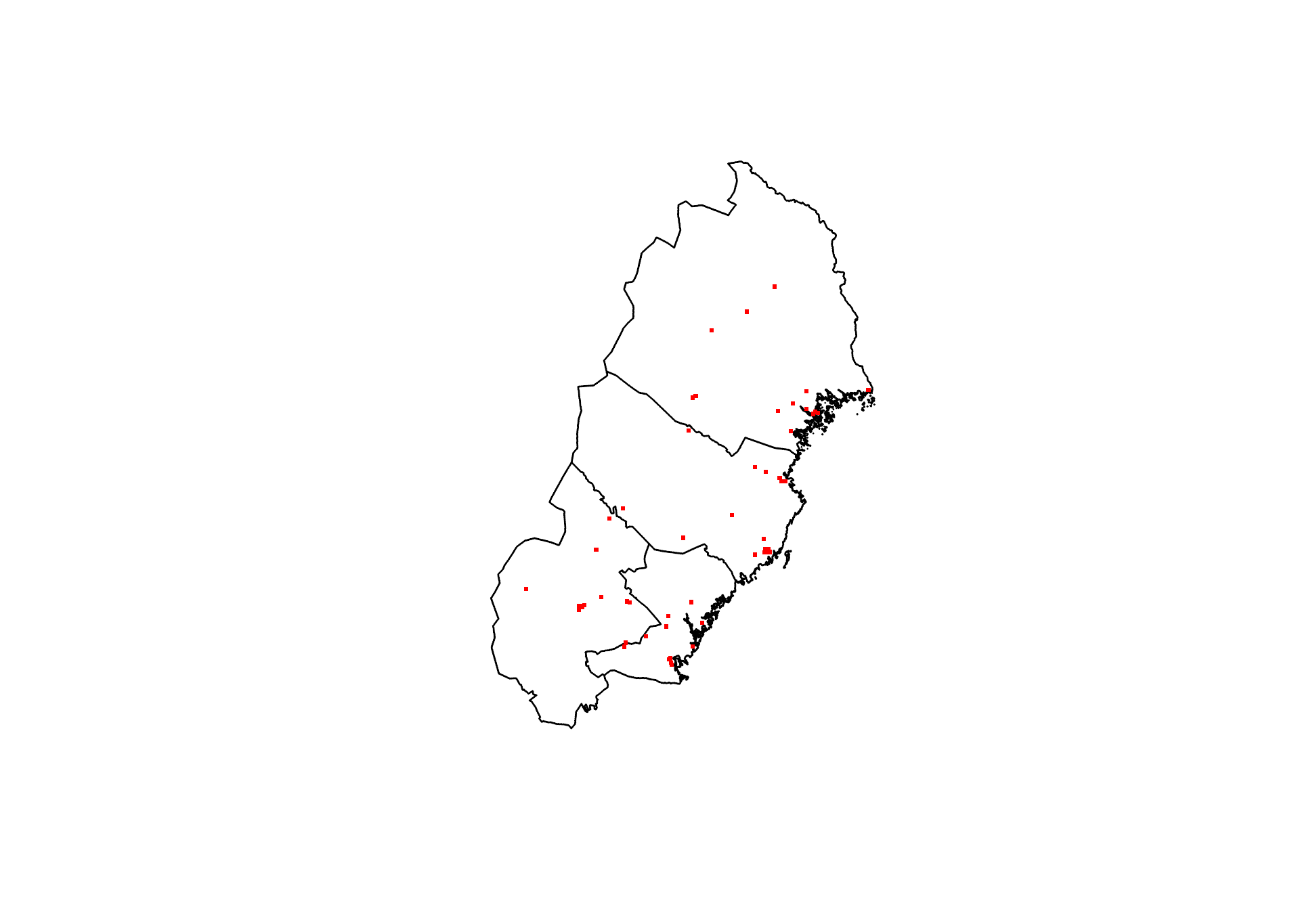}
\includegraphics[width=0.14\textwidth]{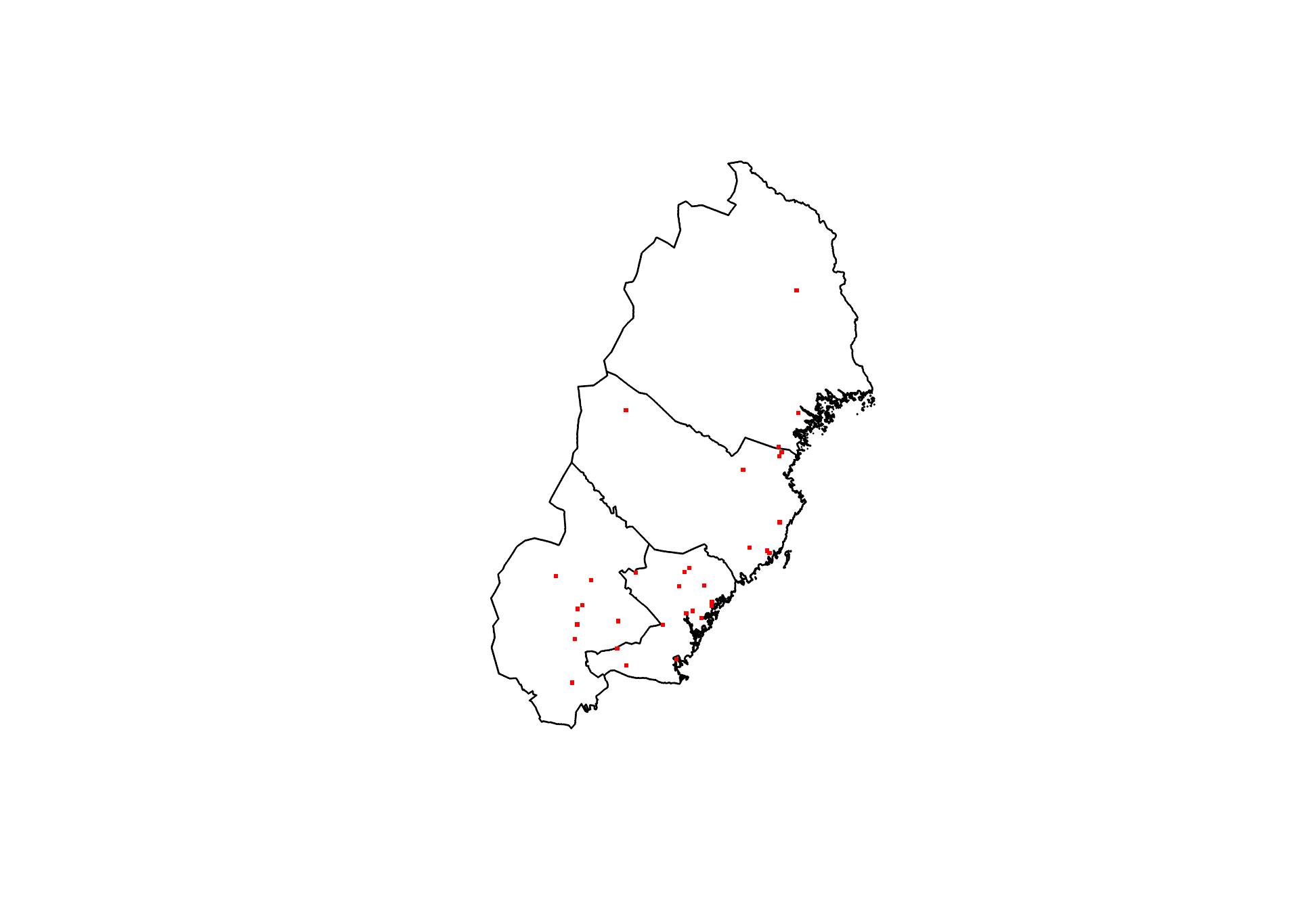}
\includegraphics[width=0.14\textwidth]{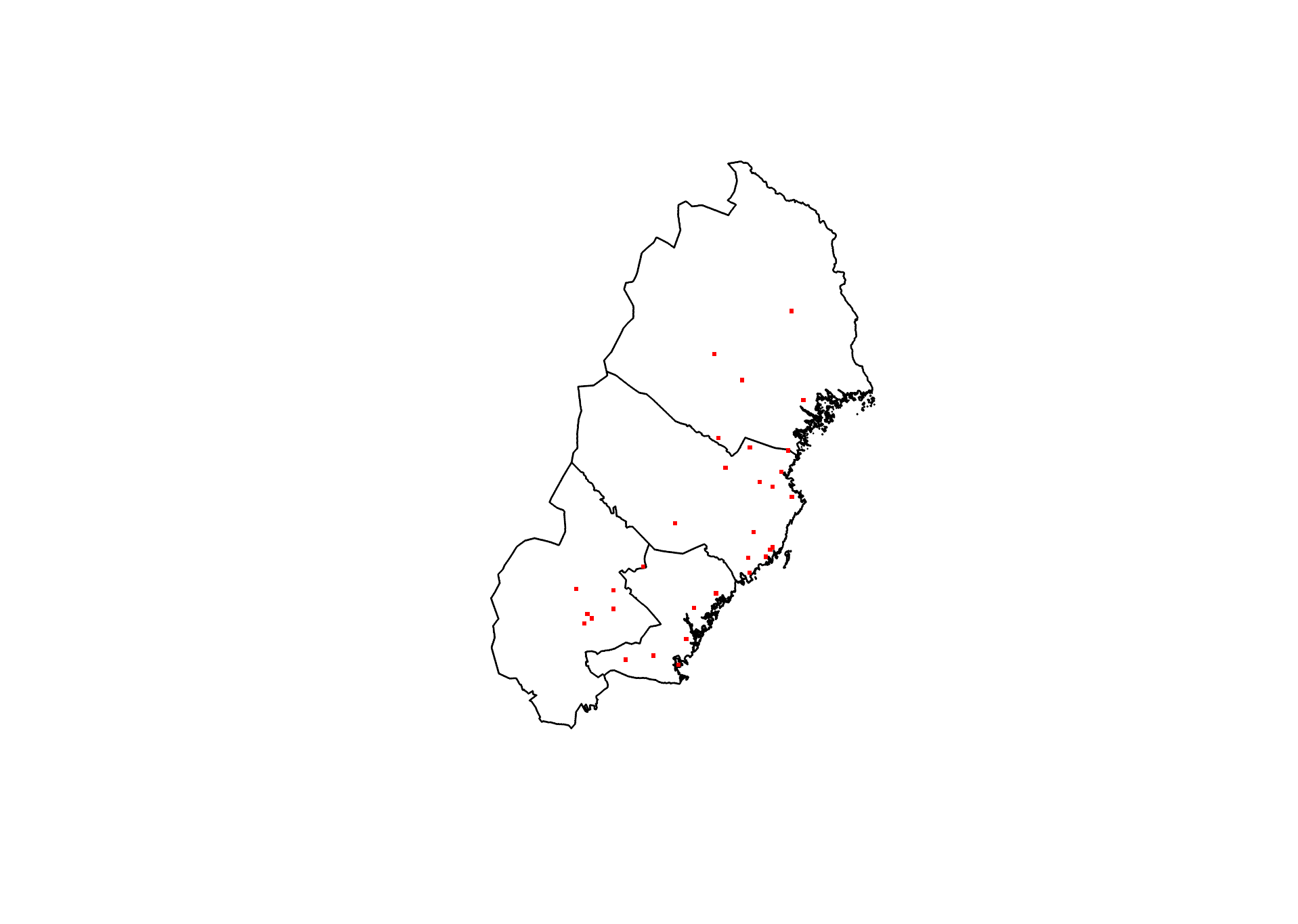}
\includegraphics[width=0.14\textwidth]{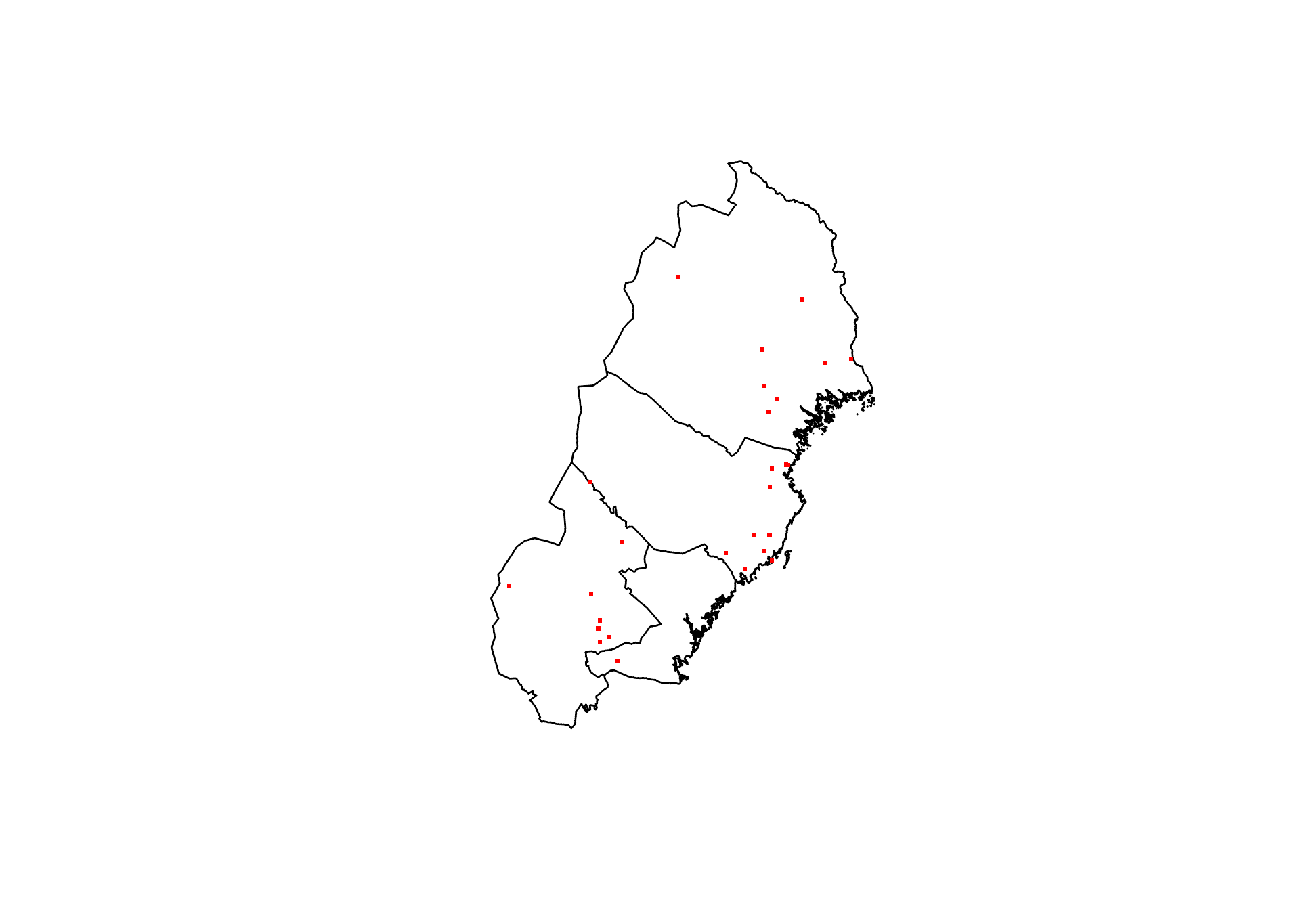}
\includegraphics[width=0.14\textwidth]{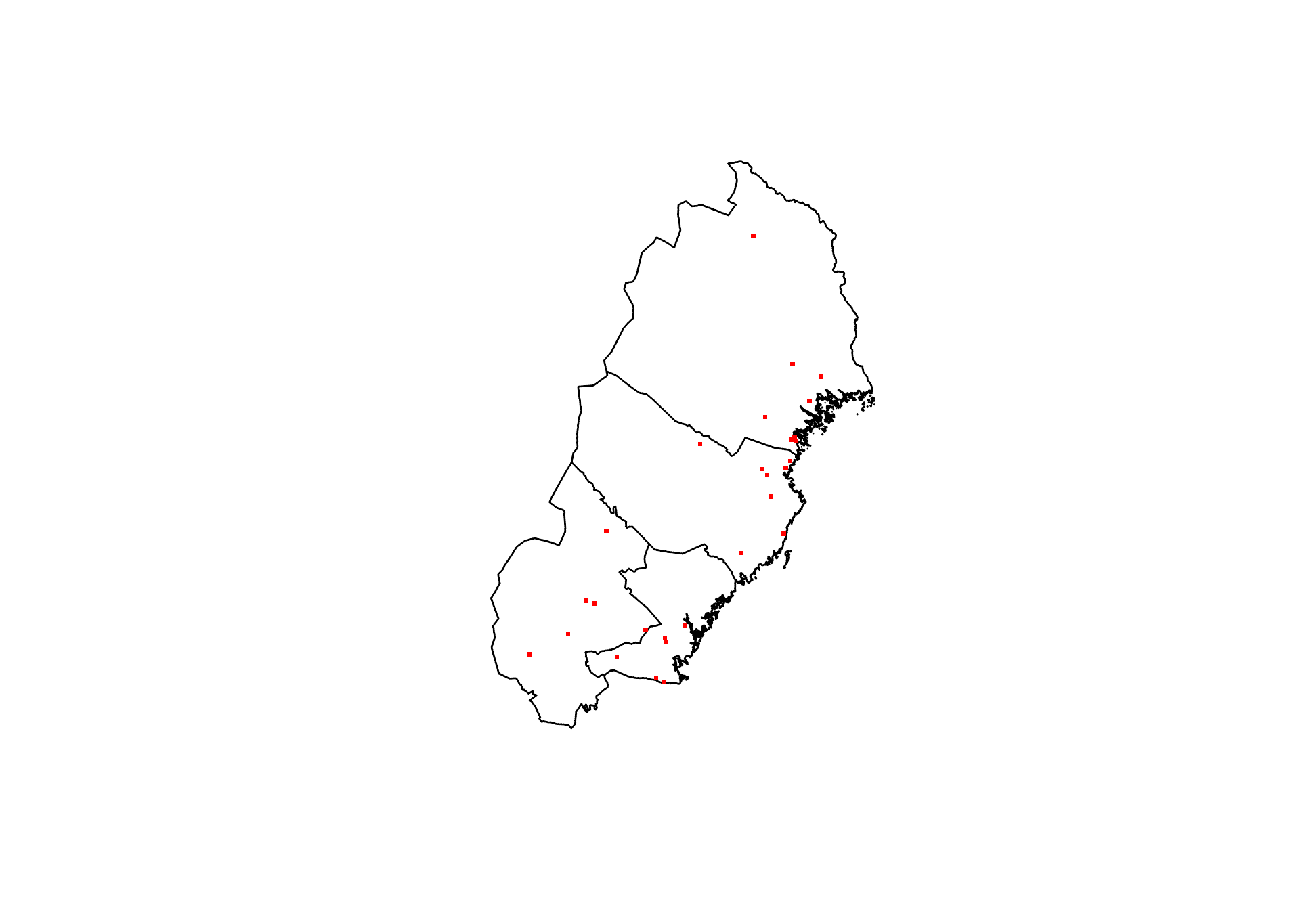}
\caption{The six test datasets i.e.~six ambulance call point patterns, which are observed on Wednesday, Thursday, Friday, Saturday, Sunday and Monday (1st row) and their corresponding simulated point patterns (2nd row).}
\label{Orgr}
\end{figure}

Next, the fitted/trained model obtained above is used to forecast the Gaussian random field to obtain forecast intensities, which have been exploited to simulate Poisson process realisations (recall the setup in Section \ref{s:Forecast}). 
The spatiotemporal LGCP-prediction has been carried out using the function \texttt{lgcpPredict} in the \textsf{R} package \texttt{lgcp}; this function computes the Monte Carlo mean and variance of the Gaussian random field and the mean and variance of the exponential of the Gaussian random field for each of the grid cells and time intervals of interest. The forecasted Poisson intensities at times beyond the last time point in our training data have been obtained using the function \texttt{lgcpForecast} in the \texttt{lgcp} package, and simulations of spatial Poisson process realisations from the obtained intensities have been generated using the function \texttt{rpoispp} in the \textsf{R} package \texttt{spatstat} \citep{baddeley2015spatial}. The first row in Figure \ref{Orgr} illustrates the observed ambulance call locations (spatial test datasets) while its second row demonstrates the corresponding simulated realisations, i.e.~simulated call locations. 
Visually, it seems that the forecasted models quite well pick up on many of the different hot-spots found in the actual data (mainly along the east coast). They do, however, not fully capture a few of the different regions where there are just a few calls, which most likely is an effect of the the nature of the spatial kernel intensity estimate (see Figure \ref{Or3567}) -- having access to relevant spatial covariates would likely result in an improved fit. Also, the counts in the simulated realisations turn out to be slightly smaller than the counts for the test datasets, which may be the result of e.g.~the fitted temporal component and the discretisation used for the random field. We stress that the simulated realisations in Figure \ref{Orgr} are representative for the typical behaviour of simulated realisations -- the above conclusions can be drawn for arbitrary simulated forecasted realisations.  

\begin{figure}[H]
\centering
\includegraphics[width=0.15\textwidth]{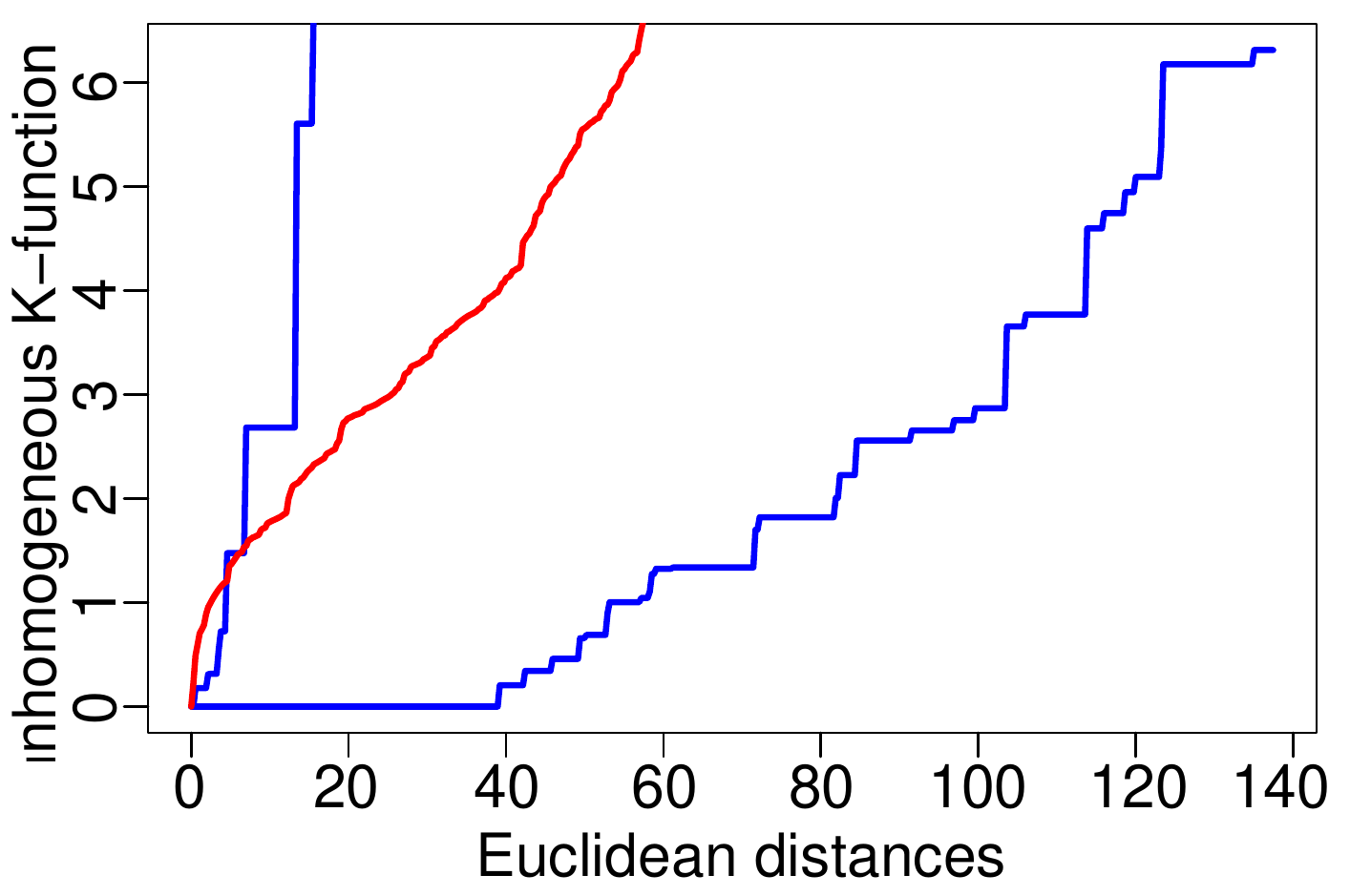}
\includegraphics[width=0.15\textwidth]{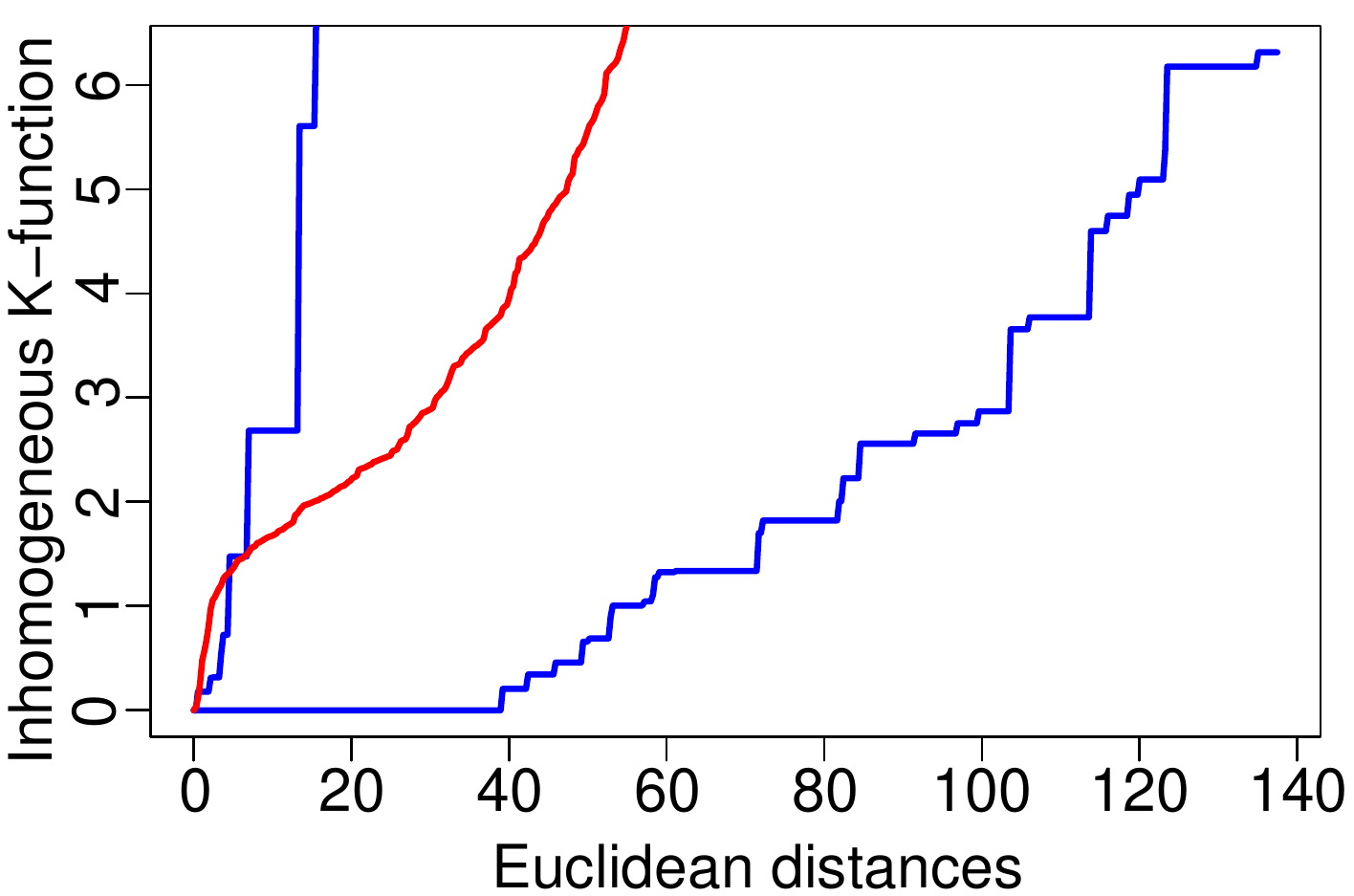}
\includegraphics[width=0.15\textwidth]{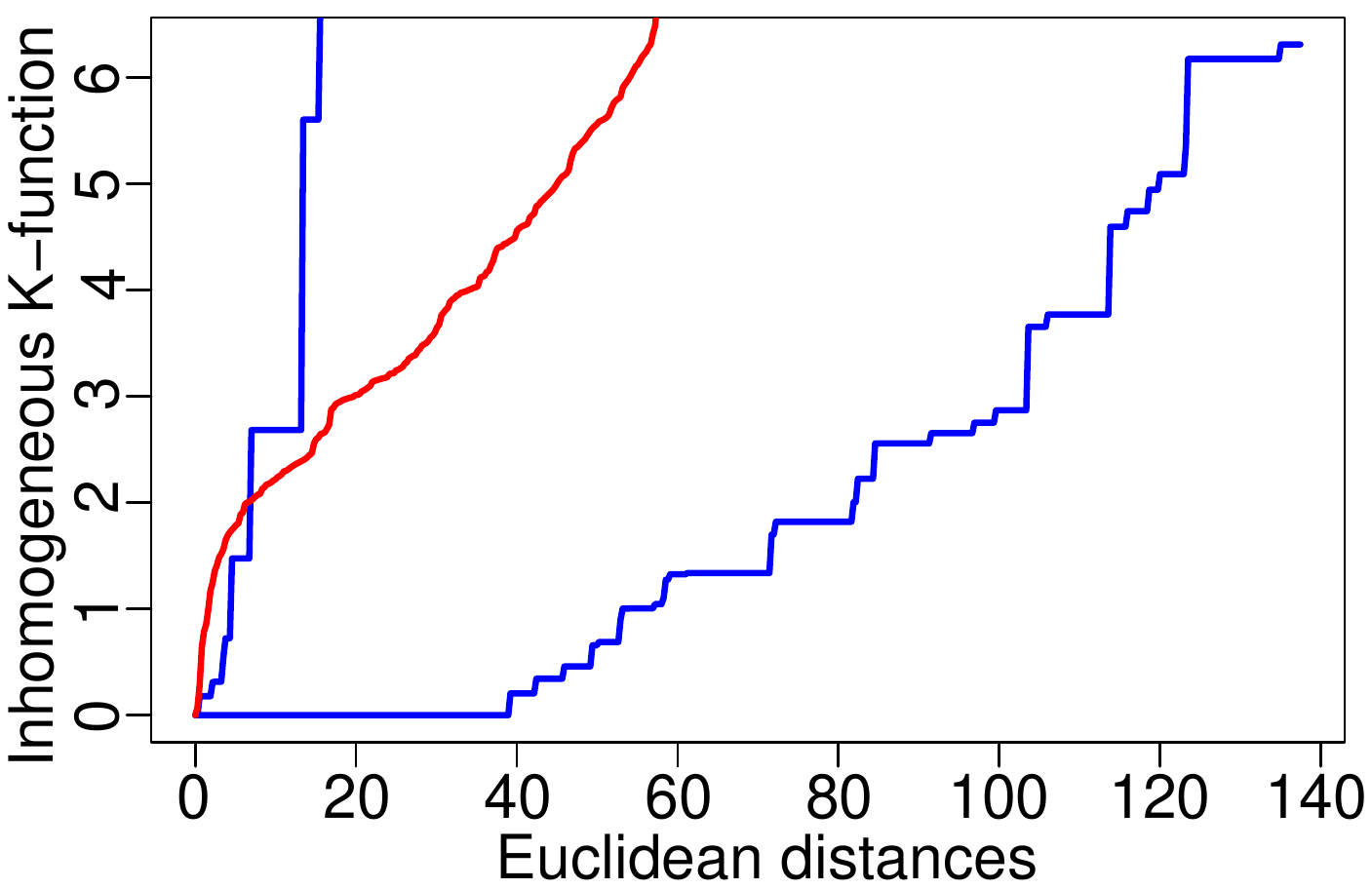}
\includegraphics[width=0.15\textwidth]{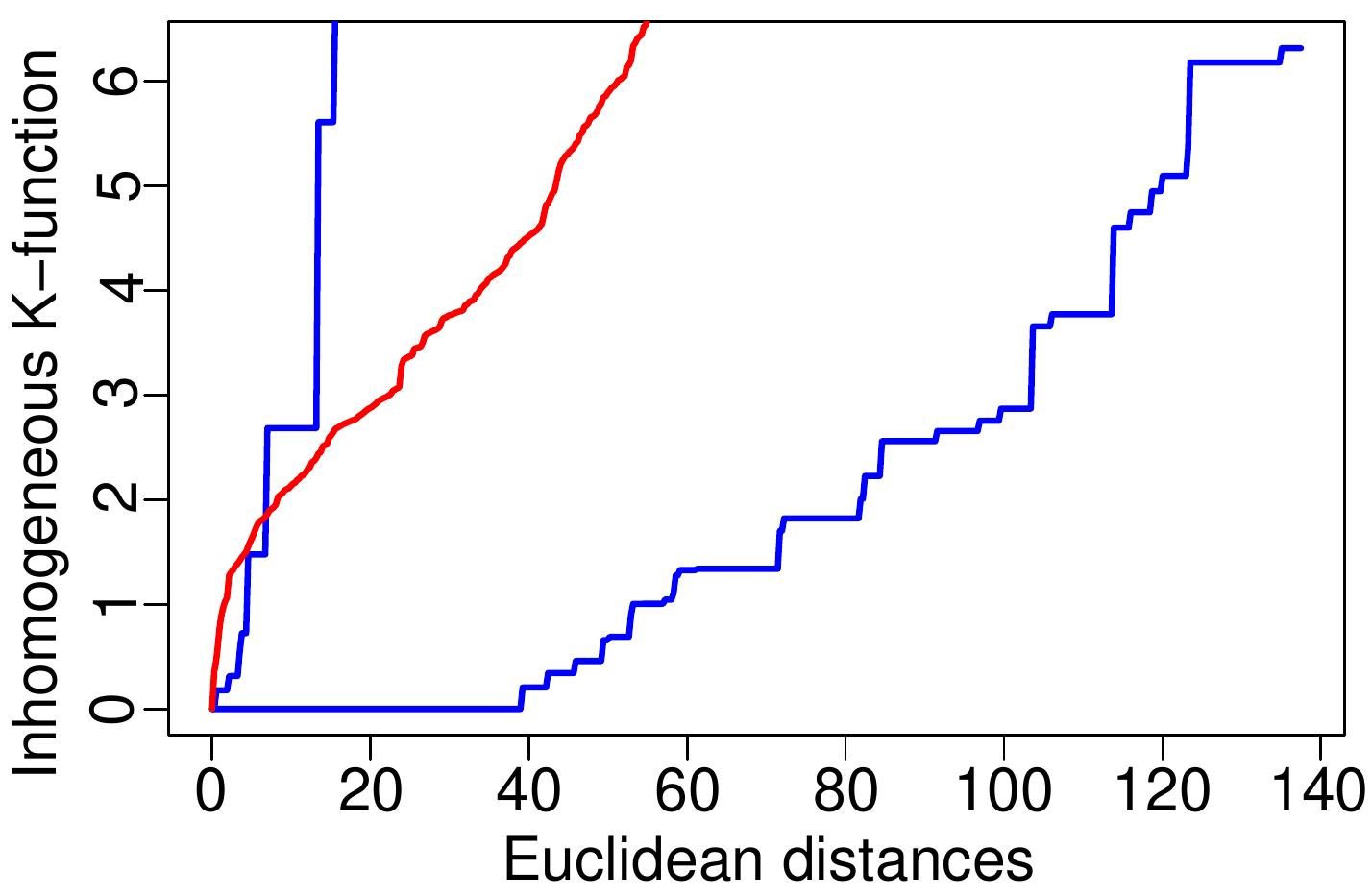}
\includegraphics[width=0.15\textwidth]{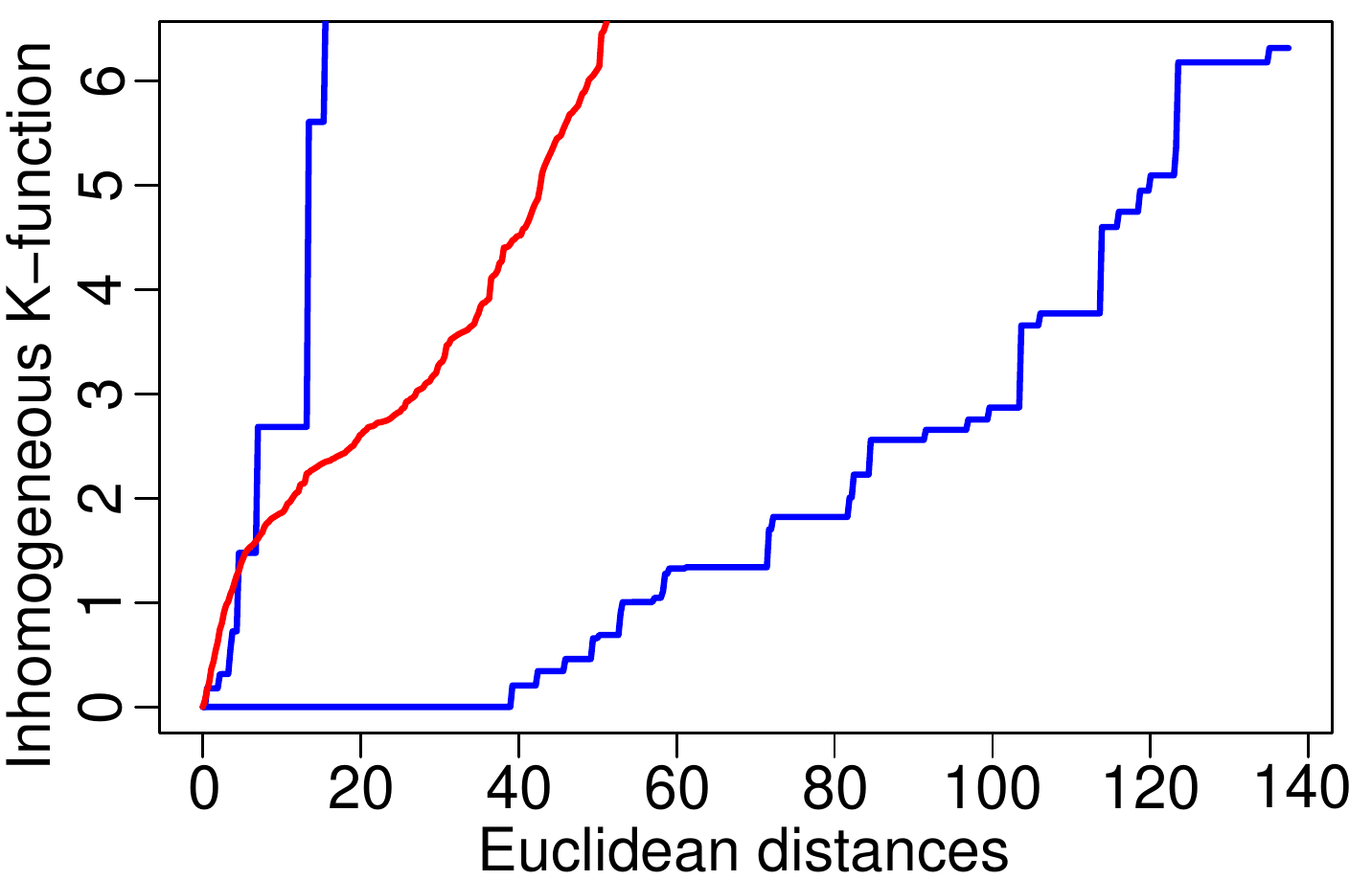}
\includegraphics[width=0.15\textwidth]{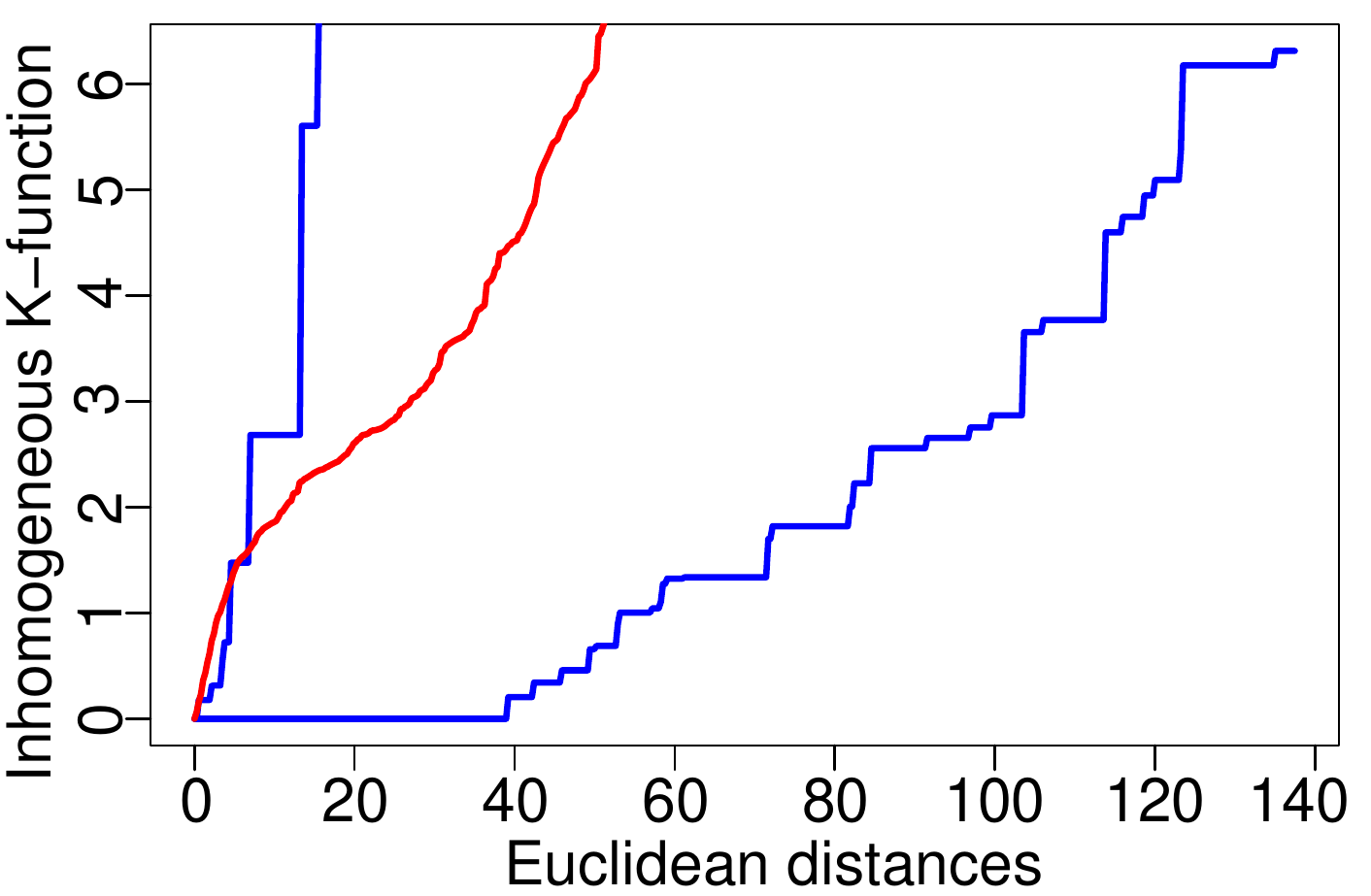}
\caption{Evaluation of the similarity between the simulated and observed point patterns. The red coloured curves represent the inhomogeneous $K$-function values for the observed point patterns while the blue coloured curves reflect the envelopes, given by the lag-wise maxima and minima of the estimated inhomogeneous $K$-functions 
generated by 200 simulated realisations of our forecast model. The values on the x- and y-axes are divided by 1000 and $10^{9}$, respectively.}
\label{OrgrKinhom}
\end{figure}

In addition to the visual assessment, we have also used $K$-function estimates to evaluate the similarity between the simulated and observed point patterns. We have simulated 200 point patterns using the forecast model and at each range we have computed the maximum and minimum values of the $K$-function values (max-min pointwise envelopes). The plots tell us whether the simulated model has a $K$-function which corresponds to that of the true data. As can be seen in Figure \ref{OrgrKinhom}, at long-ranges, the $K$-function values for the observed point patterns stay within the envelopes. However, the conclusions drawn here should be handled with care; the simulations come from a Poisson process and the figure indicates whether the data are more clustered than a Poisson process with an intensity given by that of the forecast model. This seems to be the case and we can essentially conclude that there is slight clustering in the test data for small spatial ranges.

\begin{figure}[H]
\centering
\includegraphics[width=0.14\textwidth]{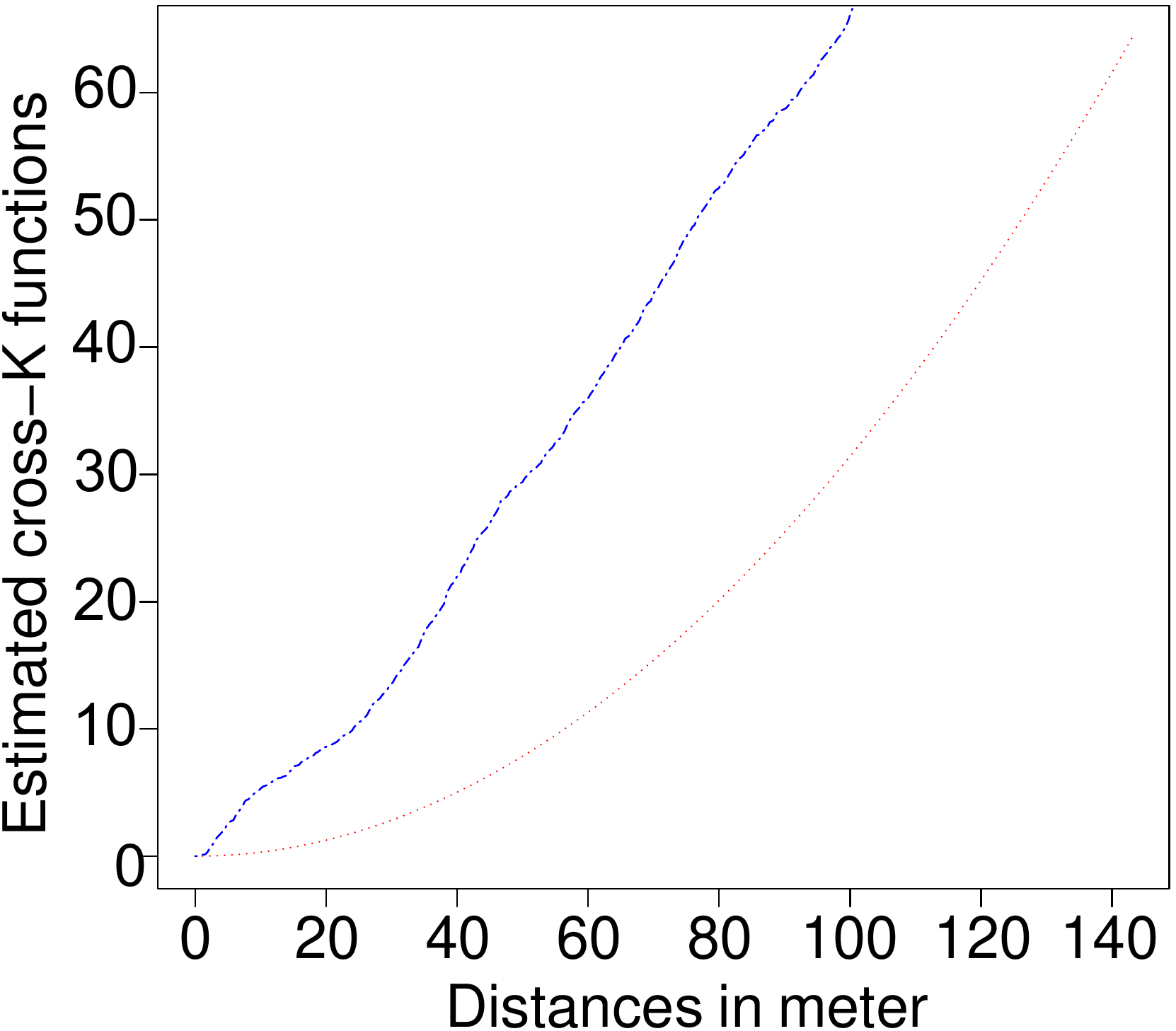}
\includegraphics[width=0.14\textwidth]{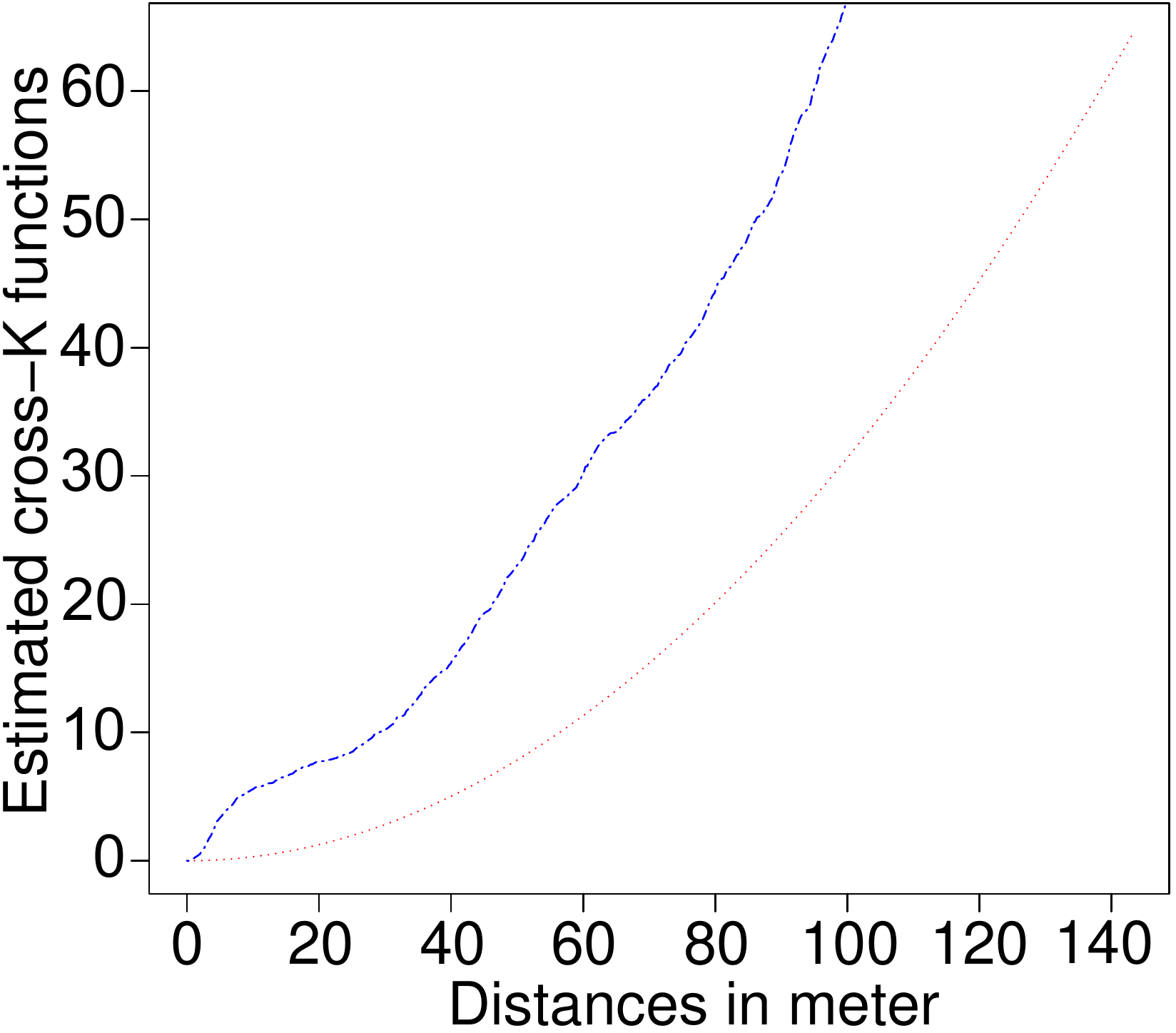}
\includegraphics[width=0.14\textwidth]{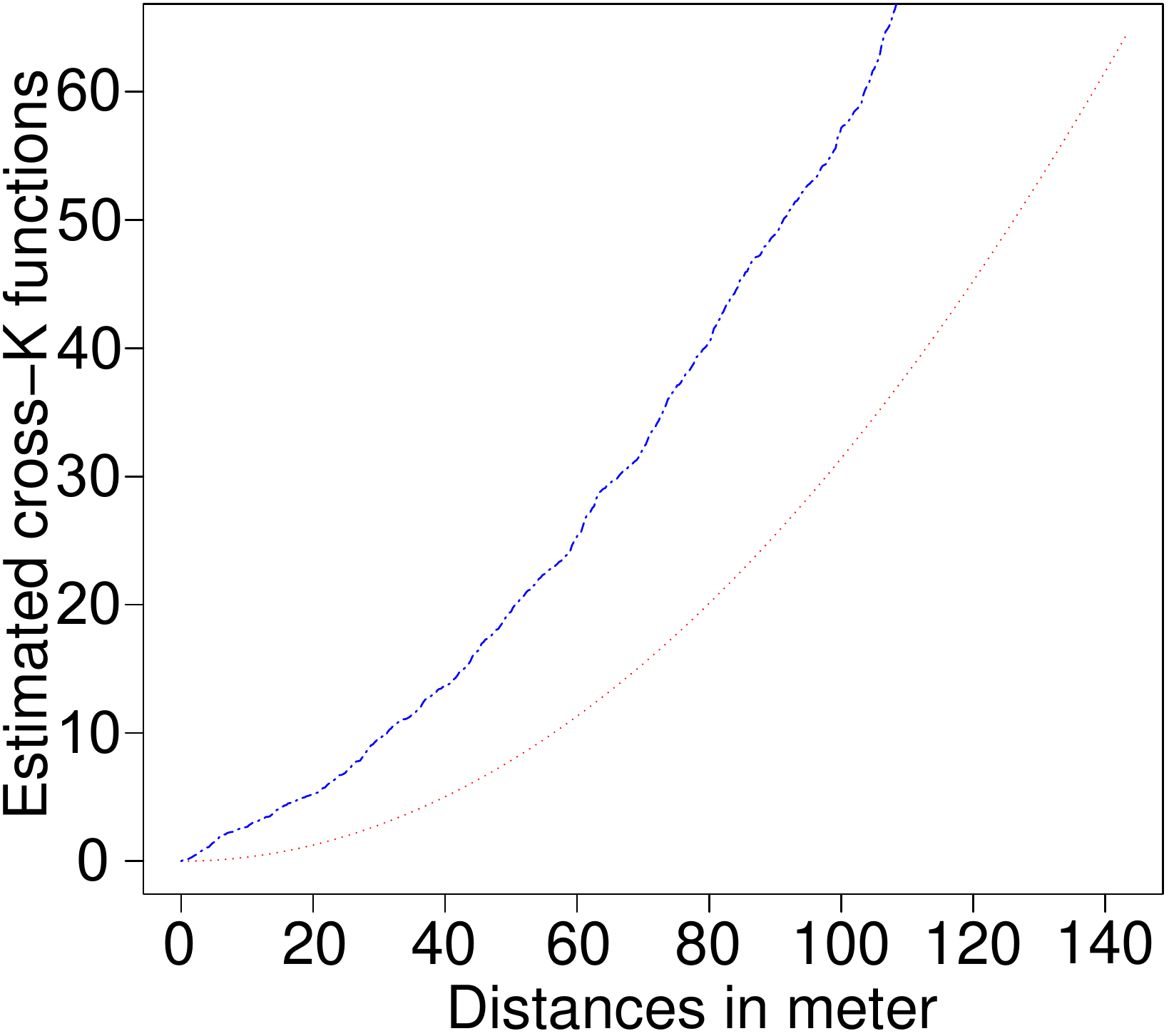}
\includegraphics[width=0.14\textwidth]{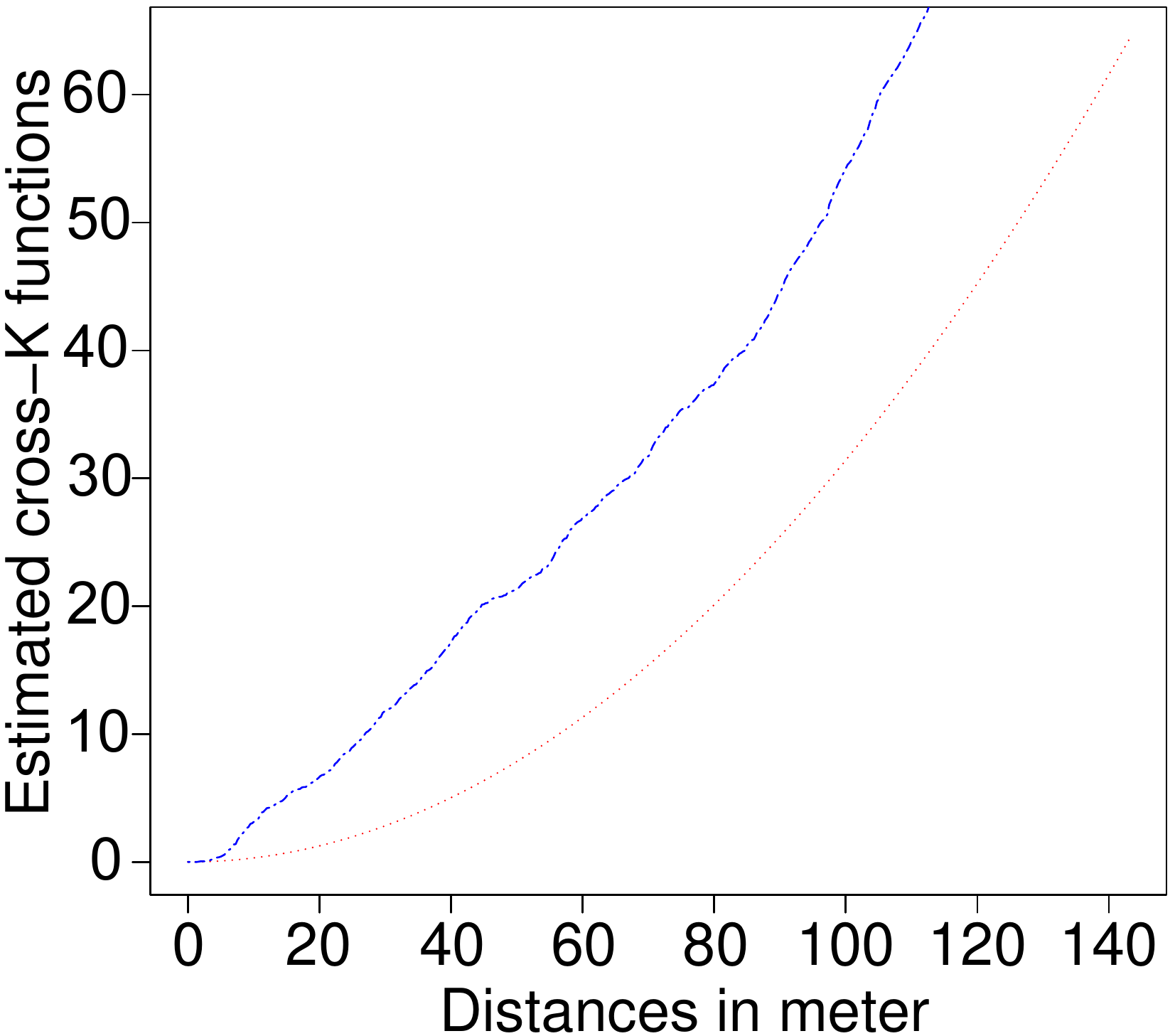}
\includegraphics[width=0.14\textwidth]{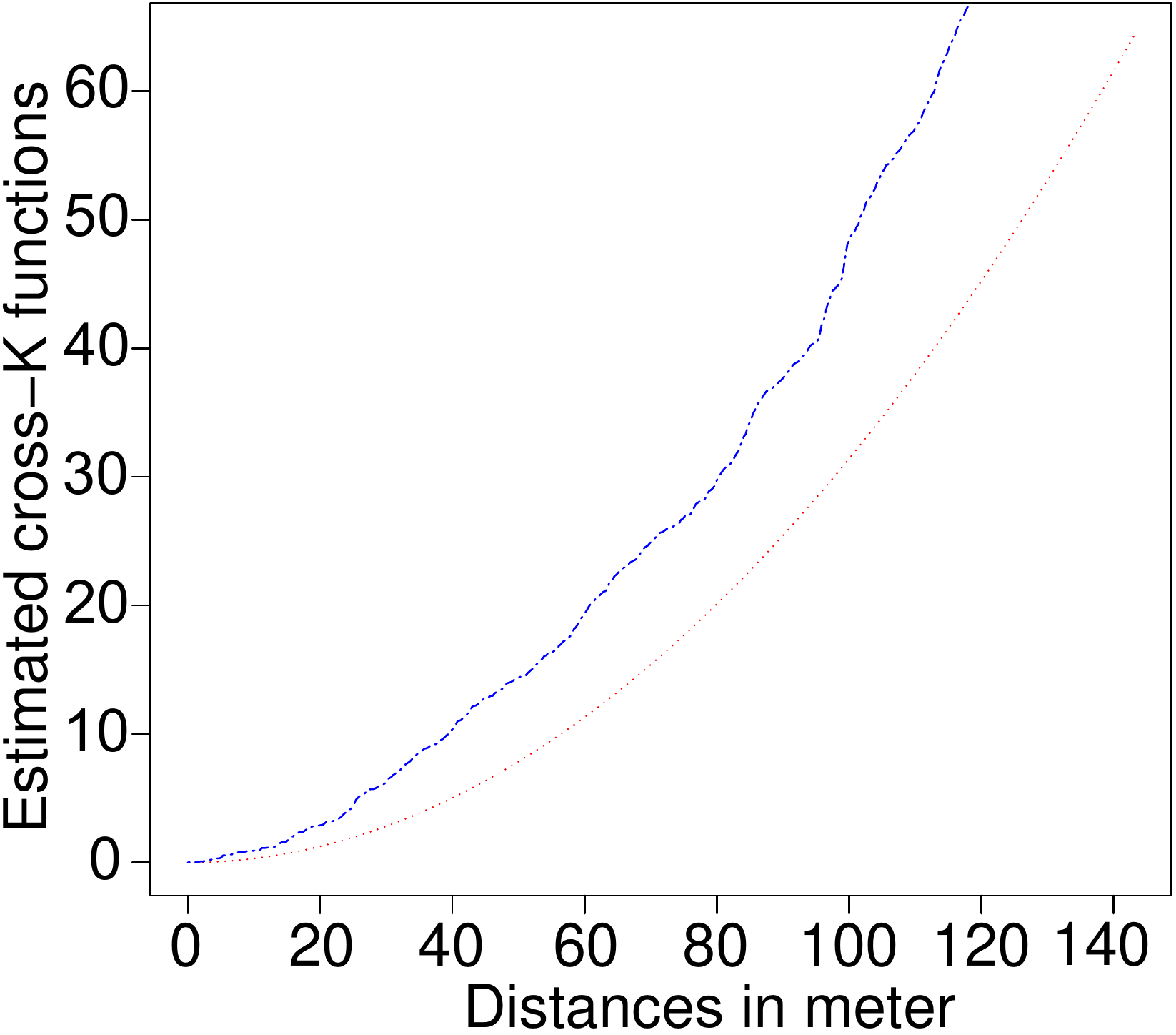}
\includegraphics[width=0.14\textwidth]{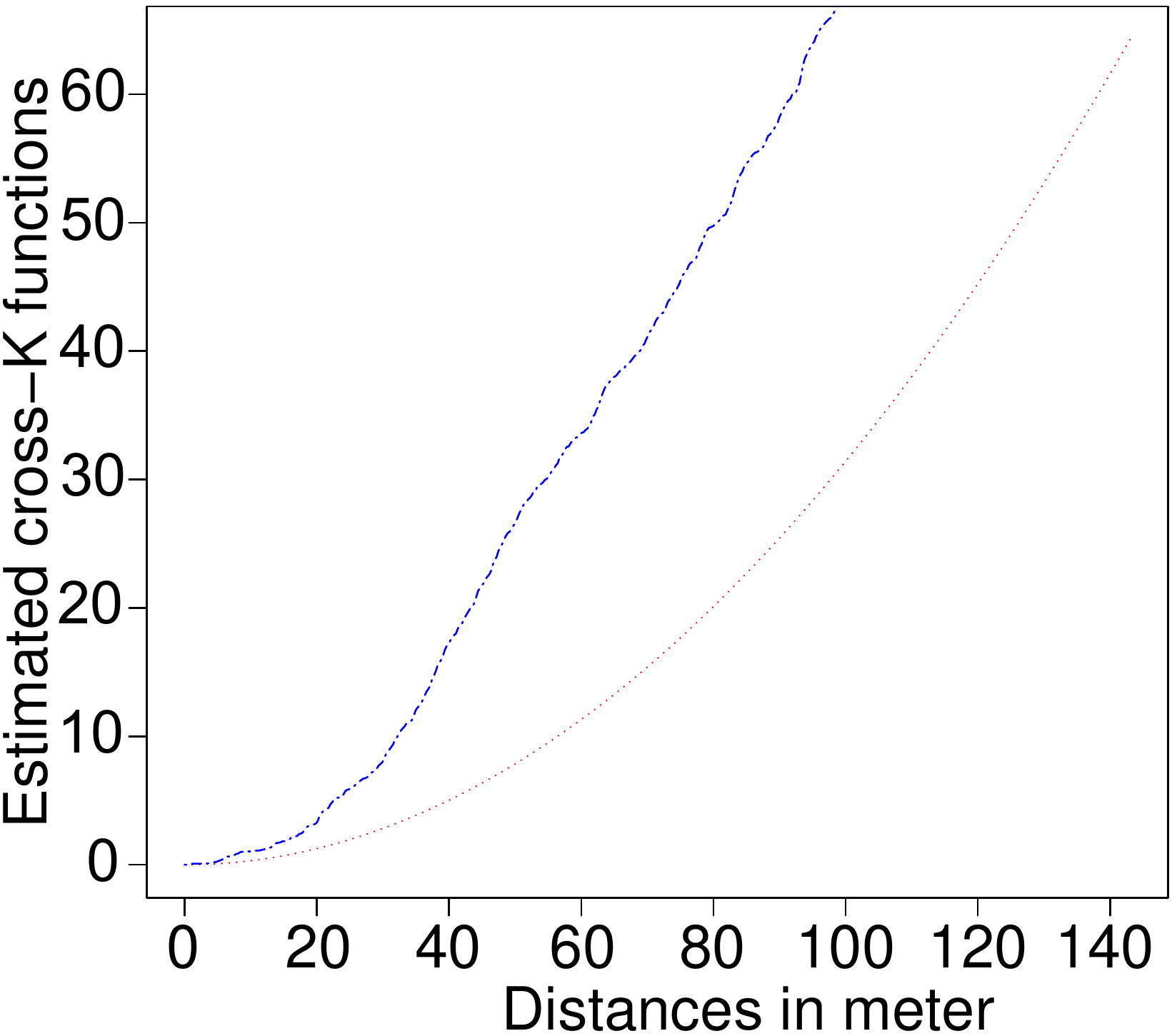}\\
\includegraphics[width=0.14\textwidth]{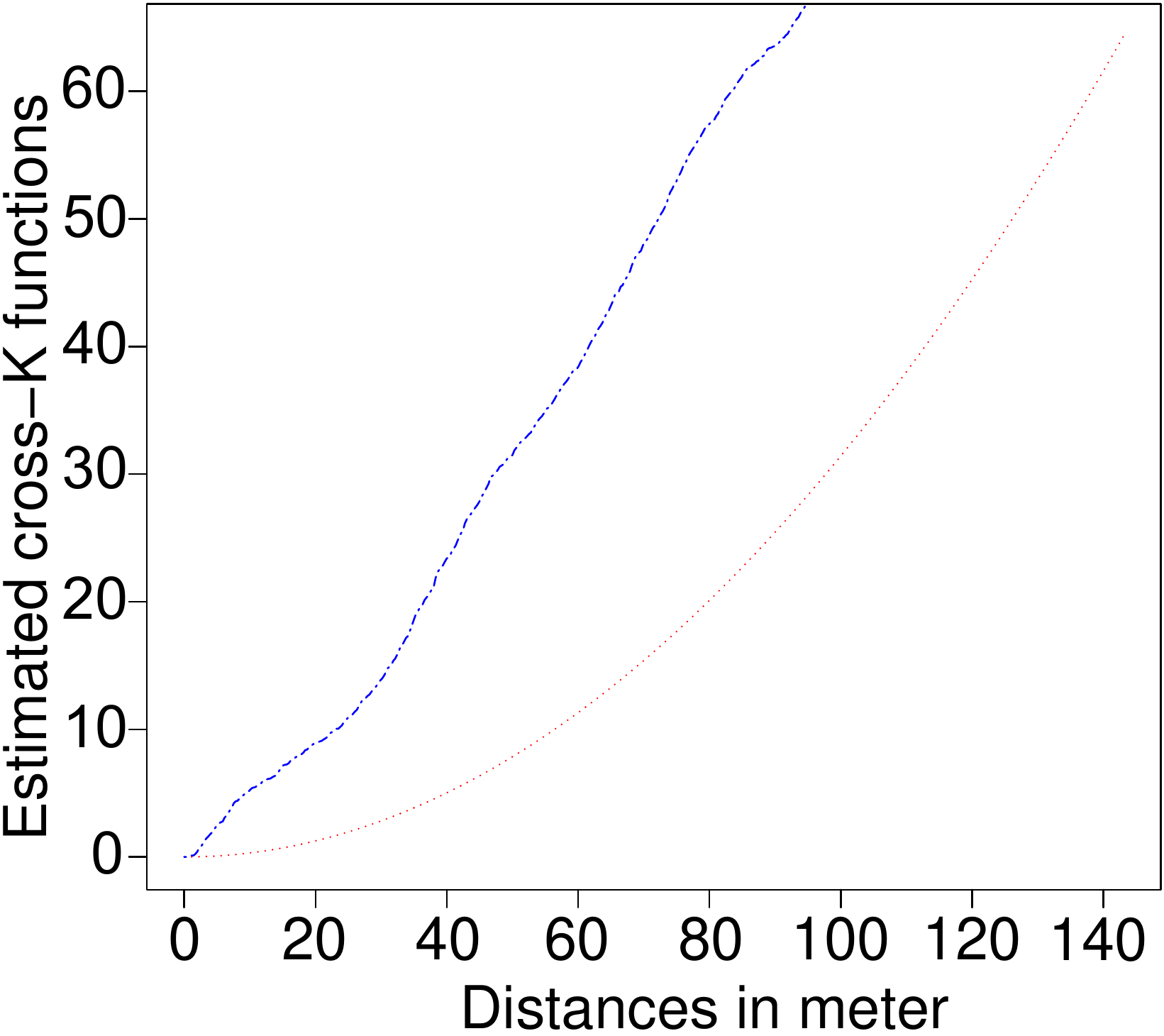}
\includegraphics[width=0.14\textwidth]{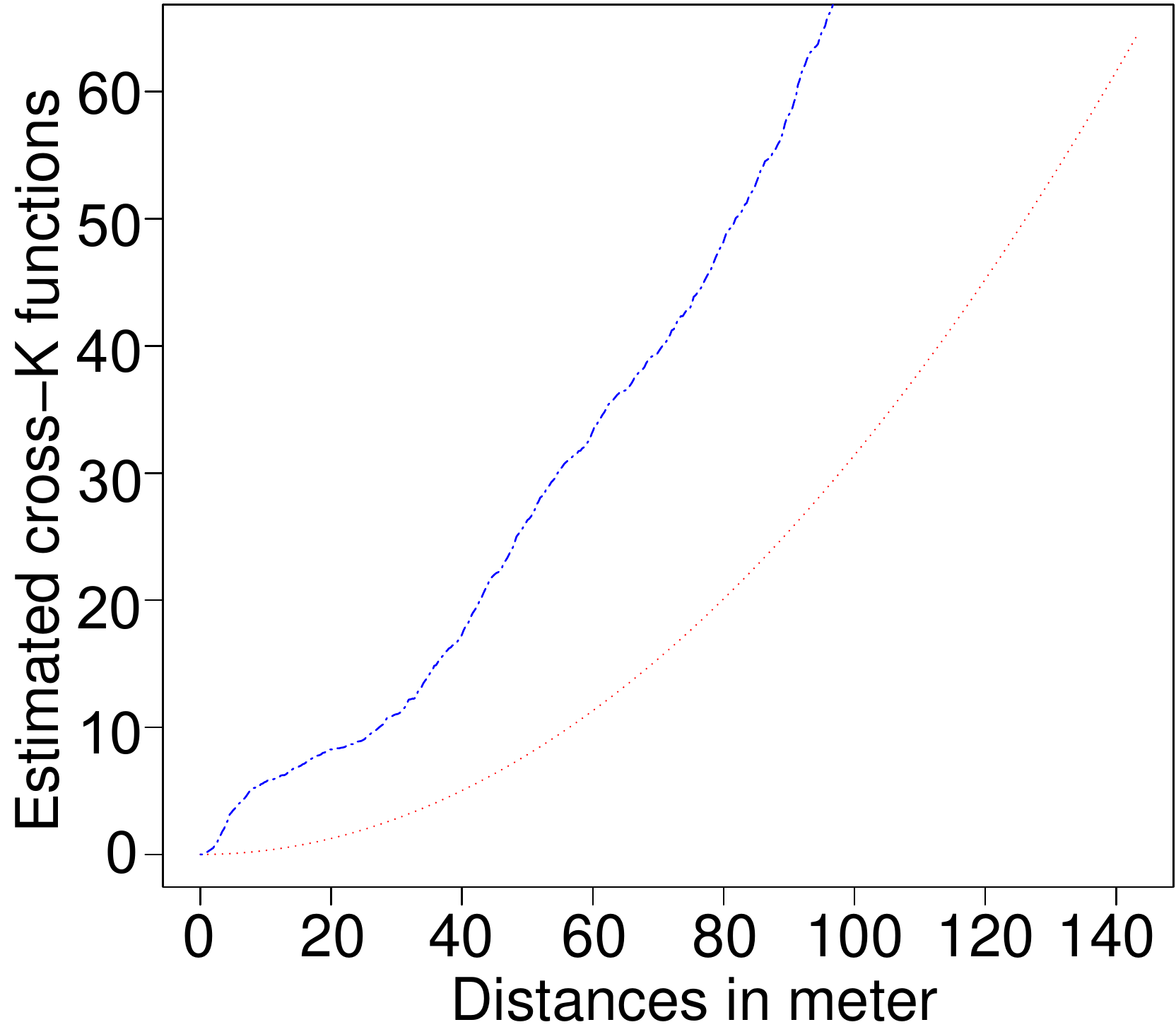}
\includegraphics[width=0.14\textwidth]{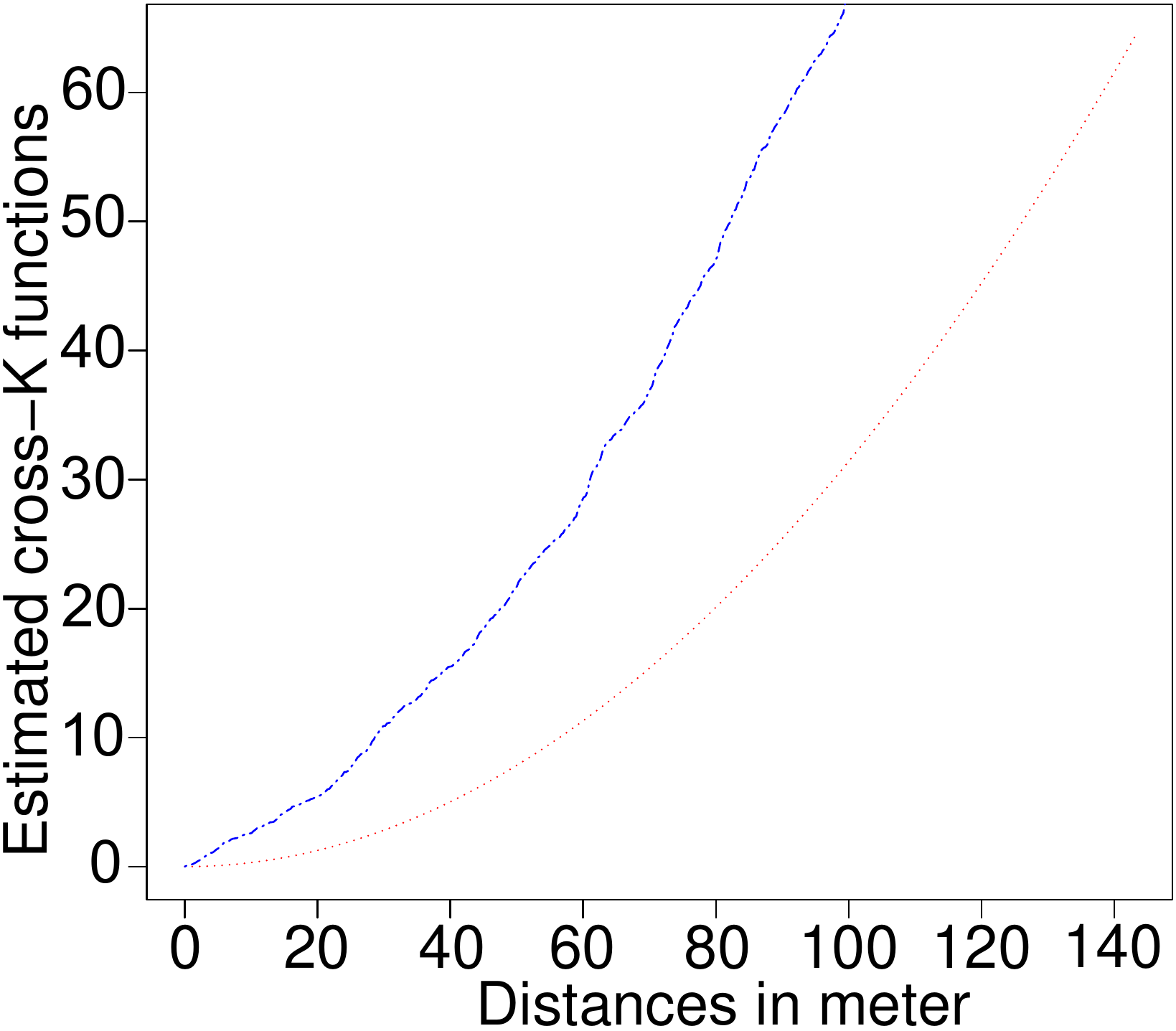}
\includegraphics[width=0.14\textwidth]{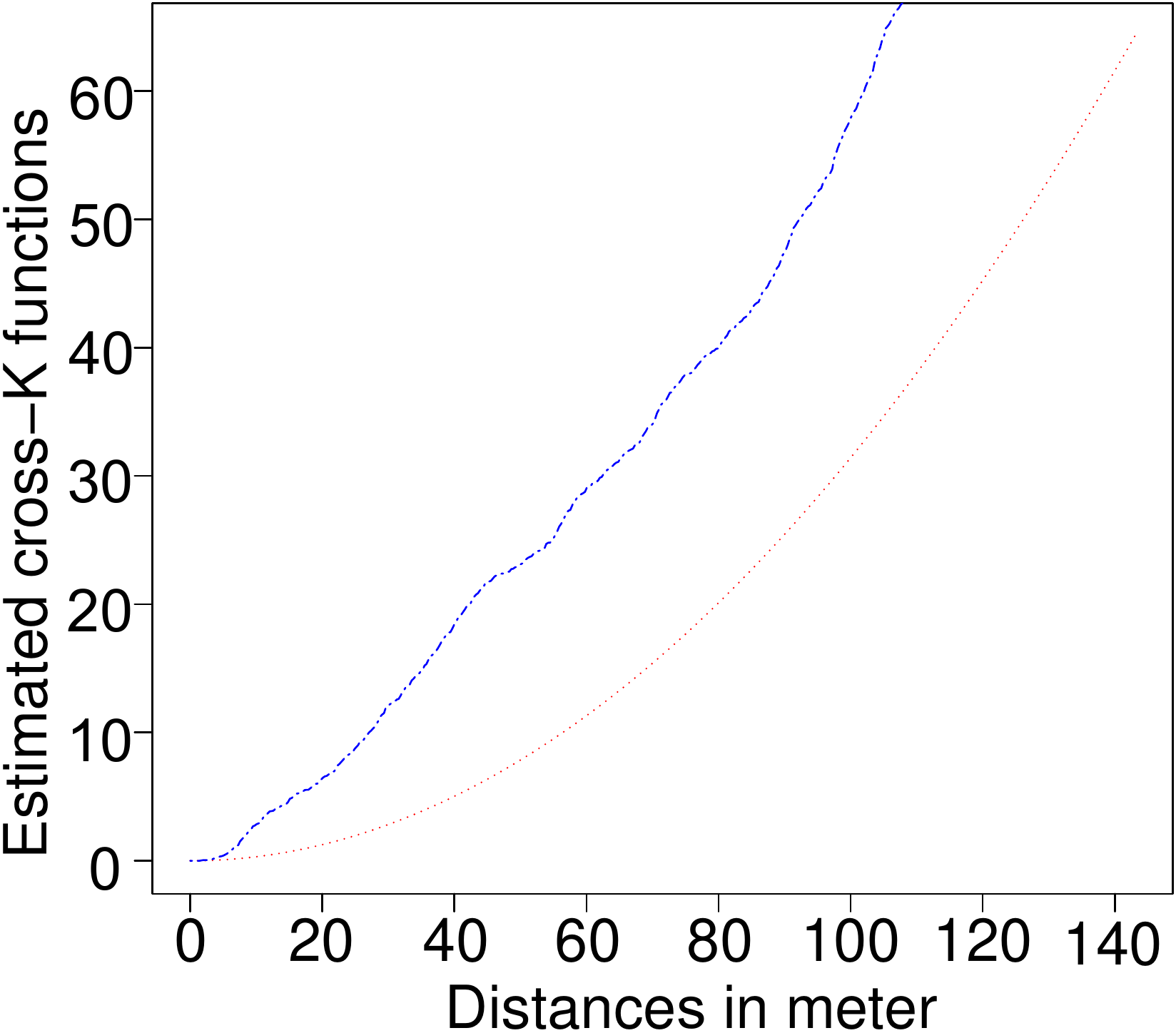}
\includegraphics[width=0.14\textwidth]{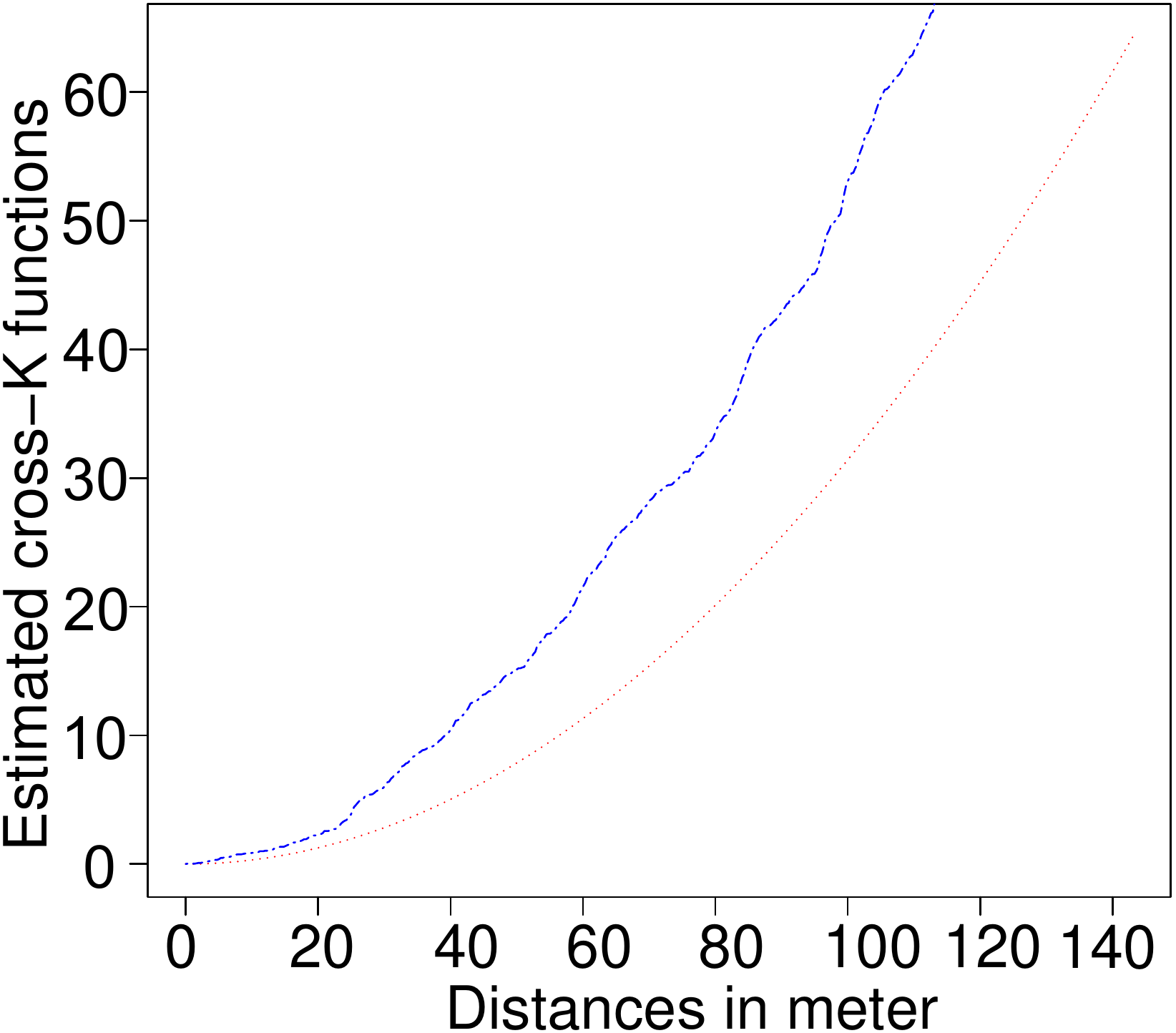}
\includegraphics[width=0.14\textwidth]{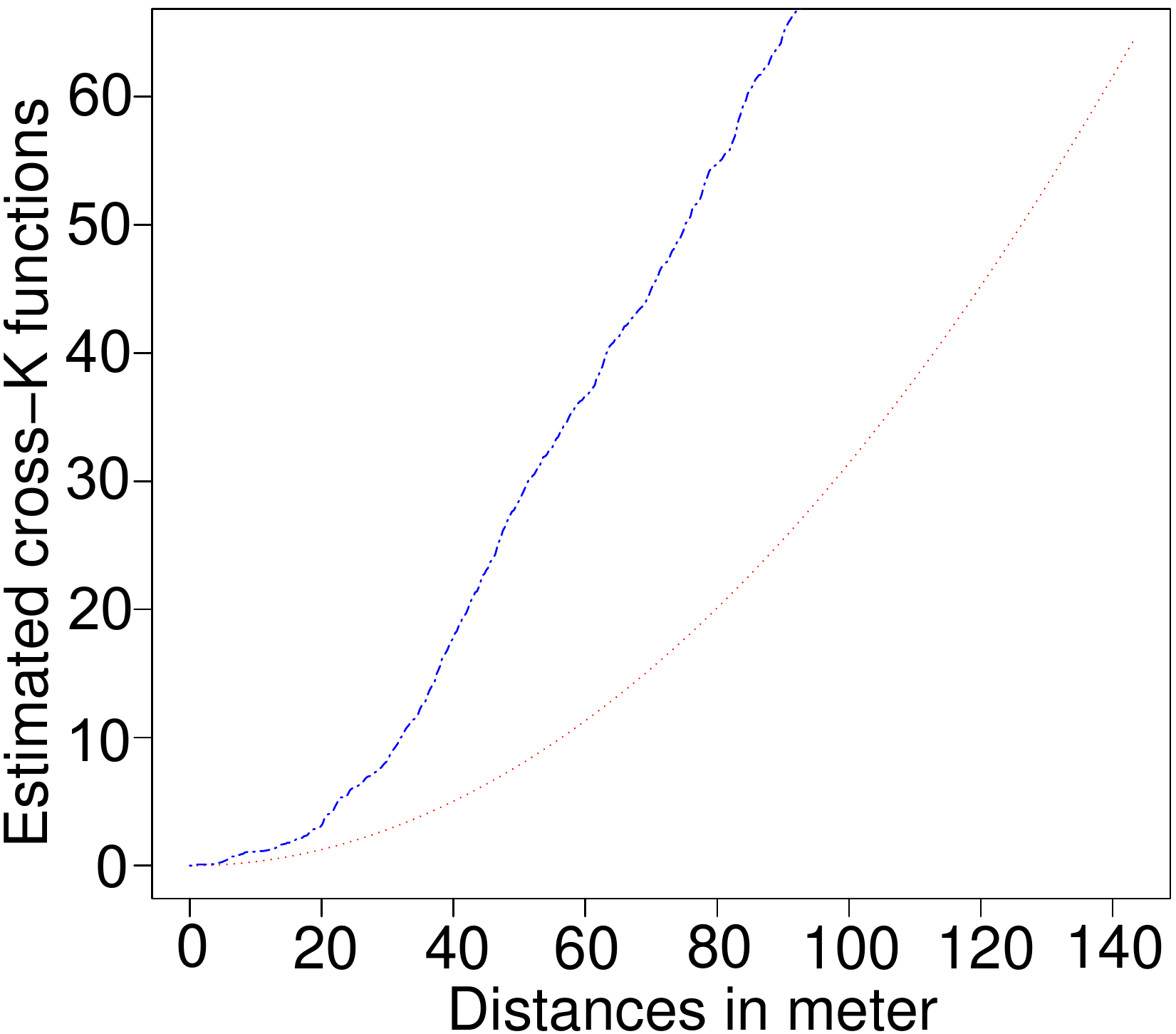}
\caption{Evaluation of the spatial interaction between the simulated spatial locations and the events on the test datasets; $K_{12}(u)$ (a typical event is a simulated event) is given on the 1st row and $K_{21}(u)$  (a typical event is a test data event) on the 2nd row. The red coloured curves represent the estimated theoretical $K_{0}(\cdot)$ functions while the blue coloured curves show the estimated $K_{12}(\cdot)$ and $K_{21}(\cdot)$ functions. The values on the x- and y- axes are divided by 1000 and $10^{9}$, respectively.}
\label{OrgrCross}
\end{figure}

Furthermore, we have also  evaluated whether the simulated points are merely randomly localised or 
tend to aggregate around data points. That is, to evaluate the existence of simulated points close to every observed point, or vice versa. Commonly, for a bivariate point process/pattern, one exploits (an estimator of) the bivariate $K$-function \citep{lotwick1982methods} $K_{12}(u)$, $u\geq0$, to analyse spatial interaction between points with mark 1 and points with mark 2. 
Under independent marking the bivariate $K$-function satisfies $K_{12}(u) = K_{21}(u) = K_{0}(u) = \pi u^2$ \citep{cronie2016summary}.  Note that $K_{12}(u)$ is equal to $K_{21}(u)$ for a stationary point process but their estimators are positively correlated but not equal when edge corrections are used \citep{DixonPM}; see \citet{iftimi2019second} for a theoretical account on this (a)symmetry issue. 
Here we have assigned the mark 1 to the points of a simulated forecast point pattern and mark 2 to the points of the observed ambulance data pattern. Then, we have exploited $K_{12}(u)$ and $K_{21}(u)$ to measure the spatial closeness of the simulated and observed points with marks 1 and 2. Note that since this is a summary statistic for homogeneous point processes, deviations from $K_{0}$ may indicate both difference in intensity as well as spatial interaction, so interpretations of departures from $K_{0}$ should be handled cautiously.  However, we here utilise such departures as a global measure of similarity between realised and simulated forecast data. The estimators of $K_{12}(u)$ and $K_{21}(u)$ here make use of Ripley's isotropic edge correction factor.  Figure \ref{OrgrCross} compares $K_{12}(u)$ to $K_{0}(u)$ and  $K_{21}(u)$ to $K_{0}(u)$, and since we have that $K_{12}(u)$ and $K_{21}(u)$ are substantially larger than $K_{0}(u)$, $u>0$ (most importantly when $u$ is small), we have indications that the simulated data correspond well with the true data (in an overall sense). 

We have chosen these simple  evaluation approaches in favour of considering prediction bands/regions (intervals) generated through \eqref{key987}, since in the latter case one quickly faces the challenging issue of choosing an appropriate non-parametric intensity estimator which should generate an estimate which, assuming a good fit, stays within these bands.

\begin{figure}[H]
\centering
\includegraphics[width=0.13\textwidth]{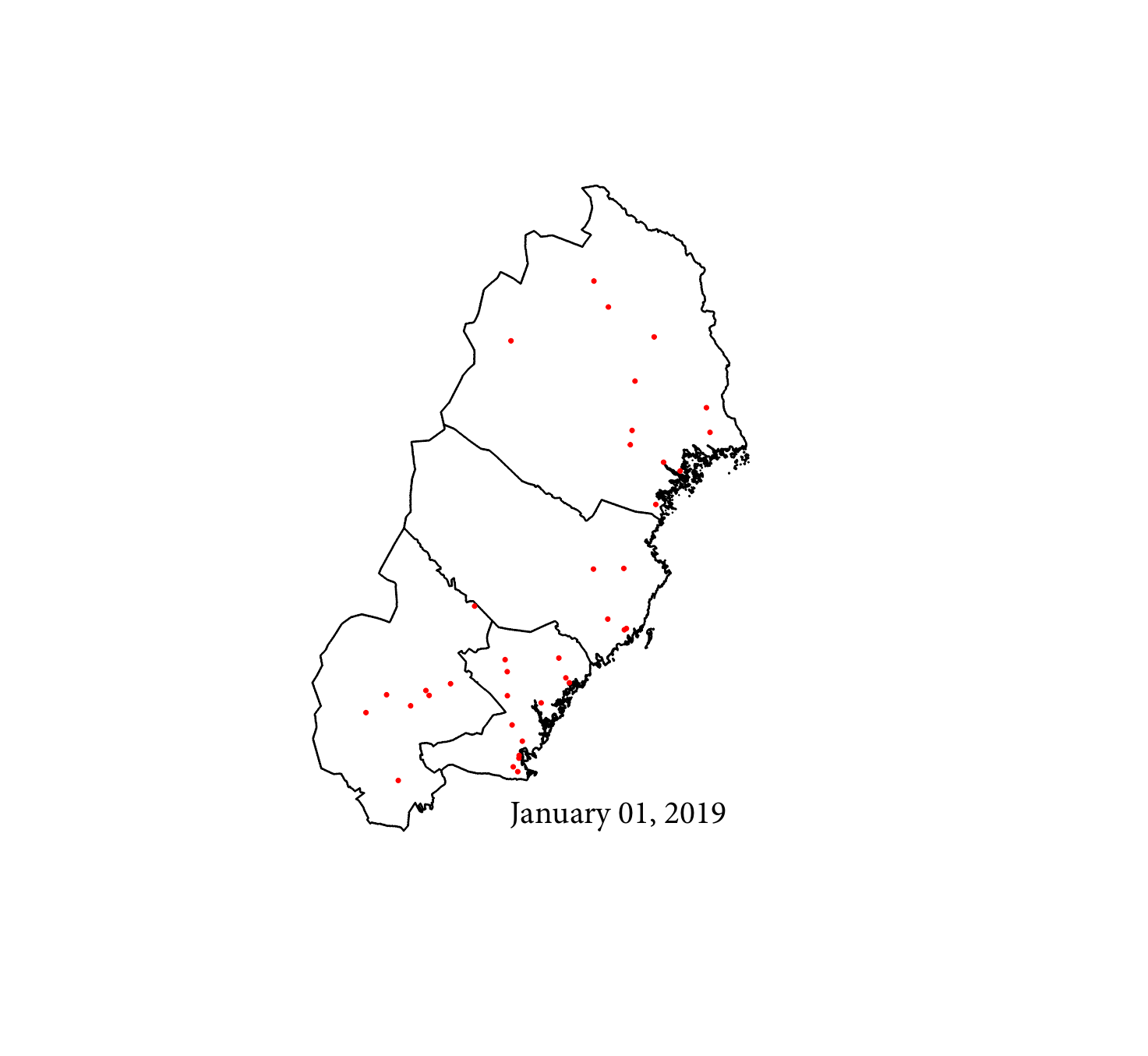}
\includegraphics[width=0.13\textwidth]{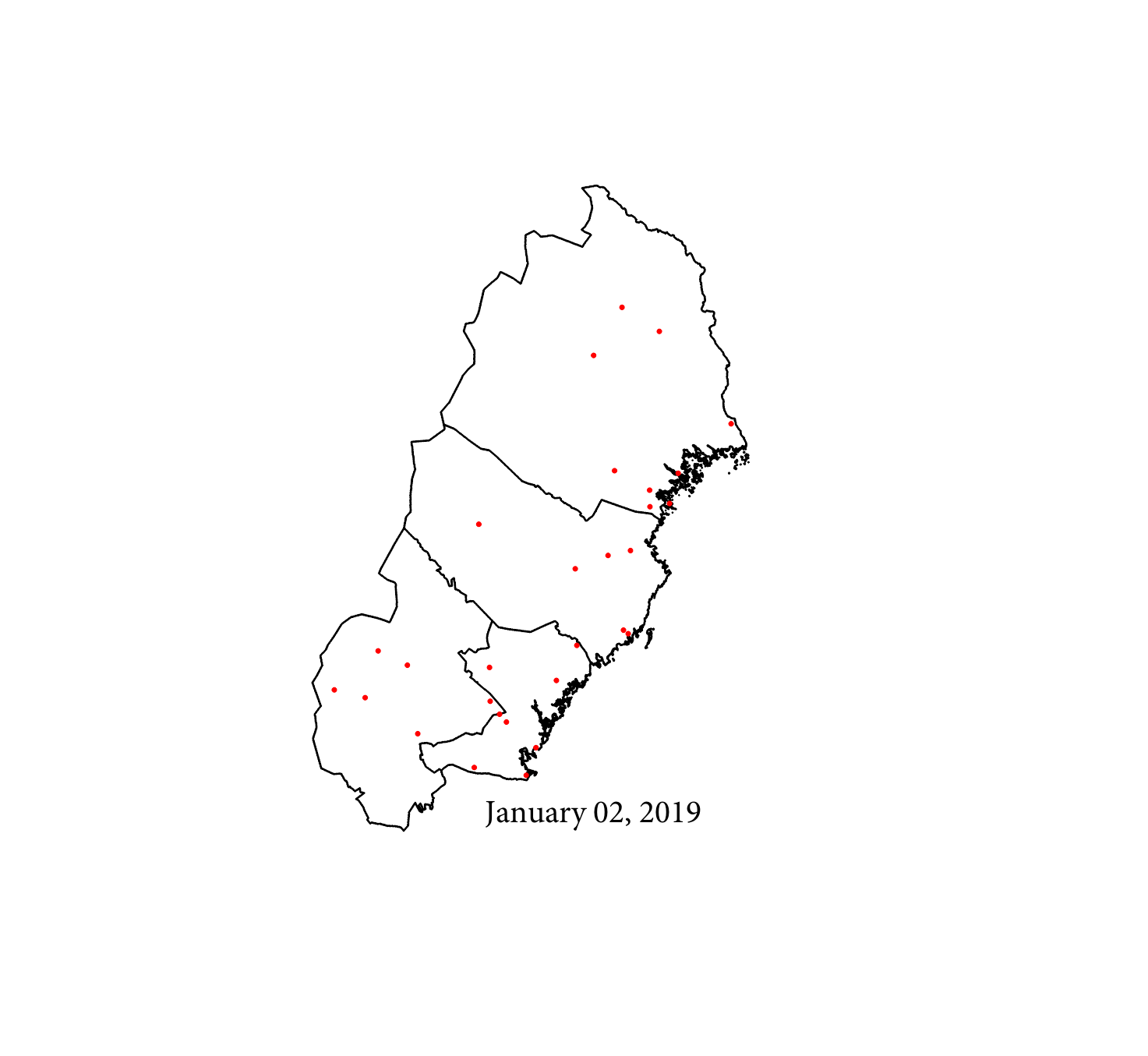}
\includegraphics[width=0.13\textwidth]{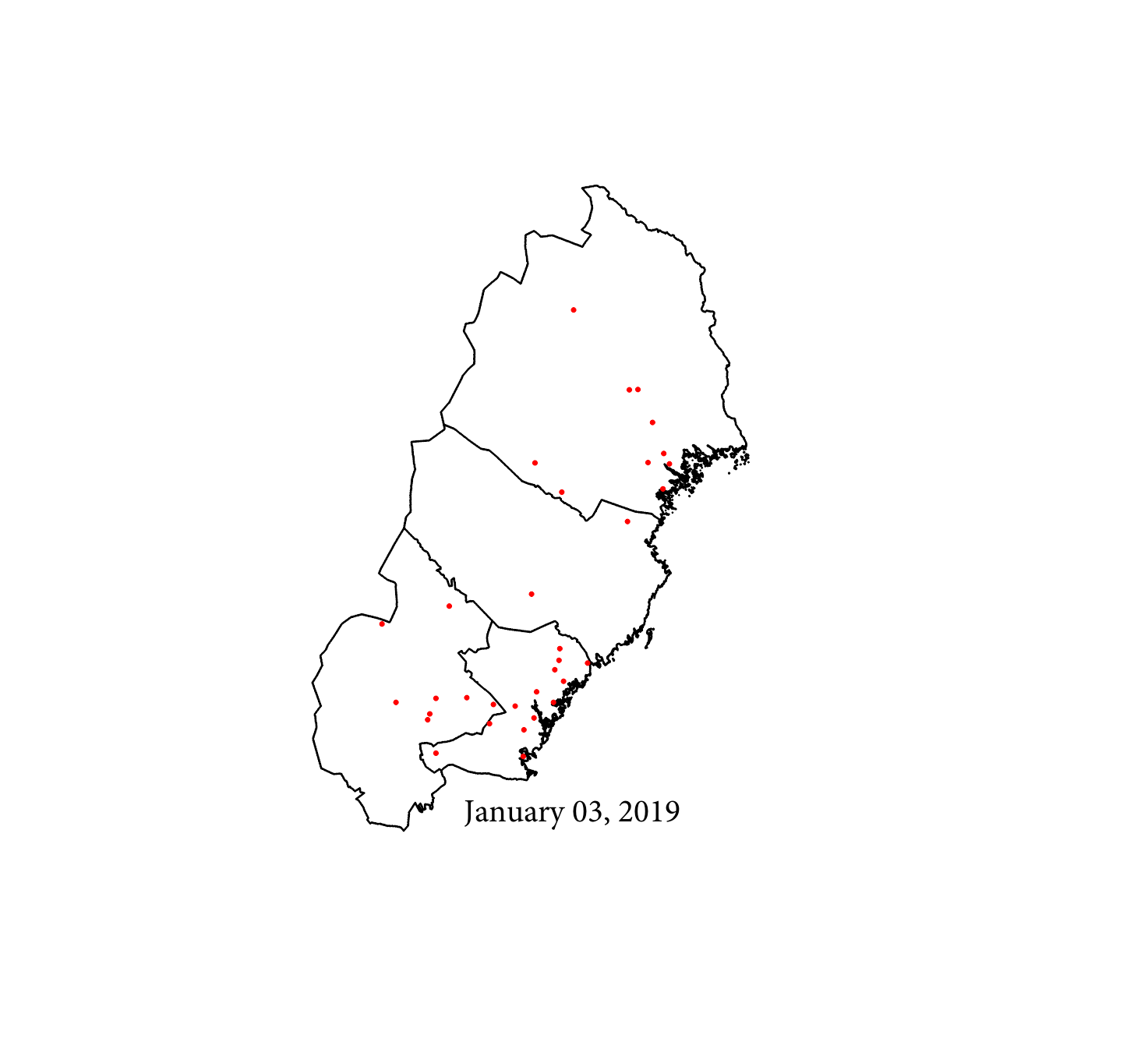}
\includegraphics[width=0.13\textwidth]{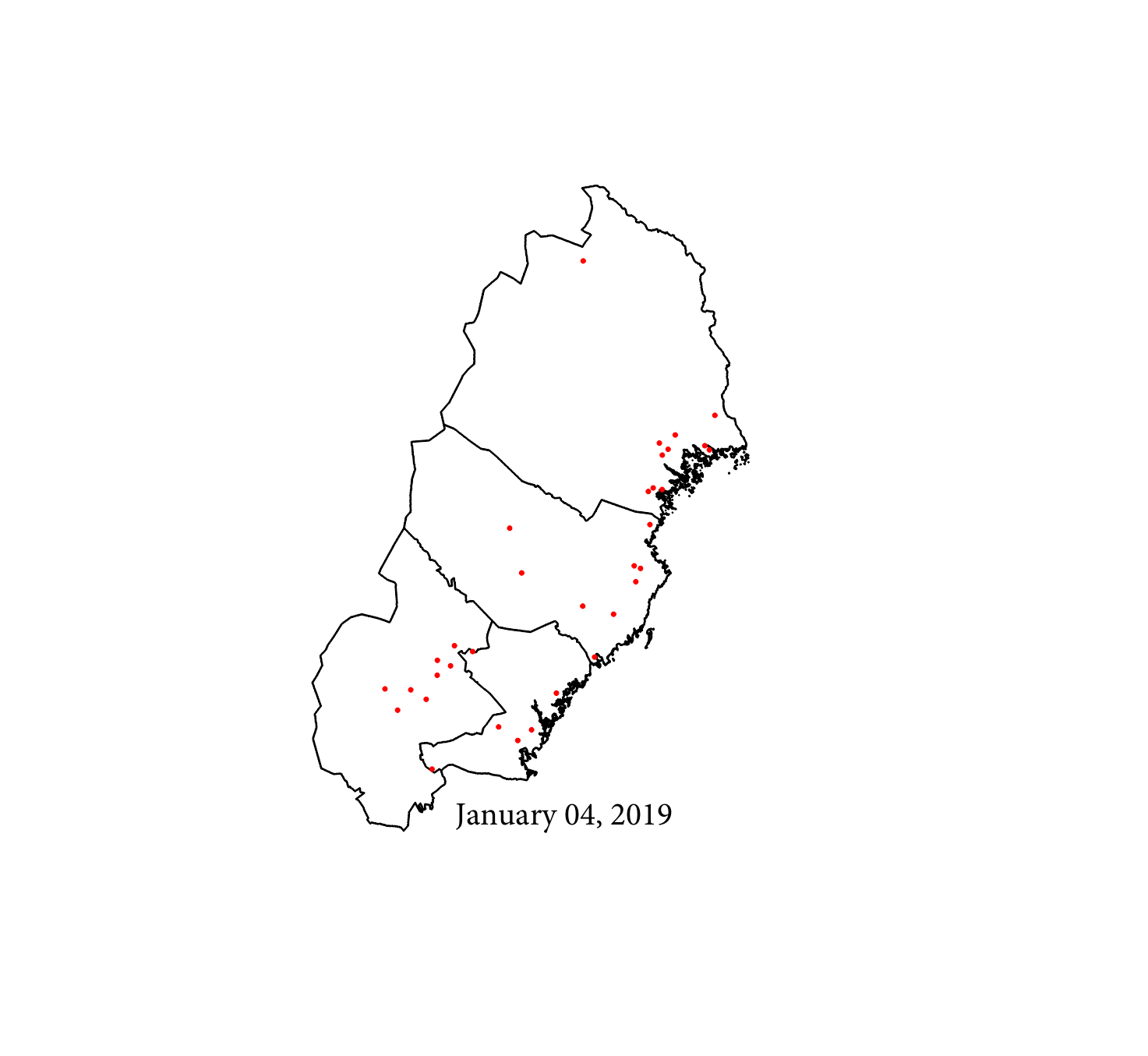}
\includegraphics[width=0.13\textwidth]{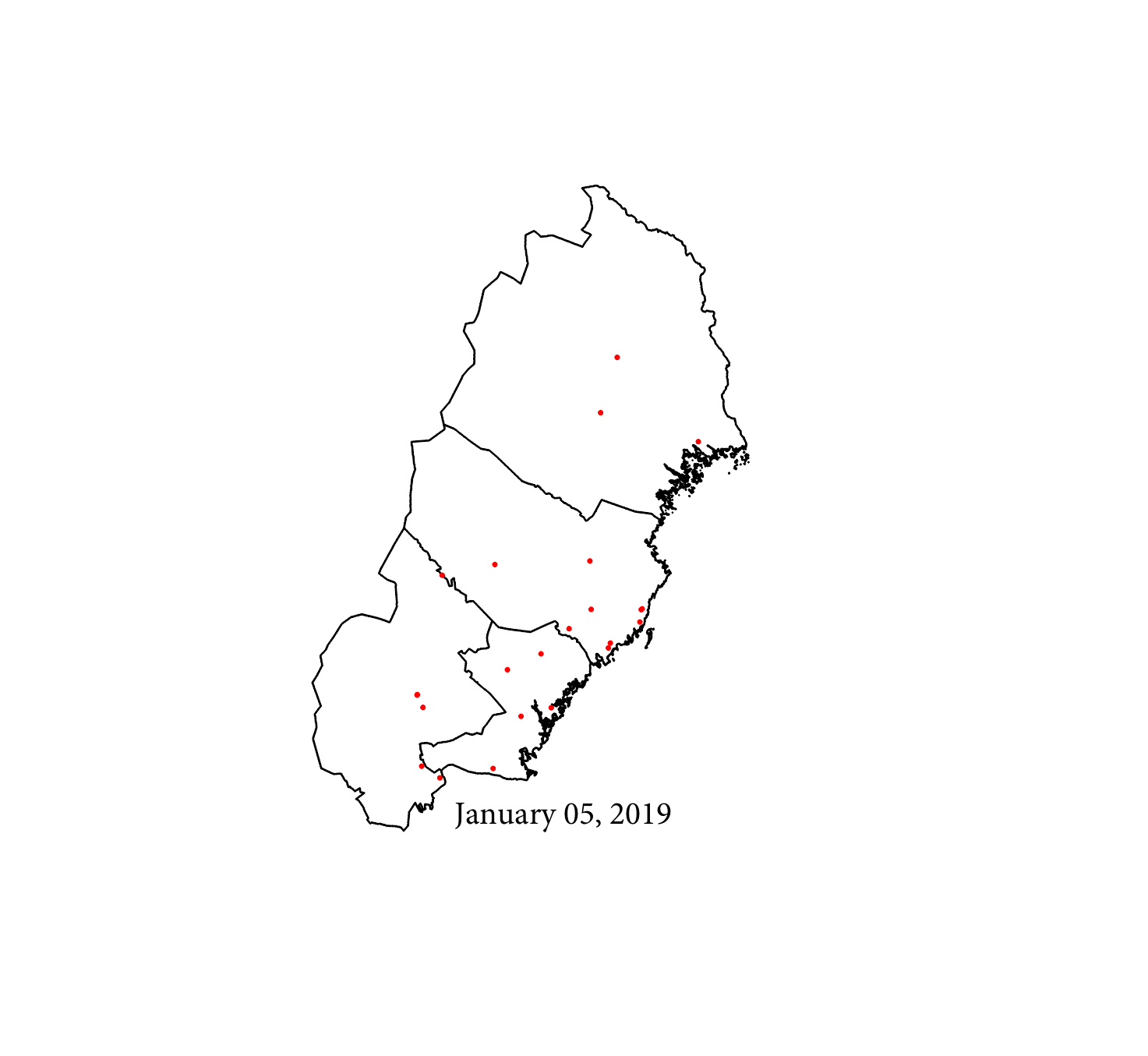}
\includegraphics[width=0.13\textwidth]{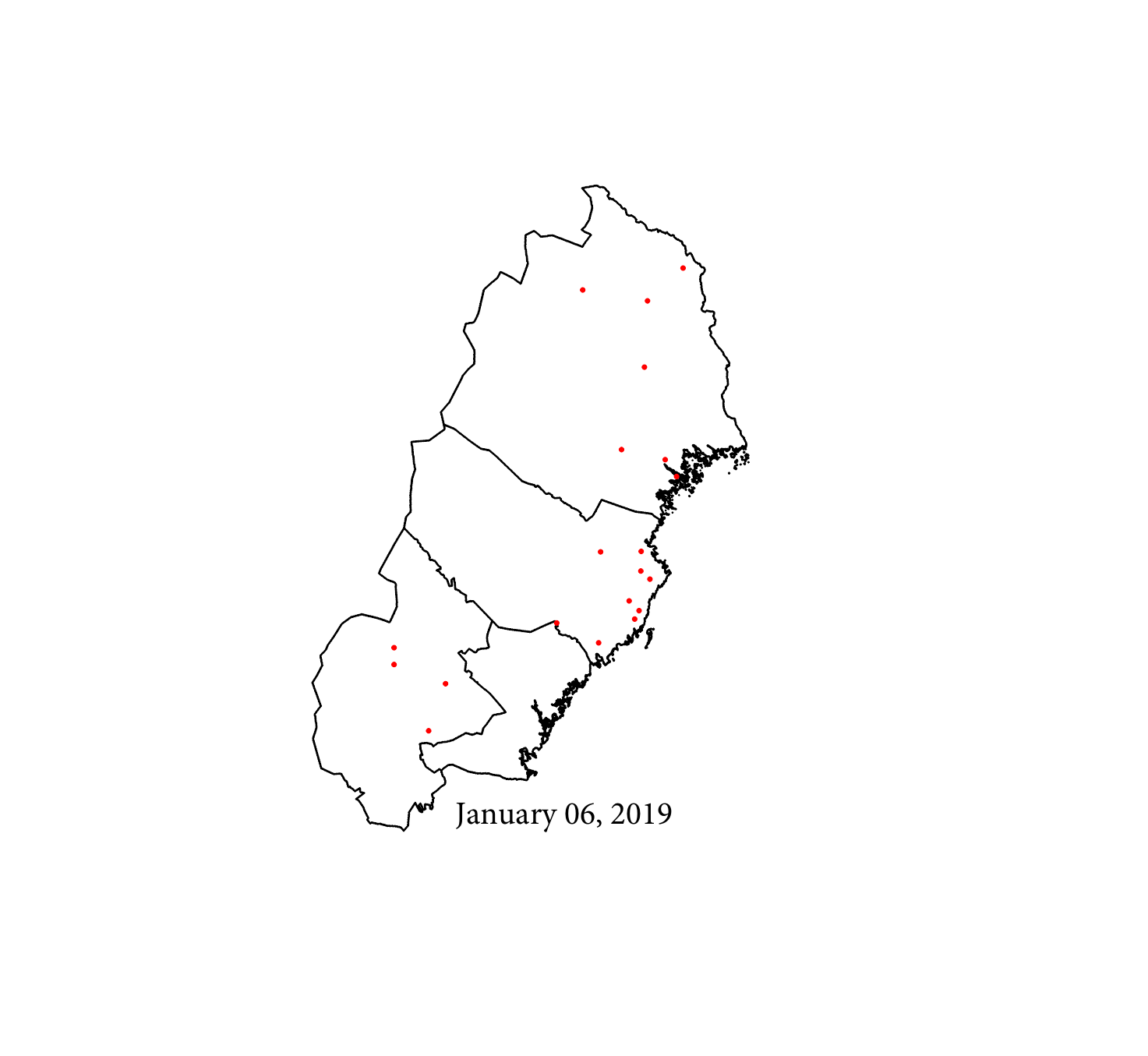}
\includegraphics[width=0.13\textwidth]{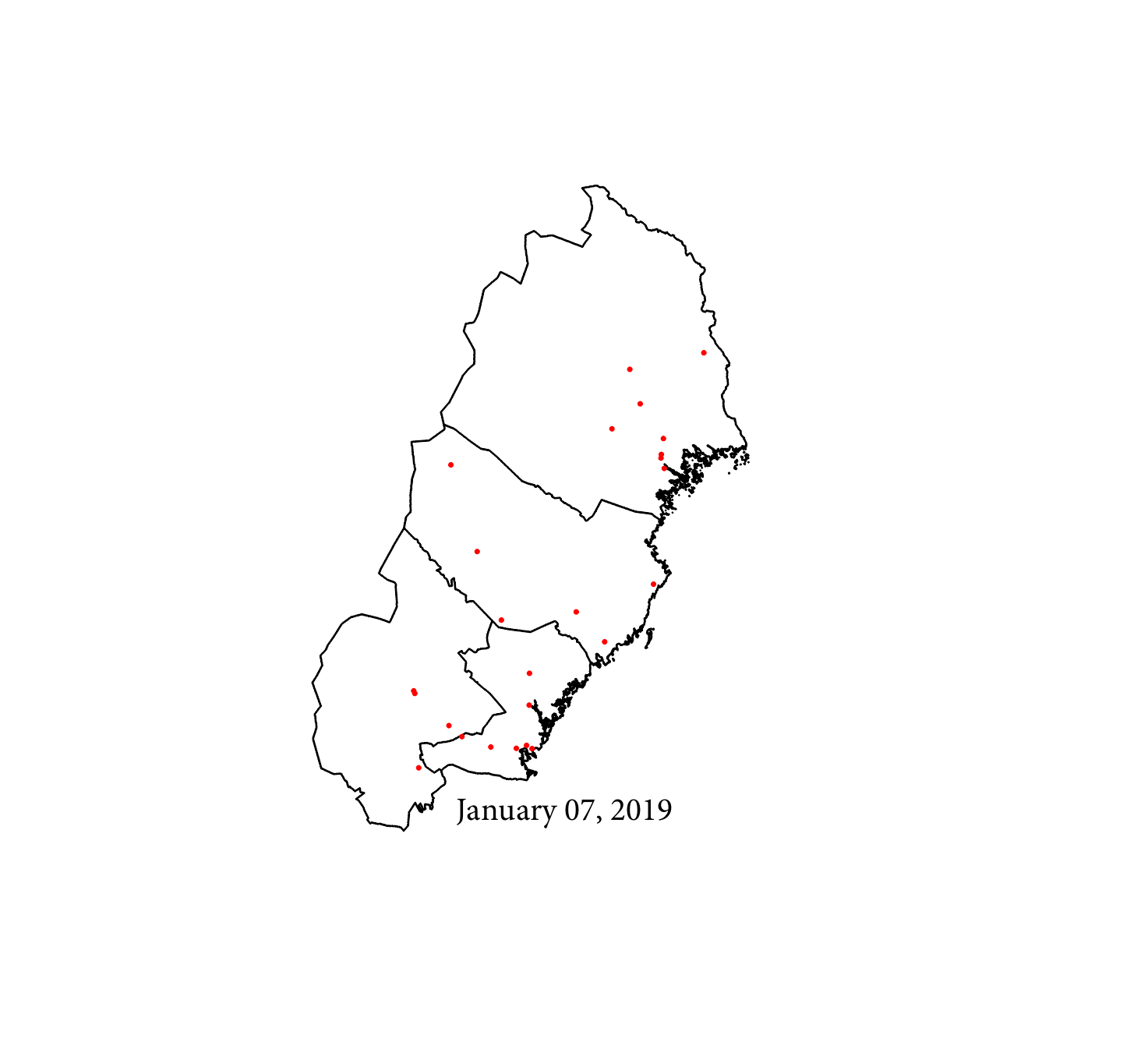}
\caption{Forecasted emergency alarm call locations for the first one week of the year 2019.}
\label{Orgr3}
\end{figure}

Taking things a step further, Figure \ref{Orgr3} shows the simulated realisations beyond the last time point of our data (training data and test datasets), for the days January 01 (a Tuesday) until January 07 (a Monday), 2019. We see that the spatial event distribution tends to change over time.

\section{Discussion}\label{dis}
The main focus of this paper is to study whether spatio-temporal log-Gaussian Cox process (LGCP) models can successfully model ambulance call data and, in addition, whether the associated spatio-temporal forecasts have the ability of generating realistic simulated future ambulance call scenarios. This is relevant e.g.~to help ambulance dispatchers in their daily activities and to design optimal ambulance dispatching rules/strategies. 
Our rather unique spatio-temporal point pattern dataset considered consists of 444 283 ambulance/emergency alarm calls, with associated gps locations and event day recordings, in the four northernmost regions of Sweden over the years 2014 to 2018.

In accordance with e.g.~\citet{diggle2005point}, we consider an LGCP which consists of three different components: a spatial component, $\lambda_{0}(s)$, a temporal component, $\lambda_{1}(t)$, and a separable log-Gaussian random intensity field, $\exp\{Z(s,t)\}$. The former two are purely deterministic components which are designed to take purely spatial and purely temporal variation of the events into account. The stochastic component $\exp\{Z(s,t)\}$, on the other hand, models dependence and spatio-temporal variation of the events. 
Moreover, computational merit can be obtained using separable covariance structure and  exponential form of correlation function for the temporal component \citep{diggle2005point}. 
In addition, assuming an exponential covariance function for the temporal part of the Gaussian random field $Z$ makes it Markovian in time.
 
The parameters of the spatial and temporal parts of the covariance function of $Z$ have been estimated using minimum contrast estimation \citep{moller1998log, davies2013assessing}, whereas quartic kernel smoothing has been utilised to estimate $\lambda_{0}(s)$ for the ambulance data. The bandwidth in the quartic kernel smoothing has been selected using $K$-means clustering, which requires selecting the number of clusters to be used during the estimation of the bandwidth. The proposed $K$-means approach better managed to capture the high concentration of events along the relatively highly populated east coast and the main roads of the region, than did many of the existing intensity estimation approaches \citep{davies2018tutorial,moradi2019resample,ogata2003modelling,ogata2004space}. 
The number of clusters has been selected by visually evaluating whether the spatial intensity generated by the quartic kernel smoothing method shows the spatial variation of the ambulance calls. 
A Poisson regression model, which incorporates different calendar covariates such as day-of-the week and season-of-the-year, has been adopted to model the temporal variation of the calls and the fitting of the parameters has been carried out using the iteratively reweighted least squares method. We find that the fit is good, and a few covariates are significantly governing the behaviour of the model. An estimate of the inhomogeneous spatio-temporal $K$-function of \citet{gabriel2009second} and a spatio-temporal interaction Monte-Carlo test have been used to verify that an LGCP is indeed a sensible choice of model -- there seems to be low/moderate spatio-temporal clustering present in the ambulance data but we hypothesise that there likely is within-day-dependence present (which we cannot quantify here since the temporal sampling of our data is per day).
Regarding the computational aspects, 
a two-dimensional discrete Fourier transform and a fast-Fourier transform have been adopted to reduce the computational cost in simulating the Gaussian random field $Z$. In addition, the Gaussian random field has been simulated using the Metropolis-adjusted Langevin algorithm. 
The fitted LGCP model can be utilised to obtain a forecast distribution of the Gaussian random field and, consequently, a Poisson intensity from which we can simulate realistic future event locations. 

To carry out the forecasting, we have excluded the last six time points in the data and carried out forecasting of the spatial structure for these time points. We find that the patterns of the observed and the simulated spatial locations are very similar. 
However, the number of spatial locations in the simulations are slightly smaller, which may be due to under estimation of the temporal component (intensity) of the model or the discretisation used for the random field.

The core challenges here stem from the nature of the current set of spatio-temporal (big) data. 
First of all, the geographic region is vast and, consequently, the changes in the data vary substantially over the spatial region. This may partly be remedied by splitting up the data set into samples over smaller sub-regions; this will, in effect, have the consequence that the final model will have a spatially varying covariance structure (which seems quite reasonable, given the scope of the data). 
Additionally, since most of the events are located in regions with a higher population density, having access to relevant spatial, e.g.~demographic, covariates (something we are currently working on obtaining) may also improve the fit substantially. A further idea, which likely requires a substantial amount of methodological development, is to transition to linear network point processes \citep{ang2012geometrically,baddeley2015spatial,baddeley2020review, cronie2020network, mateu2019spatio, moradi2019first}.

In conclusion, this preliminary study has shown that spatio-temporal log-Gaussian Cox processes are practically viable for the spatio-temporal modelling and forecasting of ambulance calls. Simulations from such forecasts may, in turn, be exploited in optimising prehospital care resources, e.g.~ by designing ambulance dispatching strategies in an optimal way.

\section*{Acknowledgements}
This work was supported by Vinnova [grant number 2018-00422] and the regions Västerbotten, Norrbotten, Västernorrland and Jämtland-Härjedalen. We are grateful to our project partners SOS Alarm and the regions Västerbotten, Norrbotten, Västernorrland and Jämtland-Härjedalen for fruitful discussions on the Swedish prehospital care. 

\bibliographystyle{apalike}
\bibliography{article2020stp}

\begin{thebibliography}{}

\bibitem[Andrieu and Thoms, 2008]{andrieu2008tutorial}
Andrieu, C. and Thoms, J. (2008).
\newblock A tutorial on adaptive \text{MCMC}.
\newblock {\em Statistics and computing}, 18(4):343--373.

\bibitem[Ang et~al., 2012]{ang2012geometrically}
Ang, Q.~W., Baddeley, A., and Nair, G. (2012).
\newblock Geometrically corrected second order analysis of events on a linear
  network, with applications to ecology and criminology.
\newblock {\em Scandinavian Journal of Statistics}, 39(4):591--617.

\bibitem[Arbia et~al., 2012]{arbia2012clusters}
Arbia, G., Espa, G., Giuliani, D., and Mazzitelli, A. (2012).
\newblock Clusters of firms in an inhomogeneous space: The high-tech industries
  in \text{Milan}.
\newblock {\em Economic Modelling}, 29(1):3--11.

\bibitem[Aringhieri et~al., 2017]{aringhieri2017emergency}
Aringhieri, R., Bruni, M.~E., Khodaparasti, S., and van Essen, J.~T. (2017).
\newblock Emergency medical services and beyond: Addressing new challenges
  through a wide literature review.
\newblock {\em Computers \& Operations Research}, 78:349--368.

\bibitem[Baddeley, 1999]{baddeley1999spatial}
Baddeley, A. (1999).
\newblock Spatial sampling and censoring.
\newblock {\em Stochastic geometry: likelihood and computation}, 2:37--78.

\bibitem[Baddeley et~al., 2000]{baddeley2000non}
Baddeley, A., M{\o}ller, J., and Waagepetersen, R. (2000).
\newblock Non-and semi-parametric estimation of interaction in inhomogeneous
  point patterns.
\newblock {\em Statistica Neerlandica}, 54(3):329--350.

\bibitem[Baddeley et~al., 2020]{baddeley2020review}
Baddeley, A., Nair, G., Rakshit, S., McSwiggan, G., and Davies, T.~M. (2020).
\newblock Analysing point patterns on networks - a review.
\newblock {\em Spatial Statistics}.

\bibitem[Baddeley et~al., 2015]{baddeley2015spatial}
Baddeley, A., Rubak, E., and Turner, R. (2015).
\newblock {\em Spatial point patterns: methodology and applications with R}.
\newblock Chapman and Hall/CRC.

\bibitem[Berman and Diggle, 1989]{berman1989estimating}
Berman, M. and Diggle, P. (1989).
\newblock Estimating weighted integrals of the second-order intensity of a
  spatial point process.
\newblock {\em Journal of the Royal Statistical Society: Series B
  (Methodological)}, 51(1):81--92.

\bibitem[Besag, 1994]{grenander1994representations}
Besag, J. (1994).
\newblock Discussion on 'representations of knowledge in complex systems' (by
  \text{U}. \text{Grenander} and \text{M. I. Miller}).
\newblock {\em Journal of the Royal Statistical Society: Series B
  (Methodological)}, 56(4):549--581.

\bibitem[Blackwell and Kaufman, 2002]{blackwell2002response}
Blackwell, T.~H. and Kaufman, J.~S. (2002).
\newblock Response time effectiveness: comparison of response time and survival
  in an urban emergency medical services system.
\newblock {\em Academic Emergency Medicine}, 9(4):288--295.

\bibitem[Brix and Diggle, 2001]{brix2001spatiotemporal}
Brix, A. and Diggle, P.~J. (2001).
\newblock Spatiotemporal prediction for \text{log-Gaussian Cox processes}.
\newblock {\em Journal of the Royal Statistical Society: Series B (Statistical
  Methodology)}, 63(4):823--841.

\bibitem[Cressie, 1993]{Cressie1993statistical}
Cressie, N. (1993).
\newblock {\em Statistics for spatial data, revised edn}.
\newblock Wiley, New York.

\bibitem[Cressie and Huang, 1999]{cressie1999classes}
Cressie, N. and Huang, H.-C. (1999).
\newblock Classes of nonseparable, spatio-temporal stationary covariance
  functions.
\newblock {\em Journal of the American Statistical Association},
  94(448):1330--1339.

\bibitem[Cronie et~al., 2020]{cronie2020network}
Cronie, O., Moradi, M., and Mateu, J. (2020).
\newblock Inhomogeneous higher-order summary statistics for point processes on
  linear networks.
\newblock {\em Statistics and computing}.

\bibitem[Cronie and S{\"a}rkk{\"a}, 2011]{cronie2011some}
Cronie, O. and S{\"a}rkk{\"a}, A. (2011).
\newblock Some edge correction methods for marked spatio-temporal point process
  models.
\newblock {\em Computational Statistics \& Data Analysis}, 55(7):2209--2220.

\bibitem[Cronie and Van~Lieshout, 2015]{cronie2015aj}
Cronie, O. and Van~Lieshout, M. N.~M. (2015).
\newblock A {$J$}-function for inhomogeneous spatio-temporal point processes.
\newblock {\em Scandinavian Journal of Statistics}, 42(2):562--579.

\bibitem[Cronie and van Lieshout, 2016]{cronie2016summary}
Cronie, O. and van Lieshout, M. N.~M. (2016).
\newblock Summary statistics for inhomogeneous marked point processes.
\newblock {\em Annals of the Institute of Statistical Mathematics},
  68(4):905--928.

\bibitem[Cronie and Van~Lieshout, 2018]{cronie2018non}
Cronie, O. and Van~Lieshout, M. N.~M. (2018).
\newblock A non-model-based approach to bandwidth selection for kernel
  estimators of spatial intensity functions.
\newblock {\em Biometrika}, 105(2):455--462.

\bibitem[Davies and Baddeley, 2018]{davies2018fast}
Davies, T.~M. and Baddeley, A. (2018).
\newblock Fast computation of spatially adaptive kernel estimates.
\newblock {\em Statistics and Computing}, 28(4):937--956.

\bibitem[Davies and Hazelton, 2013]{davies2013assessing}
Davies, T.~M. and Hazelton, M.~L. (2013).
\newblock Assessing minimum contrast parameter estimation for spatial and
  spatiotemporal \text{log-Gaussian Cox processes}.
\newblock {\em Statistica Neerlandica}, 67(4):355--389.

\bibitem[Davies et~al., 2018]{davies2018tutorial}
Davies, T.~M., Marshall, J.~C., and Hazelton, M.~L. (2018).
\newblock Tutorial on kernel estimation of continuous spatial and
  spatiotemporal relative risk.
\newblock {\em Statistics in medicine}, 37(7):1191--1221.

\bibitem[Diggle, 1985]{diggle1985kernel}
Diggle, P. (1985).
\newblock A kernel method for smoothing point process data.
\newblock {\em Journal of the Royal Statistical Society: Series C (Applied
  Statistics)}, 34(2):138--147.

\bibitem[Diggle, 2003]{Diggle2003statistical}
Diggle, P. (2003).
\newblock {\em Statistical analysis of spatial point patterns, 2nd edn}.
\newblock Edward Arnold, London.

\bibitem[Diggle et~al., 2005]{diggle2005point}
Diggle, P., Rowlingson, B., and Su, T.-l. (2005).
\newblock Point process methodology for on-line spatio-temporal disease
  surveillance.
\newblock {\em Environmetrics: The official journal of the International
  Environmetrics Society}, 16(5):423--434.

\bibitem[Diggle, 2013]{diggle2013statistical}
Diggle, P.~J. (2013).
\newblock {\em Statistical analysis of spatial and spatio-temporal point
  patterns}.
\newblock Chapman and Hall/CRC.

\bibitem[Diggle et~al., 1995]{diggle1995second}
Diggle, P.~J., Chetwynd, A.~G., H{\"a}ggkvist, R., and Morris, S.~E. (1995).
\newblock Second-order analysis of space-time clustering.
\newblock {\em Statistical methods in medical research}, 4(2):124--136.

\bibitem[Diggle and Gabriel, 2010]{diggle2010spatio}
Diggle, P.~J. and Gabriel, E. (2010).
\newblock Spatio-temporal point processes.
\newblock In {\em Handbook of Spatial Statistics}, pages 450--462. CRC Press.

\bibitem[Diggle et~al., 2007]{diggle2007second}
Diggle, P.~J., G{\'o}mez-Rubio, V., Brown, P.~E., Chetwynd, A.~G., and Gooding,
  S. (2007).
\newblock Second-order analysis of inhomogeneous spatial point processes using
  case--control data.
\newblock {\em Biometrics}, 63(2):550--557.

\bibitem[Dixon, 2014]{DixonPM}
Dixon, P.~M. (2014).
\newblock {\em Ripley's K Function}.
\newblock American Cancer Society.

\bibitem[Gabriel, 2014]{gabriel2014estimating}
Gabriel, E. (2014).
\newblock Estimating second-order characteristics of inhomogeneous
  spatio-temporal point processes.
\newblock {\em Methodology and Computing in Applied Probability},
  16(2):411--431.

\bibitem[Gabriel and Diggle, 2009]{gabriel2009second}
Gabriel, E. and Diggle, P.~J. (2009).
\newblock Second-order analysis of inhomogeneous spatio-temporal point process
  data.
\newblock {\em Statistica Neerlandica}, 63(1):43--51.

\bibitem[Geyer, 1994]{geyer1994convergence}
Geyer, C.~J. (1994).
\newblock On the convergence of \text{Monte Carlo} maximum likelihood
  calculations.
\newblock {\em Journal of the Royal Statistical Society: Series B
  (Methodological)}, 56(1):261--274.

\bibitem[Gonz{\'a}lez et~al., 2016]{gonzalez2016spatio}
Gonz{\'a}lez, J.~A., Rodr{\'\i}guez-Cort{\'e}s, F.~J., Cronie, O., and Mateu,
  J. (2016).
\newblock Spatio-temporal point process statistics: a review.
\newblock {\em Spatial Statistics}, 18:505--544.

\bibitem[Goreaud and P{\'e}lissier, 1999]{goreaud1999explicit}
Goreaud, F. and P{\'e}lissier, R. (1999).
\newblock On explicit formulas of edge effect correction for \text{Ripley's
  K-function}.
\newblock {\em Journal of Vegetation Science}, 10(3):433--438.

\bibitem[Griffith, 1980]{griffith1980towards}
Griffith, D.~A. (1980).
\newblock Towards a theory of spatial statistics.
\newblock {\em Geographical Analysis}, 12(4):325--339.

\bibitem[Haase, 1995]{haase1995spatial}
Haase, P. (1995).
\newblock Spatial pattern analysis in ecology based on \text{Ripley's
  K-function}: Introduction and methods of edge correction.
\newblock {\em Journal of vegetation science}, 6(4):575--582.

\bibitem[Iftimi et~al., 2019]{iftimi2019second}
Iftimi, A., Cronie, O., and Montes, F. (2019).
\newblock Second-order analysis of marked inhomogeneous spatiotemporal point
  processes: Applications to earthquake data.
\newblock {\em Scandinavian Journal of Statistics}, 46(3):661--685.

\bibitem[Illian et~al., 2008]{illian2008statistical}
Illian, J., Penttinen, A., Stoyan, H., and Stoyan, D. (2008).
\newblock {\em Statistical analysis and modelling of spatial point patterns},
  volume~70.
\newblock John Wiley \& Sons.

\bibitem[Jacobsen et~al., 1993]{jacobsen1993brief}
Jacobsen, M., Niemi, H., Hognas, G., Shiryaev, A., and Melinkov, A. (1993).
\newblock A brief account of the theory of homogeneous \text{Gaussian}
  diffusions in finite dimensions.
\newblock {\em Front. Pure Appl. Probab}, 1:86--94.

\bibitem[Law et~al., 2009]{law2009ecological}
Law, R., Illian, J., Burslem, D.~F., Gratzer, G., Gunatilleke, C., and
  Gunatilleke, I. (2009).
\newblock Ecological information from spatial patterns of plants: insights from
  point process theory.
\newblock {\em Journal of Ecology}, 97(4):616--628.

\bibitem[Li and Zhang, 2007]{li2007comparison}
Li, F. and Zhang, L. (2007).
\newblock Comparison of point pattern analysis methods for classifying the
  spatial distributions of spruce-fir stands in the north-east \text{USA}.
\newblock {\em Forestry}, 80(3):337--349.

\bibitem[Liu et~al., 2007]{liu2007characterizing}
Liu, D., Kelly, M., Gong, P., and Guo, Q. (2007).
\newblock Characterizing spatial--temporal tree mortality patterns associated
  with a new forest disease.
\newblock {\em Forest Ecology and Management}, 253(1-3):220--231.

\bibitem[Loader, 1999]{Load99}
Loader, C. (1999).
\newblock {\em Local Regression and Likelihood}.
\newblock Springer, New York.

\bibitem[Lotwick and Silverman, 1982]{lotwick1982methods}
Lotwick, H. and Silverman, B. (1982).
\newblock Methods for analysing spatial processes of several types of points.
\newblock {\em Journal of the Royal Statistical Society: Series B
  (Methodological)}, 44(3):406--413.

\bibitem[Mateu et~al., 2019]{mateu2019spatio}
Mateu, J., Moradi, M., and Cronie, O. (2019).
\newblock Spatio-temporal point patterns on linear networks: Pseudo-separable
  intensity estimation.
\newblock {\em Spatial Statistics}.

\bibitem[M{\o}ller et~al., 1998]{moller1998log}
M{\o}ller, J., Syversveen, A.~R., and Waagepetersen, R.~P. (1998).
\newblock \text{Log-Gaussian Cox processes}.
\newblock {\em Scandinavian journal of statistics}, 25(3):451--482.

\bibitem[Moller and Waagepetersen, 2003]{moller2003statistical}
Moller, J. and Waagepetersen, R.~P. (2003).
\newblock {\em Statistical inference and simulation for spatial point
  processes}.
\newblock Chapman and Hall/CRC.

\bibitem[Moradi et~al., 2019]{moradi2019resample}
Moradi, M.~M., Cronie, O., Rubak, E., Lachieze-Rey, R., Mateu, J., and
  Baddeley, A. (2019).
\newblock Resample-smoothing of \text{Voronoi} intensity estimators.
\newblock {\em Statistics and Computing}, pages 1--16.

\bibitem[Moradi and Mateu, 2019]{moradi2019first}
Moradi, M.~M. and Mateu, J. (2019).
\newblock First-and second-order characteristics of spatio-temporal point
  processes on linear networks.
\newblock {\em Journal of Computational and Graphical Statistics}, pages 1--21.

\bibitem[Mrkvi{\v{c}}ka et~al., 2014]{mrkvivcka2014two}
Mrkvi{\v{c}}ka, T., Mu{\v{s}}ka, M., and Kube{\v{c}}ka, J. (2014).
\newblock Two step estimation for \text{Neyman-Scott} point process with
  inhomogeneous cluster centers.
\newblock {\em Statistics and Computing}, 24(1):91--100.

\bibitem[Ogata, 2004]{ogata2004space}
Ogata, Y. (2004).
\newblock Space-time model for regional seismicity and detection of crustal
  stress changes.
\newblock {\em Journal of Geophysical Research: Solid Earth}, 109(B3).

\bibitem[Ogata et~al., 2003]{ogata2003modelling}
Ogata, Y., Katsura, K., and Tanemura, M. (2003).
\newblock Modelling heterogeneous space--time occurrences of earthquakes and
  its residual analysis.
\newblock {\em Journal of the Royal Statistical Society: Series C (Applied
  Statistics)}, 52(4):499--509.

\bibitem[O'keeffe et~al., 2011]{o2011role}
O'keeffe, C., Nicholl, J., Turner, J., and Goodacre, S. (2011).
\newblock Role of ambulance response times in the survival of patients with
  out-of-hospital cardiac arrest.
\newblock {\em Emergency medicine journal}, 28(8):703--706.

\bibitem[Pell et~al., 2001]{pell2001effect}
Pell, J.~P., Sirel, J.~M., Marsden, A.~K., Ford, I., and Cobbe, S.~M. (2001).
\newblock Effect of reducing ambulance response times on deaths from out of
  hospital cardiac arrest: cohort study.
\newblock {\em Bmj}, 322(7299):1385--1388.

\bibitem[Pommerening and Stoyan, 2006]{pommerening2006edge}
Pommerening, A. and Stoyan, D. (2006).
\newblock Edge-correction needs in estimating indices of spatial forest
  structure.
\newblock {\em Canadian Journal of Forest Research}, 36(7):1723--1739.

\bibitem[Ripley, 1988]{RipleyBrian1988statistical}
Ripley, B. (1988).
\newblock {\em Statistical inference for spatial processes}.
\newblock Cambrige University Press, Cambridge.

\bibitem[Ripley, 1977]{ripley1977modelling}
Ripley, B.~D. (1977).
\newblock Modelling spatial patterns.
\newblock {\em Journal of the Royal Statistical Society: Series B
  (Methodological)}, 39(2):172--192.

\bibitem[Roberts and Rosenthal, 1998]{roberts1998optimal}
Roberts, G.~O. and Rosenthal, J.~S. (1998).
\newblock Optimal scaling of discrete approximations to \text{Langevin}
  diffusions.
\newblock {\em Journal of the Royal Statistical Society: Series B (Statistical
  Methodology)}, 60(1):255--268.

\bibitem[Roberts and Stramer, 2002]{roberts2002langevin}
Roberts, G.~O. and Stramer, O. (2002).
\newblock Langevin diffusions and metropolis-hastings algorithms.
\newblock {\em Methodology and computing in applied probability},
  4(4):337--357.

\bibitem[Roberts et~al., 1996]{roberts1996exponential}
Roberts, G.~O., Tweedie, R.~L., et~al. (1996).
\newblock Exponential convergence of \text{Langevin} distributions and their
  discrete approximations.
\newblock {\em Bernoulli}, 2(4):341--363.

\bibitem[Stoyan and Stoyan, 1994]{stoyan1994fractals}
Stoyan, D. and Stoyan, H. (1994).
\newblock {\em Fractals, random shapes, and point fields: methods of
  geometrical statistics}, volume 302.
\newblock John Wiley \& Sons Inc.

\bibitem[Taylor et~al., 2013]{taylor2013lgcp}
Taylor, B.~M., Davies, T.~M., Rowlingson, B.~S., Diggle, P.~J., et~al. (2013).
\newblock lgcp: an r package for inference with spatial and spatio-temporal
  log-gaussian cox processes.
\newblock {\em Journal of Statistical Software}, 52(4):1--40.

\bibitem[Terrell, 1990]{terrell1990maximal}
Terrell, G.~R. (1990).
\newblock The maximal smoothing principle in density estimation.
\newblock {\em Journal of the American Statistical Association},
  85(410):470--477.

\bibitem[van Lieshout, 2012]{van2012estimation}
van Lieshout, M.-C.~N. (2012).
\newblock On estimation of the intensity function of a point process.
\newblock {\em Methodology and Computing in Applied Probability},
  14(3):567--578.

\bibitem[Wood and Chan, 1994]{wood1994simulation}
Wood, A.~T. and Chan, G. (1994).
\newblock Simulation of stationary \text{Gaussian} processes in [0, 1] d.
\newblock {\em Journal of computational and graphical statistics},
  3(4):409--432.

\bibitem[Yamada and Rogerson, 2003]{yamada2003empirical}
Yamada, I. and Rogerson, P.~A. (2003).
\newblock An empirical comparison of edge effect correction methods applied to
  \text{K-function} analysis.
\newblock {\em Geographical Analysis}, 35(2):97--109.

\bibitem[Zhou et~al., 2016]{zhou2016predicting}
Zhou, Z., Matteson, D.~S., et~al. (2016).
\newblock Predicting melbourne ambulance demand using kernel warping.
\newblock {\em The Annals of Applied Statistics}, 10(4):1977--1996.

\bibitem[Zhou et~al., 2015]{zhou2015spatio}
Zhou, Z., Matteson, D.~S., Woodard, D.~B., Henderson, S.~G., and Micheas, A.~C.
  (2015).
\newblock A spatio-temporal point process model for ambulance demand.
\newblock {\em Journal of the American Statistical Association},
  110(509):6--15.

\end{thebibliography}

\clearpage
\appendix
\begin{center}
    {\Large Supplementary material}
\end{center}

\section*{Fast Fourier transform for covariance matrix computation}
To implement methods for log-Gaussian Cox processes, it is necessary to discretise the spatial study region $R$. The most common such discretisation strategy is to construct an approximation grid over an encapsulating rectangle $W$ of the study region since $R$ can be an irregular region in $\R^2$. Here, the cell-centroids are given by
\begin{eqnarray*}
C = \left\lbrace \left(s^{x}_{min} + \left(i-0.5\right) \Delta_{x}, s^{y}_{min} + \left( j-0.5\right) \Delta_{y}\right)\mid i = 1, 2, \cdots m; j = 1, 2,\cdots, p \right\rbrace, 
\end{eqnarray*}
where $\Delta_{x}$ is the maximum range spanning $R$ along the x-axis divided by the number of disjoint cells $m$ along the x-axis,  $\Delta_{y}$ is the maximum range spanning $R$ along the y-axis divided by the number of disjoint cells $p$ on the y-axes, $s^{x}_{min}$ and $s^{y}_{min}$ are the minimum x-coordinate and y-coordinate of $W$ and the index $\left(i, j\right)$ represents the cell-centroids of the regular spatial grid over $W$. For any given time, the spatially continuous Gaussian process $Z$ can be approximated by a collection of random variables at cell-centroids of the regular spatial grid over $W$ covering the observation window $R$. This collection of random variables can be ordered lexicographically to obtain a long column vector of random variables $\mathbf{z} = \text{lexo}\{\{z_{i,j}\}^{m,  p}_{i=1,j=1}\} $. In this way, the simulation of the spatially continuous Gaussian process $Z$ can be approximately translated to the simulation of a high-dimensional Gaussian random vector $\mathbf{z}$. Under this assumption, the stationary Gaussian random field $\mathbf{z}$ has mean $\boldsymbol{\mu}$ and covariance structure $\boldsymbol{\Sigma}$ with dimensions $mp \times 1$ and  $mp \times mp$, respectively. The $\left(i, j\right)^{th}$ entry of the covariance matrix $\boldsymbol{\Sigma}$ may be obtained as follows: 
\begin{eqnarray*}
\boldsymbol{\Sigma}_{ij}  = \sigma^{2}r_{\phi}\left(d_{ij} \right),
\end{eqnarray*}
where $i, j\in I$ = $\left\lbrace 1, 2, \cdots,  mp\right\rbrace$,  $d_{ij} = \norm{C_{\mathbf{z}\left(i\right)}-C_{\mathbf{z}\left(j\right)}}$,  and $r_{\phi}$ is the spatial correlation function with parameter $\phi$. The quantities  $C_{\mathbf{z}\left(i\right)}$ and $C_{\mathbf{z}\left(j\right)} \in C$ are the cell-centroids corresponding to the $i^{th}$ and $j^{th}$ element of $\mathbf{z}$ in the lexicographic ordering. A massive computation is required to simulate the Gaussian random field directly even for a coarse regular grid approximation of the study area. Given the discretization, the size of the covariance structure grows dramatically and it is necessary to consider an eigendecomposition of the covariance structure to simulate a correlated multivariate Gaussian random field. 
A method using a two-dimensional discrete Fourier transform and a fast-Fourier transform to reduce the computational cost in simulating a Gaussian random field has been proposed \citep{moller1998log,wood1994simulation}.  A fast-Fourier transform is a quick evaluation algorithm that helps to obtain the eigenvalues of block circulant matrices at a dramatically reduced computational cost. In its current form, the covariance structure $\boldsymbol{\Sigma}$ is not a block circulant matrix. However, a fast-Fourier transform requires the matrix to be block circulant. Therefore, circulant embedding has to be performed using an extended encapsulating rectangular grid, which must be at least twice the original size along both axes of the study region. The cell-centroids of the extended encapsulating regular spatial grid of the study region can be given by
\begin{eqnarray}\label{extendedcentriod}
C^{ext} = \left\lbrace \left(s^{x}_{min} + \left(i-0.5\right) \Delta_{x}, s^{y}_{min} + \left( j-0.5\right) \Delta_{y}\right)\mid i = 1, 2, \cdots M; j = 1, 2,\cdots, N \right\rbrace, 
\end{eqnarray}
where $M \geq 2m$ and  $N \geq 2p$. The number of equally-spaced disjoint cells of the extended encapsulating rectangle of the study region is $MN$. The extended cell-centroid based Euclidean distance definition must correspond to wrapping the extended lattice on a torus and then computing the shortest Euclidean distances to satisfy the block circulant requirement of the covariance structure. Let $\left(u, v\right) = C^{ext}_{\mathbf{z}^{ext}\left(i\right)}$ be an index in $C^{ext}$ corresponding to the $i^{th}$ entry in $\mathbf{z}^{ext}$, which is a column vector obtained by a lexicographic ordering of the Gaussian random fields at the cell-centroids in $C^{ext}$. Denote the shortest distance on the torus between $\left(u_{1}, v_{1}\right) = C^{ext}_{\mathbf{z}^{ext}\left(i\right)}$  and $\left(u_{2}, v_{2}\right) = C^{ext}_{\mathbf{z}^{ext}\left(j\right)}$ by
\begin{align*}
d\left\lbrace \left(u_{1}, v_{1}\right), \left(u_{2}, v_{2}\right) \right\rbrace  = \sqrt{d^{2}_{u_{1}u_{2}}  + d^{2}_{v_{1}v_{2}} },
\end{align*}
where $d_{u_{1}u_{2}}  = \min\left\lbrace \abs{u_{1}-u_{2}}, R_{x} - \abs{u_{1}-u_{2}}\right\rbrace$ and  $d_{v_{1}v_{2}}  = \min\left\lbrace \abs{v_{1}-v_{2}}, R_{y} - \abs{v_{1}-v_{2}}\right\rbrace$. The ranges in the x- and y-directions $R_{x} =$ range$\left(Index^{x}\right)$ and $R_{y} =$ range$\left(Index^{y}\right)$  are obtained from the extended index set 
\begin{align*}
Index = \left\lbrace \left(s^{x}_{min} + i\Delta_{x}, s^{y}_{min} + j\Delta_{y}\right)\mid i = 0, 1, 2, \cdots M; j = 0, 1, 2,\cdots, N \right\rbrace.
\end{align*}
Accordingly, the $\left(i, j\right)^{th}$ entry of the covariance matrix $\boldsymbol{\Sigma}$ can be obtained as follows: 
\begin{eqnarray}\label{centriodbc}
\boldsymbol{\Sigma}^{ext}_{ij}  = \sigma^{2}r_{\phi}\left(d\left\lbrace \left(u_{1}, v_{1}\right), \left(u_{2}, v_{2}\right) \right\rbrace  \right),
\end{eqnarray}
where $i, j\in I$ = $\left\lbrace 1, 2, \cdots, MN\right\rbrace$. The covariance structure obtained from equation \eqref{centriodbc} is a block circulant matrix and therefore, its spectral decomposition can be computed efficiently using a fast-Fourier transform. Let $\mathbf{F}_{M}$ represents a normalised $M\times M$ discrete Fourier transform matrix with entries given by
\begin{eqnarray*}
\mathbf{F}_{L}\left(i, j\right)  = \frac{\exp\left\lbrace -2\pi ij\sqrt{-1}/M\right\rbrace}{\sqrt{M}},
\end{eqnarray*}
where  $i, j \in \left\lbrace 0, 1, \cdots, M-1\right\rbrace$.  The block circulant matrix property of $\boldsymbol{\Sigma}^{ext}$ helps to drive its eigenvalues from an $M \times N$ matrix $\boldsymbol{\Sigma}_{r}$, which is obtained only from the first row of $\boldsymbol{\Sigma}^{ext}$. The eigenvalues corresponding to $\boldsymbol{\Sigma}^{ext}$ can be quickly computed as follows:
\begin{eqnarray*}
\boldsymbol\Omega  = \sqrt{MN}\bar{\mathbf{F}}_{M}\boldsymbol{\Sigma}_{r}\bar{\mathbf{F}}_{N},
\end{eqnarray*}
where $\bar{\mathbf{F}}_{M}$ and $\bar{\mathbf{F}}_{N}$ are $M \times M$ and $N \times N$ complex conjugates of $F_{M}$ and $F_{N}$, and $\boldsymbol{\Sigma}_{r}$ is an  $M \times N$ matrix obtained only from the first row of $\boldsymbol{\Sigma}^{ext}$. On the extended lattice, compute an $M \times N$ matrix $\boldsymbol\Gamma$ as follows: 
\begin{eqnarray*}
\boldsymbol\Gamma=\bar{\mathbf{F}}_{M}\left[ \left[\mathbf{F}_{M}\mathbf{U}\mathbf{F}_{N}\right] \odot \boldsymbol{\Omega}^{\cdot 1/2}\right]\bar{\mathbf{F}}_{N}, 
\end{eqnarray*}
where $\mathbf{U}$ is an $M \times N$ matrix containing independent standard normal variables, the symbol $\odot$ denotes the Hadamard product, and $\boldsymbol{\Omega}^{\cdot 1/2}$ represents element-wise power of the eigenvalues in eigenvalue matrix $\boldsymbol{\Omega}$ to the scalar $1/2$. Using a lexicographically ordered form $\boldsymbol\Gamma^{ext}$ of $\boldsymbol\Gamma$, a realisation of the Gaussian random field $\mathbf{z}^{ext}$ on the extended lattice can be obtained by
\begin{eqnarray*}
\mathbf{z}^{ext}  = \boldsymbol\Gamma^{ext} + \boldsymbol{\mu}^{ext}, 
\end{eqnarray*}
Since the square root of the eigenvalues in $\boldsymbol{\Omega}^{\cdot 1/2}$ is required, the successfulness of the discrete Fourier transform depends on the positive semidefiniteness of $\boldsymbol{\Sigma}^{ext}$. In practice, there is no guarantee for the positive semidefiniteness of $\boldsymbol{\Sigma}^{ext}$. However, this problem occurs rarely provided that suitable values of $M$ and $N$ are used \citep{moller1998log}. In terms of computational speed and efficiency, the fast-Fourier transform algorithms perform their best when the grid cell resolutions $M$ and $N$ are factorised as powers of two  \citep{moller1998log}. To obtain the realised Gaussian random field on the study region, we simply discard the elements of $\mathbf{z}^{ext}$ that correspond to the centroids falling outside the study region $R$.

\section*{The Metropolis-adjusted Langevin algorithm}
The Metropolis-adjusted Langevin algorithm has been suggested by \cite{grenander1994representations} and it was later studied in detail by \cite{roberts1996exponential}. The Metropolis-adjusted Langevin algorithm is a Metropolis-Hasting type of Markov chain Monte Carlo algorithm and it differs from Metropolis-Hasting algorithm in that its proposal distribution for each transition exploits gradient information about the target distribution. At each iteration of Metropolis-adjusted Langevin algorithm, multiple spatial intensities are considered in the spatio-temporal log-Gaussian Cox process due to the effect of the temporal correlation on the spatial variation. Monte Carlo simulation from \eqref{rwe} can be made faster by working with a linear transformation of $\mathbf{z}$. As was stated earlier, fast-Fourier transform algorithms perform their best when the extended grid cell resolution are $M = 2m$ and $N = 2p$, where $m$ and $p$ are some positive integer powers of two. Let $D$ be the rank of the covariance matrix $\boldsymbol{\Sigma}^{ext}$. Set 
\begin{eqnarray*}
	\mathbf{z}_{t}^{ext}  = \mathbf{A}\boldsymbol\gamma_{t} + \boldsymbol\upsilon^{ext},
\end{eqnarray*}
where $\boldsymbol\gamma_{t}$ follows a D-dimensional standard multivariate normal distribution at time $t$, $\mathbf{A}$ is the matrix that diagonalizes the covariance matrix $\boldsymbol{\Sigma}^{ext}$, and  the restriction of $\boldsymbol\upsilon^{ext}$ to $C$ agrees with the mean of $\mathbf{z}_{t}$.

Our goal is to simulate ambulance call locations at the last time point $T$ in the data, given the observed data up to that time. 
In comparison to $\left[\mathbf{z}^{ext}_{T}\mid \mathbf{x}^{ext}_{1:T}\right]$, it is easier to simulate from $\left[\mathbf{z}^{ext}_{1:T}\mid \mathbf{x}^{ext}_{1:T}\right]$ as the distribution is known up to a proportionality constant \citep{brix2001spatiotemporal}. In big data settings, such as in our case, computational efficiency problems can be encountered when simulating $T$ Gaussian random fields at each iteration of the chain. This problem can be minimised by considering only data corresponding to the most recent times \citep{brix2001spatiotemporal}. Let $\zeta = T - \nu + 1$ be a reasonably small positive integer. To overcome the computational efficiency problem, the spatial data at $\zeta$ time points can be utilised to simulate Gaussian random fields from the distribution  $\left[\mathbf{z}^{ext}_{\nu:T}\mid \mathbf{x}^{ext}_{\nu:T}\right]$. The strength of the estimated temporal correlation parameter $\theta$ can be used to suggest the number of time points $\zeta$ to be exploited in the simulation.  Under the assumption that $\left\lbrace \mathbf{z}^{ext}_{t}:t\in \left\lbrace \nu, \cdots, T \right\rbrace \right\rbrace $ is a Markov process and  $\mathbf{x}^{ext}_{t}$ is conditionally independent of $\mathbf{x}^{ext}_{\varrho}$ and $\mathbf{z}^{ext}_{\varrho}$ given  $\mathbf{z}^{ext}_{t}$ for all $t \neq \varrho$,  the target density using $\boldsymbol\gamma_{t}$ can be given by  
\begin{eqnarray}\label{key123}
\left[\boldsymbol\gamma_{\nu:T}\mid \mathbf{x}^{ext}_{\nu:T}\right]
&\propto& \left[\mathbf{x}^{ext}_{\nu:T}\mid \boldsymbol\gamma_{\nu:T}\right]\left[\boldsymbol\gamma_{\nu:T}\right]
= \prod_{t=\nu}^{T}\left[\mathbf{x}^{ext}_{t}\mid \mathbf{z}^{ext}_{t}\right]\left[ \left[\boldsymbol\gamma_{\nu}\right]\prod_{t=\nu+1}^{T}\left[\boldsymbol\gamma_{t}\mid\boldsymbol\gamma_{t-1}\right]\right].
\end{eqnarray}
Since $\boldsymbol\gamma_{t}$ is a Gaussian random field, the probability distributions $\left[\boldsymbol\gamma_{t}\right]$ and $\left[\boldsymbol\gamma_{t}\mid\boldsymbol\gamma_{t-1}\right]$
are Gaussian for each time $t$. The logarithm of the target density $\left[\boldsymbol\gamma_{\nu:T}\mid \mathbf{x}^{ext}_{\nu:T}\right]$ can be expressed explicitly using equation \eqref{key123}. Taking the gradient information of the target density into account during simulation of the Gaussian random field can assist in directing the proposal candidates toward dense regions of the target density. Besides, it can help to achieve superior convergence rates of the Markov chains \citep{moller1998log}. In the Metropolis-Hastings algorithm, \cite{roberts1996exponential} suggested a gradient information based proposal density given by
\begin{eqnarray}\label{key4}
\left[\boldsymbol\gamma_{\nu:T}\mid \boldsymbol\gamma^{(k)}_{\nu:T}\right] = \mathcal{N}\left(\cdot\mid \boldsymbol\gamma^{(k)}_{\nu:T} + 0.5\xi^{2}\nabla\log\left[\boldsymbol\gamma^{(k)}_{\nu:T}\mid \mathbf{x}^{ext}_{\nu:T}\right], \xi^{2}\mathbf{I}\right),
\end{eqnarray}
where $\mathcal{N}\left(\cdot\right)$ represents a Gaussian distribution, $\xi^{2}$ denotes the variance of a mutually independent multivariate Gaussian distribution, $\nabla\log\left[\boldsymbol\gamma^{(k)}_{\nu:T}\mid \mathbf{x}^{ext}_{\nu:T}\right]$ is a $D\times \zeta$ dimensional gradient of the logarithm of the target density, and  $\boldsymbol\gamma^{(k)}_{\nu:T}$ is the current state of the chain. A realisation of  $\boldsymbol\gamma_{\nu:T}$ can be generated from the Gaussian distribution in equation \eqref{key4} to exploit it as a proposal candidate in gradient information based Metropolis-Hastings algorithm. The Metropolis-adjusted Langevin algorithm can be summarized as follows:

\begin{algorithm}[H]
	\caption{The Metropolis-adjusted Langevin algorithm}\label{Oromia2022}
	\begin{algorithmic}[1]	
		\State Select the current state of the chain $\boldsymbol\gamma^{(k)}_{\nu:T}$,  k =0,
		
		\State Generate a proposal candidate from the proposal distribution: \label{Oromia2025}
		\begin{eqnarray*}
		\boldsymbol\gamma' \sim \left[\boldsymbol\gamma_{\nu:T}\mid \boldsymbol\gamma^{(k)}_{\nu:T}\right] = \mathcal{N}\left(\cdot\mid \boldsymbol\gamma^{(k)}_{\nu:T} + 0.5\xi^{2}\nabla\log\left[\boldsymbol\gamma^{(k)}_{\nu:T}\mid \mathbf{x}^{ext}_{\nu:T}\right], \xi^{2}\mathbf{I}\right),
		\end{eqnarray*}
	    \State Compute the quantity:
	     \begin{eqnarray*}
                   \alpha\left(\boldsymbol\gamma', \boldsymbol\gamma^{(k)}\right) = \min\left\lbrace 1, \frac{\left[\boldsymbol\gamma'_{\nu:T}\mid \mathbf{x}^{ext}_{\nu:T}\right]\left[\boldsymbol\gamma^{(k)}_{\nu:T}\mid \boldsymbol\gamma'_{\nu:T}\right]}{\left[\boldsymbol\gamma^{(k)}_{\nu:T}\mid \mathbf{x}^{ext}_{\nu:T}\right]\left[\boldsymbol\gamma'_{\nu:T}\mid \boldsymbol\gamma^{(k)}_{\nu:T}\right]}\right\rbrace,
	    \end{eqnarray*}
      \State Draw $\ell\sim U\left[0, 1\right]$, a uniform distribution on the interval $\left[0, 1\right]$,
		\State Accept or reject the proposal candidate according to:\label{Oromia2w}
		\begin{eqnarray*}
		\boldsymbol\gamma^{(k+1)}= \begin{cases} \boldsymbol\gamma'\hspace{0.1in} \text{if}  \hspace{0.1in} \ell\leq \alpha\left(\boldsymbol\gamma', \boldsymbol\gamma^{(k)}\right),\\
		\boldsymbol\gamma^{(k)}, \hspace{0.1in} \text{otherwise},
		\end{cases}
       \end{eqnarray*}
		\State Repeat steps \ref{Oromia2025} and \ref{Oromia2w} until the desired length of the chain is reached. 
	\end{algorithmic}
\end{algorithm}
It is not easy to set the tuning parameter $\xi^{2}$ in the Metropolis-adjusted Langevin algorithm. There is evidence how to choose the variance $\xi^{2}$ of the proposal distribution to obtain the fastest rate of convergence and an overall acceptance rate of around 0.574 \citep{roberts1998optimal}. By examining some shorter preliminary runs of the algorithm, we can set the value of the tuning parameter.  An algorithm of \cite{andrieu2008tutorial} can be exploited for automatic choice of  $\xi^{2}$ so that the acceptance rate  0.574 can be achieved without disturbing the ergodic property of the chain.
\end{document}